\documentclass[a4paper,10pt]{article}
\usepackage{jheppub} 
\usepackage{hyperref} 
\usepackage[T1]{fontenc} 
\usepackage{graphicx} 
\usepackage{subcaption} 
\usepackage{feynarts} 
\usepackage{float} 
\usepackage{mathrsfs} 

\graphicspath{{./}{./figs/}}

\newcommand{\mr}[1]{\mathrm{#1}}
\newcommand{\mb}[1]{\mathbf{#1}}
\newcommand{\mc}[1]{\mathcal{#1}}
\newcommand{\ms}[1]{\mathscr{#1}}

\newcommand\MSbar{$\overline{\text{MS}}$}
\newcommand\sumint[1]{\hbox{$\sum$}\!\!\!\!\!\!\!\int_{#1}}
\newcommand\ep{\alpha_3}
\newcommand\cond[1]{\langle\bar{\phi}_3^{#1}\rangle}
\newcommand\gE{\gamma_{\rm E}}
\newcommand\Vol{\mc{V}_{3}}
\newcommand{\rmi}[1]{{\mbox{\scriptsize #1}}}

\newcommand\eVac{\varepsilon_{\rm vac}}
\newcommand\rr{r}
\newcommand\rs{r_*}
\newcommand\eqend{\\}
\newcommand\eqendlab[1]{\label{eq:#1}\\}
\newcommand\figwidthfudge{0.5}

\title{Real scalar phase transitions: a nonperturbative analysis}
\author{Oliver Gould}
\affiliation{School of Physics and Astronomy, University of Nottingham, Nottingham NG7 2RD, United Kingdom}
\affiliation{Helsinki Institute of Physics, University of Helsinki, FI-00014, Finland}
\emailAdd{oliver.gould@nottingham.ac.uk}
\date{\today}
\abstract{
We study the thermal phase transitions of a generic real scalar field, without a $Z_2$-symmetry, referred to variously as an inert, sterile or singlet scalar, or $\phi^3+\phi^4$ theory.
Such a scalar field arises in a wide range of models, including as the inflaton, or as a portal to the dark sector.
At high temperatures, we perform dimensional reduction, matching to an effective theory in three dimensions, which we then study both perturbatively to three-loop order and on the lattice.
For strong first-order transitions, with large tree-level cubic couplings, our lattice Monte-Carlo simulations agree with perturbation theory within error.
However, as the size of the cubic coupling decreases, relative to the quartic coupling, perturbation theory becomes less and less reliable, breaking down completely in the approach to the $Z_2$-symmetric limit, in which the transition is of second order.
Notwithstanding, the renormalisation group is shown to significantly extend the validity of perturbation theory.
Throughout, our calculations are made as explicit as possible so that this article may serve as a guide for similar calculations in other theories.
}
\preprint{HIP-2021-2/TH}

\begin{document}

\maketitle

\section{Introduction}

Since the Hot Big Bang, the universe may have passed through a number of different phases.
In the Standard Model (SM) of particle physics, electroweak symmetry breaking and colour confinement took place at temperatures of approximately $\sim$160~GeV~\cite{DOnofrio:2015gop} and $\sim$155~MeV~\cite{Aoki:2006we,Aoki:2009sc,Bazavov:2018mes} respectively, though both these transitions are smooth crossovers.
Extensions of the Standard Model may lead to a wide variety of phases and phase transitions in the early universe.
Such phase transitions may have an importance for baryogenesis~\cite{Kuzmin:1985mm,Shaposhnikov:1986jp,Shaposhnikov:1987tw,Morrissey:2012db}, and may lead to a detectable signal of gravitational waves~\cite{Caprini:2019egz}, allowing the possibility to probe particle physics in a completely new way.
The gravitational waves produced by first-order phase transitions even offer the possibility of studying dark sectors that are uncoupled to the Standard Model~\cite{Schwaller:2015tja,Croon:2018erz,Breitbach:2018ddu}.

Scalar fields, whether fundamental or effective, often lie at the heart of phase transitions, acting as order parameters which take different expectation values in different phases.
In this article, we study the simplest scalar theory which gives rise to a first-order phase transition.
Its Lagrangian is given by,
\begin{align}
 \ms{L}_{\rm singlet} &= \frac{1}{2}(\partial_\mu \phi)^2 - V(\phi)
 \;, \label{eq:lagrangian}\\
 V(\phi) &= \sigma \phi + \frac{1}{2}m^2 \phi^2 + \frac{1}{3!}g \phi^3 + \frac{1}{4!}\lambda \phi^4
 \;, \label{eq:potential}
\end{align}
where $\phi$ is the scalar field, $\mu$ runs over 0,1,2,3 and we have used the mostly minus metric signature.
It has been referred to variously as an inert, sterile or singlet scalar, and in the most part we will refer to it as a singlet scalar.

The singlet scalar may couple to the SM, or other fields, in which case the full Lagrangian takes the form
\begin{equation}
\ms{L} = \ms{L}_{\rm SM} + \ms{L}_{\rm singlet} + \ms{L}_{\rm portal}
\;, \label{eq:portal}
\end{equation}
where the portal sector contains SM-singlet interactions.
The singlet extended SM is referred to as the xSM~\cite{Barger:2007im}.
Indeed couplings to gravity are always present, couplings to the Higgs field are generically expected on effective field theory grounds, and likewise for couplings to sterile neutrinos~\cite{Dodelson:1993je} in minimal extensions of the Standard Model such as the $\nu$MSM~\cite{Asaka:2005pn,Asaka:2005an,Shaposhnikov:2006xi}.
On the other hand, being a single real scalar, there can be no (renormalisable) couplings to gauge fields, or to the charged chiral fermions of the Standard Model.
In this article, we will consider phase transitions in the singlet direction, focusing on the case where the effects of all other fields can be accounted for by modifying the effective couplings of the infrared modes of the singlet.

We are motivated to study this model, in part, because it arises in a wide range of cosmological and particle-physics model building:
providing a possible dark matter candidate~\cite{Silveira:1985rk,Burgess:2000yq,Steele:2013fka,Athron:2017kgt,Ghorbani:2018yfr,Baker:2019ndr},
acting as the inflaton~\cite{Linde:1983gd,Ratra:1987rm,Shaposhnikov:2006xi,Martin:2013tda,Martins:2020oxv}
and providing for electroweak baryogenesis~\cite{Cline:2012hg,Vaskonen:2016yiu}.
As a consequence, it has been the focus of many collider searches~\cite{Patt:2006fw,OConnell:2006rsp,Barger:2007im,Robens:2015gla,Chen:2017qcz,Lewis:2017dme,Fuchs:2020cmm}.
An additional recent attraction to this work has been the possibility of observing a gravitational wave background from such a first-order phase transition~\cite{Espinosa:1993bs,Choi:1993cv,Ham:2004cf,Ashoorioon:2009nf,Espinosa:2011ax,Huang:2016cjm,Vaskonen:2016yiu,Beniwal:2017eik,Alves:2018jsw,Beniwal:2018hyi,Gould:2019qek,Kozaczuk:2019pet} at future gravitational wave detectors, such as LISA~\cite{Audley:2017drz}, DECIGO~\cite{Kawamura:2011zz}, BBO~\cite{Harry:2006fi} and Taiji~\cite{Guo:2018npi}.
Finally the simplicity of the real scalar theory is itself a motivation for its study, as it offers the possibility of carrying out relatively high order calculations explicitly, in turn allowing us to test the convergence and reliability of perturbation theory.

This last point, to test the convergence and reliability of perturbation theory, is another important motivation for this work.
The gravitational wave spectrum produced by a first-order phase transition depends sensitively on the thermodynamics of the transition, which in turn is difficult to calculate reliably.
A recent work~\cite{Croon:2020cgk} investigating this in a minimal extension of the Standard Model found that typical (one-loop) calculations suffer from huge multiplicative uncertainties in the gravitational wave peak amplitude, of order $O(10^{2}-10^3)$ depending on the strength of the transition.
In addition, Ref.~\cite{Croon:2020cgk} identified uncertainties of unknown, but potentially large, numerical importance.
In light of all this, progress in the theoretical methodology is necessary to make robust predictions for future gravitational wave experiments.

Fundamentally, studies of high temperature physics are hampered by the strong coupling of light bosonic modes, which arises as a collective effect of their high occupancies. If at zero temperature $\lambda$ is the perturbative expansion parameter,%
\footnote{Here $\lambda$ stands for a generic loop expansion parameter, such as $e^2$ for a gauge coupling $e$, and not merely the coupling in this specific scalar theory.}
at high temperatures, $T$, the effective expansion parameter for light modes, with mass $m\ll T$, is modified as,
\begin{equation} \label{eq:thermal_coupling}
\lambda \to \lambda\frac{T}{m}
\;.
\end{equation}
For such light modes, the effective coupling constant is much larger than the corresponding zero temperature coupling constant, and for sufficiently light modes the perturbative expansion breaks down altogether~\cite{Linde:1980ts}, as it does for non-Abelian gauge bosons in the symmetric phase.

For scalar fields at sufficiently high temperatures, thermal corrections to the effective mass grow quadratically and always dominate over the zero temperature mass.
At such temperatures the lightest scalar modes have an effective mass $m^2\sim \lambda T^2$, implying that the effective coupling constant of such modes is reduced from $\lambda\to \sqrt{\lambda}$.
By utilising effective theory techniques, such expansions in $\sqrt{\lambda}$ have been carried out to relatively high orders in several theories (see for example Refs.~\cite{Zhai:1995ac,Kajantie:2002wa,Andersen:2004fp,Andersen:2009ct}).

However, in the vicinity of a phase transition, the situation is somewhat more difficult.
Near the critical temperature, there is an approximate cancellation between the tree-level and thermal contributions to the mass of the field undergoing the transition, so that $m^2\lesssim \lambda T^2$ and perhaps even $m^2\ll \lambda T^2$.
This means that the effective coupling constant can be larger even that $\sqrt{\lambda}$, and the perturbative expansion for these light modes can break down altogether.
In this case, the only reliable method of calculation is lattice Monte-Carlo simulation.

Importantly, the potentially nonperturbative physics of the light bosonic modes is universal.
Due to the hierarchy of scales, one can construct an effective field theory (EFT) for just the light modes~\cite{Farakos:1994kx,Braaten:1995cm,Kajantie:1995dw}.
This EFT is defined in 3d, and hence the construction is called {\em (high temperature) dimensional reduction}.
The EFT depends on physics at shorter scales only through its effective parameters.
Thus by studying the EFT with generic parameters, one arrives at results applicable to a wide range of 4d particle physics models.
In this way, one needs only perform lattice Monte-Carlo simulations once, and the results can be recycled again and again;
see for example Refs.~\cite{Andersen:2017ika,Gorda:2018hvi,Niemi:2018asa,Gould:2019qek} where simulations of the SU(2)-Higgs 3d EFT were recycled.

In this article, we make use of dimensional reduction to perform a comprehensive study of phase transitions in which the infrared dynamics is governed by the singlet scalar.
We perform dimensional reduction explicitly to next-to-leading order (NLO), or $O(\lambda^2)$, starting from the 4d singlet scalar theory, and additionally we demonstrate the leading contributions due to couplings to other fields.
In utilising dimensional reduction for carrying out the high temperature resummations, we differ from the widely adopted approach of (one-loop) daisy resummation~\cite{Parwani:1991gq,Arnold:1992rz} (see Ref.~\cite{Senaha:2020mop} for a recent review).
We do so because dimensional reduction offers several advantages over daisy resummation (see Ref.~\cite{Croon:2020cgk} for a recent discussion), in addition to isolating the universal dynamics of the light bosonic modes.

Within the 3d EFT, on the one hand we perform a state-of-the-art perturbative calculation, computing the equilibrium properties of the low-energy effective theory to three-loop order, and utilising renormalisation group improvement.
Together with the dimensional reduction, our perturbative calculation for the minimal model is accurate to $O(\lambda^{5/2})$.
On the other hand, we put the 3d EFT on the lattice and calculate its properties at the critical temperature with relatively high statistics and utilising exact lattice-continuum relations.
Our lattice Monte-Carlo simulations are carried out for parameter choices ranging over two orders of magnitude, allowing us to determine precisely where perturbation theory is reliable, and where it breaks down.

In outline, for a complete, nonperturbative calculation%
\footnote{
In fact only the high temperature EFT is treated nonperturbatively.
We will assume that the full 4d theory is weakly coupled and well defined at zero temperature, in that there is a UV fixed point~\cite{Bond:2017wut,Christiansen:2017gtg}.
Note that an interacting real scalar field, uncoupled to any other fields, suffers from the triviality problem~\cite{Callaway:1988ya}.
}
of the thermodynamics of a given particle-physics model, the following steps are performed:
\begin{enumerate}
\item[(I)] Vacuum renormalisation: matching relations between physical observables and Lagrangian (\MSbar) parameters~\cite{Fleischer:1980ub}. \label{enum:overview_step_1}
\item[(II)] Dimensional reduction: matching relations between 4d theory parameters and effective 3d theory parameters~\cite{Kajantie:1995dw,Braaten:1995cm,Braaten:1995jr} \label{enum:overview_step_2}
\item[(III)] Perturbative study in 3d effective theory: computation of thermodynamic quantities such as the jump of the order parameter and the latent heat~\cite{Braaten:1995cm,Farakos:1994kx,Farakos:1994xh}
\item[(IV)] Framework for lattice Monte-Carlo simulations in 3d: lattice-continuum relations~\cite{Farakos:1994xh,Laine:1995np,Laine:1997dy} and possible $O(a)$ improvement~\cite{Moore:1996bf,Moore:1997np,Moore:2001vf,Arnold:2001ir}
\item[(V)] Lattice Monte-Carlo simulations of thermodynamic quantities in 3d: computations on fixed lattices, statistical analysis and taking the continuum limit \cite{Kajantie:1995kf}
\end{enumerate}
We perform all five steps in this article, as was also done in Ref.~\cite{Laine:2012jy} for an MSSM-like model. 
One of our aims in this is simultaneously to give a bird's eye view of the technical steps required, and to flesh out all the relevant details, so as to offer a concrete guide to performing such calculations in other models. For relevant reviews, see Refs.~\cite{Andersen:2004fp,Jakovac:2016zkg,Laine:2016hma,Ghiglieri:2020dpq}.

Step~(I) does not involve any thermal physics, and hence is rather separate to the others.
As the calculation is standard, we present it in Appendix~\ref{appendix:zero_temperature}.
The following section, Sec.~\ref{sec:dr}, consequently begins with Step~(II), dimensional reduction.
In Sec.~\ref{sec:phase_diagram} we analyse the phase diagram of the real singlet model at the broadest level: using only symmetries and results from the literature regarding the second-order phase transition in the $Z_2$-symmetric model.
In Sec.~\ref{sec:perturbation_theory} we carry out Step~(III), a perturbative analysis of the thermodynamics of the phase transition.
In Sec.~\ref{sec:lattice} we perform Steps~(IV) and (V), carrying out Monte-Carlo simulations for six different parameter points.
Finally in Sec.~\ref{sec:discussion} we discuss the implications of our results, in particular, we compare lattice and perturbation theory to better understand the limits of the validity of perturbation theory.

For completeness, we should note that in this article we do not study bubble nucleation or bubble growth, but content ourselves with studying purely equilibrium properties of the phase transition.
Within dimensional reduction, a framework for studying bubble nucleation on the lattice has been developed in Refs.~\cite{Moore:2000jw,Moore:2001vf}.
As argued in Ref.~\cite{Croon:2020cgk}, following Refs.~\cite{Langer:1969bc,langer1974metastable}, dimensional reduction also offers a natural framework for semiclassical calculations of bubble nucleation.
We plan a follow-up paper to this in which we study the bubble nucleation rate in this model.

\section{Dimensional reduction} \label{sec:dr}

The equilibrium thermodynamics of quantum field theories can be studied in the imaginary time formalism (for a review, see Refs.~\cite{Kapusta:2006pm,Laine:2016hma}).
In this case the fields live in $\mathbb{R}^3\times S^1$, i.e. three infinite spatial directions and one compact `Euclidean time' direction, with length $1/T$.
In the compact direction bosons, such as the scalar $\phi$, satisfy periodic boundary conditions and hence may be expanded in Fourier modes as,%
\footnote{
Fermions satisfy anti-periodic boundary conditions so their corresponding Fourier modes are $2\pi T (n + \tfrac{1}{2})$.
As such, the masses of all fermionic modes scale with $T$ at high temperatures and hence are always integrated out in dimensional reduction.
}
\begin{equation}
\phi(\tau,\mb{x}) = \sum_{n=-\infty}^{\infty} \varphi_n(\mb{x}) \mr{e}^{i 2\pi T n \tau}\;.
\end{equation}
In this context the Fourier modes are referred to as Matsubara modes~\cite{Matsubara:1955ws}. Here $\tau\in (-1/2T,1/2T]$ parameterises the compact direction, $\mb{x}\in \mathbb{R}^3$ parameterises the spatial directions and $i$ runs over the spatial indices, 1, 2, 3. In terms of these modes, the quadratic part of the action for this scalar field takes the form,
\begin{equation}
S_0 = \frac{1}{T}\int d^3\mb{x} \sum_{n=-\infty}^{\infty}\left[ \frac{1}{2}\left(\partial_i \varphi_n\right)^2 + \frac{1}{2}(2\pi T n)^2\varphi_n^2 + \frac{1}{2}m^2\varphi_n^2\right].
\end{equation}
The rest of the scalar potential also provides interactions between Matsubara modes. In sum, one can view the equilibrium thermodynamics of a theory in 3+1 dimensions as the vacuum dynamics of a theory in 3 Euclidean dimensions containing infinitely many fields, $\varphi_n$, with squared masses $m^2 + (2\pi T n)^2$.

At high temperatures, when $2\pi T \gg m$, there is a hierarchy between the masses of the $n=0$ and the $n\neq 0$ modes.
This hierarchy causes some Feynman diagrams to become parametrically larger than their loop counting would suggest, hence requiring resummation.
Dimensional reduction is a means to carry out such resummations in a systematically improvable way.
It consists of integrating out the heavy (or ultraviolet (UV)) $n\neq 0$ modes, to derive an effective theory for the light (or infrared (IR)) $n=0$ mode.

The practical steps of dimensional reduction were worked out independently in Refs.~\cite{Farakos:1994kx,Kajantie:1995dw} and Ref.~\cite{Braaten:1995cm}.
Here we follow the approach of Refs.~\cite{Braaten:1995cm,Braaten:1995jr} more closely, performing the matching in \emph{strict perturbation theory}%
\footnote{
By this we mean that the tadpole and mass terms are treated as interactions in perturbation theory, in addition to the usual cubic and quartic terms.
The free Lagrangian in strict perturbation theory consists of only the kinetic terms.
}
for the generic real scalar theory.
In outline the steps are:
\begin{enumerate}
 \item Write down the most general 3d theory obeying the same internal and spatial symmetries and containing the same number of light bosonic field degrees-of-freedom as the original 4d theory. \label{enum:dr_step_1}
 \item Calculate the static correlation functions for the operators in the Lagrangian in both the original theory and the effective theory. In both cases use strict perturbation theory. \label{enum:dr_step_2}
 \item Determine the coefficients of the 3d effective theory by matching the results of the two theories for momenta $p\sim \sqrt{\lambda}T$. \label{enum:dr_step_3}
\end{enumerate}
In the following we will outline these steps, first for the pure, real singlet scalar model.
Afterwards we will consider the effect of interactions with Standard Model, and other, fields.

Note that the approach we have adopted makes use of the high-temperature approximation in computing thermal loop integrals.
While the validity of this follows naturally from the hierarchy of scales assumed in dimensional reduction,
it is possible to avoid this approximation if necessary~\cite{Laine:2000kv,Laine:2019uua}.

\subsection{Minimal model} \label{sec:dr_minimal}

We start by considering the scenario where the real singlet scalar field is uncoupled to any other field, i.e.\ the Lagrangian is purely $\ms{L}_{\rmi{singlet}}$ given in Eq.~\eqref{eq:lagrangian}.
This will allow us to set out the method, as well as to show many explicit details, given the simplicity of the model.

The accuracy of the matching procedure can be assessed by counting powers of coupling constants.
The couplings of the 3d effective theory will consist of a sum of vacuum contributions and thermal contributions from the $n\neq 0$ modes which have been integrated out.
We will restrict ourselves to sufficiently high temperatures such that the vacuum contributions are not parametrically larger than the thermal contributions. This amounts to the following power counting prescription,
\begin{align*}
\sigma \sim g T^2\;, \qquad m^2 \sim \lambda T^2\;.
\end{align*}
These conditions will generally be satisfied in the vicinity of the critical temperature, at which point thermal and vacuum contributions are balanced.
There remains a final scaling relation between $g$, $\lambda$ and $T$.
For the theory to be perturbative at zero temperature it must be that $g \ll m$, and hence we have that $g \ll \sqrt{\lambda}T$.
As a default assumption for the dimensional reduction, we will take $g \sim \lambda T$.
In this case, at high temperature our power counting prescription agrees with the counting of factors of $1/(4\pi)$.
However we will leave this final scaling relation more freedom than the others, because the order of the phase transition will depend on it.

We perform the matching at NLO,
in which all effective couplings are calculated up to $O(\lambda^2)$ multiplied by appropriate powers of $T$ to make up the dimensions.
This amounts to one-loop matching for the cubic and quartic couplings,
and two-loop matching for the tadpole coupling and mass~\cite{Kajantie:1995dw,Brauner:2016fla}, though depending on the scaling relation between $g$ and $\lambda$ some $Z_2$-breaking terms may be dropped as subdominant.
Reaching this order is crucial for cancelling the leading renormalisation scale dependence.

For the pure real scalar theory, Step~\ref{enum:dr_step_1} is rather straightforward.
The 3d effective theory is simply,
\begin{align}
 \ms{L}_3 &= \frac{1}{2}(\partial_i \phi_3)^2 + V_3(\phi_3)
 \;, \label{eq:lagrangian_3d} \\
 V_3(\phi_3) &= \sigma_3 \phi_3 + \frac{1}{2}m_3^2 \phi_3^2 + \frac{1}{3!}g_3 \phi_3^3 + \frac{1}{4!}\lambda_3 \phi_3^4
 \;. \label{eq:potential_3d}
\end{align}
We have used the subscript 3 to denote quantities pertaining to the 3d effective theory.
Note this theory is a Euclidean field theory, hence the positive sign between derivative and potential terms.

For Step~\ref{enum:dr_step_2}, the philosophy is the following.
The coupling constants of the low energy effective theory account for the effect of the UV modes of the full theory, which have been integrated out.
Thus, in matching the two theories, we are free to treat the IR contributions in any way we choose, as long as we do so in the same way in both theories.
That is because such IR contributions do not contribute to the coupling constants of the effective theory, and by treating the IR contributions in the same way in both theories, their contributions will cancel exactly.
Given this freedom, the simplest way to treat the IR contributions is simply to cut them off in dimensional regularisation.
Further, we are free to expand around any constant background field, including $\phi=0$, as the choice of background will only affect the IR.
These choices significantly simplify practical calculations.

Strict perturbation theory in the full 4d theory is defined by the following split,
\begin{align}
\ms{L}_{\rm free} &= \frac{1}{2}\left(\partial_\tau \phi \right)^2 + \frac{1}{2}\left(\partial_i \phi \right)^2
\;, \label{eq:strict_free} \\
\ms{L}_{\rm int} &= (\sigma + \delta \sigma) \phi + \frac{1}{2}(m^2+\delta m^2)\phi^2 + \frac{1}{3!}(g + \delta g) \phi^3 + \frac{1}{4!}(\lambda+\delta\lambda) \phi^4
\;, \label{eq:strict_interactions}
\end{align}
where we have explicitly shown the counterterms, but not the powers of the scale $\Lambda$, which make up the dimensions in dimensional regularisation.
Note that we adopt the \MSbar\ renormalisation scheme, in which case there is no need for a field renormalisation counterterm, as it receives no divergent contributions.
The omission of the terms proportional to $\sigma$ and $m^2$ from the free Lagrangian is justified, within strict perturbation theory, by the smallness of these coefficients in comparison with the scale $2\pi T$, which characterises the free Lagrangian.

Strict perturbation theory in the low energy effective theory is defined by the following split,
\begin{align}
\ms{L}_{\rm 3, free} &= \frac{1}{2}\left(\partial_i \phi_3 \right)^2
\;, \label{eq:strict_free_3d}\\
\ms{L}_{\rm 3, int} &= (\sigma_3+\delta \sigma_3) \phi_3 + \frac{1}{2}(m_3^2+\delta m_3^2)\phi_3^2 + \frac{1}{3!}g_3 \phi_3^3 + \frac{1}{4!}\lambda_3 \phi_3^4
\;. \label{eq:strict_interactions_3d}
\end{align}
Due to the superrenormalisable nature of this theory in 3d, there are only a finite number of divergent diagrams; see for example Ref.~\cite{Braaten:1995cm}.
The counterterms needed are those shown here, which appear only at two-loop order.
There are further divergent diagrams at four-loop order, but these are independent of the field configuration, contributing only to the cosmological constant, which we omit.

To derive the matching relations, we must choose a suitable set of observables to match between the full theory and the effective theory.
Following Ref.~\cite{Kajantie:1995dw}, we choose these to be the connected, one-particle irreducible (1PI) correlation functions, denoted by $\Gamma^{(k)}$ in the full theory and $\Gamma_3^{(k)}$ in the effective theory.%
\footnote{
In principle the philosophy is to match observables computed in the full theory and in the EFT.
However, it is computationally simpler to match just the 1PI correlation functions, and in doing this no information is lost as long as all tadpoles can be generated by taking derivatives of the 1PI correlation functions within the 3d EFT.
This is possible except if there are heavy scalars present in the full theory which are not present in the 3d EFT.
If such heavy scalars are present, 1PR diagrams of the heavy scalars must be included in the matching~\cite{Brauner:2016fla,Manohar:2020nzp}.
}
These are equal to minus the sum of all connected, 1PI Feynman diagrams with $k$ legs.
We expand around zero background field, not the minimum of the potential, as the difference is anyway projected out by the IR cut off.
The correlation functions are evaluated with soft external momenta: zero Matsubara modes and small spatial momenta, $p \sim \sqrt{\lambda}T$.

\begin{figure}
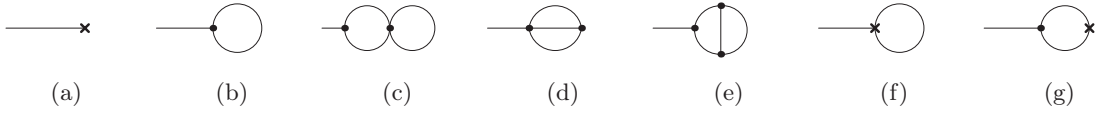

\begin{feynartspicture}(432,75)(7,1)

\FADiagram{(a)} 
\FAProp(2.,10.)(12.5,10.)(0.,){Straight}{0}
\FAVert(12.5,10.){1}

\FADiagram{(b)}
\FAProp(0.,10.)(7.5,10.)(0.,){Straight}{0}
\FAProp(7.5,10.)(7.5,10.)(14.,10.){Straight}{0}
\FAVert(7.5,10.){0}

\FADiagram{(c)}
\FAProp(0.,10.)(3.,10.)(0.,){Straight}{0}
\FAProp(3.,10.)(9.,10.)(1.,){Straight}{0}
\FAProp(3.,10.)(9.,10.)(-1.,){Straight}{0}
\FAProp(9.,10.)(9.,10.)(15.,10.){Straight}{0}
\FAVert(3.,10.){0}
\FAVert(9.,10.){0}

\FADiagram{(d)}
\FAProp(0.,10.)(5.5,10.)(0.,){Straight}{0}
\FAProp(12.5,10.)(5.5,10.)(0.857,){Straight}{0}
\FAProp(12.5,10.)(5.5,10.)(0.,){Straight}{0}
\FAProp(12.5,10.)(5.5,10.)(-0.857,){Straight}{0}
\FAVert(12.5,10.){0}
\FAVert(5.5,10.){0}

\FADiagram{(e)}
\FAProp(0.,10.)(5.5,10.)(0.,){Straight}{0}
\FAProp(9.,13.)(5.5,10.)(0.378,){Straight}{0}
\FAProp(9.,6.5)(9.,13.)(1.077,){Straight}{0}
\FAProp(9.,6.5)(9.,13.)(0.,){Straight}{0}
\FAProp(9.,6.5)(5.5,10.)(-0.434,){Straight}{0}
\FAVert(9.,6.5){0}
\FAVert(9.,13.){0}
\FAVert(5.5,10.){0}

\FADiagram{(f)}
\FAProp(0.,10.)(7.5,10.)(0.,){Straight}{0}
\FAProp(7.5,10.)(7.5,10.)(14.,10.){Straight}{0}
\FAVert(7.5,10.){1}

\FADiagram{(g)}
\FAProp(0.,10.)(7.5,10.)(0.,){Straight}{0}
\FAProp(14.,10.)(7.5,10.)(0.923,){Straight}{0}
\FAProp(14.,10.)(7.5,10.)(-0.923,){Straight}{0}
\FAVert(7.5,10.){0}
\FAVert(14.,10.){1}

\end{feynartspicture}
\caption{Feynman diagrams contributing to the one-point function up to two-loop order. They are shown here in the same order that they appear in Eqs.~\eqref{eq:one-point} and \eqref{eq:one-point-3d-integrals}.}
\label{fig:one-point}
\end{figure}

The one-point correlation function is given, up to two-loop order, by the sum of diagrams shown in Fig.~\ref{fig:one-point}, generated using \emph{FeynArts}~\cite{Hahn:2000kx}.
The results of the relevant loop integrals can be found in the literature and are listed in Appendix~\ref{appendix:loop_integrals}.
Evaluating the diagrams in the strict perturbative expansion, we find,
\begin{align}
\Gamma^{(1)}(\mb{0}) &\approx \sigma + \delta \sigma 
+ \frac{1}{2}g\ \sumint{P}\frac{1}{P^2} 
- \frac{1}{4}g\lambda\ \sumint{PQ}\frac{1}{P^2 Q^4} 
- \frac{1}{6}g\lambda\ \sumint{PQ}\frac{1}{P^2 Q^2 (P+Q)^2} \nonumber \\
&\quad + \frac{1}{4}g^3\ \sumint{PQ}\frac{1}{P^4 Q^2 (P+Q)^2}
+ \frac{1}{2}\delta g\ \sumint{P}\frac{1}{P^2}
- \frac{1}{2} g (m^2 + \delta m^2)\ \sumint{P}\frac{1}{P^4}
\;, \label{eq:one-point} \\
&\approx \sigma
+ \frac{g T^2}{24}
+ \frac{1}{(4\pi)^2}\bigg[
\frac{1}{6}g\lambda T^2 
\left(
\frac{1}{4 \epsilon } 
+ \frac{1}{8}L_b(\Lambda)
+ 6 \log (A)
-\frac{1}{2}\gE
\right) \nonumber \\
&\quad
-\frac{1}{2}g m^2 L_b(\Lambda)
+ \frac{g^3}{(4\pi)^2} \left(
\frac{1}{8}L_b^2(\Lambda)
+\frac{1}{4}L_b(\Lambda)
+\frac{3}{8}
\right)
\bigg]
\;, \label{eq:one-point-result}
\end{align}
where our notation for momenta and loop integration follows Refs.~\cite{Braaten:1995cm,Braaten:1995jr}, and is given in Appendix~\ref{appendix:loop_integrals}.
We use the symbol $\approx$ to denote an equality which holds only in strict perturbation theory up to some loop order.
In Eq.~\eqref{eq:one-point-result} $\gE$ is the Euler-Mascheroni constant and $A$ is the Glaisher-Kinkelin constant.
In order to simplify the formulae, and following Ref.~\cite{Kajantie:1995dw}, we have introduced the notation,
\begin{equation}
L_b(\Lambda) \equiv 2\log \left(\frac{e^{\gE } \Lambda }{4 \pi  T}\right)
\;.
\end{equation}
In going from Eq.~\eqref{eq:one-point} to Eq.~\eqref{eq:one-point-result} we have used the counterterms given in Appendix~\ref{appendix:zero_temperature}, which cancel the temperature-independent divergences.
Being temperature independent, these are the same counterterms that one would find in a calculation at $T=0$.

Performing the same calculation in the effective theory, we find,
\begin{align}
\Gamma_3^{(1)}(\mb{0}) &\approx \sigma_3 + \delta \sigma_3 
+ \frac{1}{2} g_3 \int_p \frac{1}{p^2} 
- \frac{1}{4} g_3 \lambda_3 \int_{pq} \frac{1}{p^2 q^4}  
- \frac{1}{6} g_3\lambda_3 \int_{pq} \frac{1}{p^2 q^2 (\mb{p}+\mb{q})^2} \nonumber \\
&\quad + \frac{1}{4} g_3^3 \int_{pq} \frac{1}{p^4 q^2 (\mb{p}+\mb{q})^2}
+ \frac{1}{2} \delta g_3 \int_p \frac{1}{p^2} 
- \frac{1}{2} g_3 (m_3^2 + \delta m_3^2) \int_p \frac{1}{p^4}
\;,  \label{eq:one-point-3d-integrals}\\
&\approx \sigma_3 + \delta \sigma_3
\;. \label{eq:one-point-3d}
\end{align}
Again, the notation follows Refs.~\cite{Braaten:1995cm,Braaten:1995jr} and is defined in Appendix~\ref{appendix:loop_integrals}.
In going from Eq.~\eqref{eq:one-point-3d-integrals} to \eqref{eq:one-point-3d} we have used that scaleless integrals vanish identically in dimensional regularisation.
Note that the pure zero Matsubara parts of Eq.~\eqref{eq:one-point} give exactly the same scaleless integrals as in Eq.~\eqref{eq:one-point-3d}.
Thus, one can see that the cancellation of the purely IR physics would also occur for other regularisation schemes.
Possible contributions from IR-UV cross terms are essentially projected out by the strict perturbative expansion, which alternatively can be shown to cancel upon performing resummation~\cite{Kajantie:1995dw}.

Calculations of the remaining correlation functions, those with two, three and four external legs, are very similar and are given in Appendix~\ref{appendix:correlation_functions}.
The only real change is that for the two-point function, the leading $O(p^2)$ momentum dependence must also be calculated.
With the results for the correlation functions in hand, we can turn to Step~\ref{enum:dr_step_3} of dimensional reduction, matching.

For Step~\ref{enum:dr_step_3}, we must match to find the field, $\phi_3$, as well as the four parameters of the effective theory, $\sigma_3$, $m^2_3$, $g_3$ and $\lambda_3$.
Naively, one might think that equating the four correlation functions would not give enough conditions, but in fact an extra condition arises from the leading momentum dependence of the two-point function.
This extra condition essentially matches the kinetic operators, while the other four conditions match the potential terms.

To match the momentum dependence of the correlation functions, we must allow for a field normalisation factor between the zero Matsubara mode, $\varphi_0$, and the low energy field operator, $\phi_3$~\cite{Kajantie:1995dw}.
Demanding that the $p^2$ part of the two-point functions agree, once this normalisation is taken into account, implies
\begin{align}
\phi_3^2 &= \frac{1}{T}\frac{\partial\Gamma^{(2)}(\mb{p},-\mb{p})}{\partial p^2}\bigg|_{p^2=0} \varphi_0^2\;,\label{eq:matching_field_1} \\
&= \frac{1}{T}\left(1 + \Pi '(\mb{p},-\mb{p})|_{p^2=0}\right) \varphi_0^2\;, \label{eq:matching_field_2}
\end{align}
where $\Pi$ is the self-energy of the zero Matsubara modes in the full theory, as computed in strict perturbation theory, and on the second line we have introduced the dash to denote the derivative with respect to $p^2$.
The power of $T$ arises so that $\phi_3$ is canonically normalised in three dimensions.

Once the normalisation of the effective field operator is taken care of, matching the Lagrangian parameters simply amounts to equating the correlation functions at zero momentum and taking into account overall powers of $T$ to make up the dimensions.
Doing this leads to the following matching equations,
\begin{align}
\Gamma_3^{(1)}(\mb{0}) &= T^{-1/2}\left(1 - \tfrac{1}{2}\Pi '(\mb{p},-\mb{p})|_{p^2=0}\right)\Gamma^{(1)}(\mb{0})
\;, \label{eq:matching_correlator_1}\\
\Gamma_3^{(2)}(\mb{0}) &= T^{0}\left(1 - \Pi '(\mb{p},-\mb{p})|_{p^2=0}\right)\Gamma^{(2)}(\mb{0},\mb{0})
\;, \label{eq:matching_correlator_2}\\
\Gamma_3^{(3)}(\mb{0},\mb{0},\mb{0}) &= T^{1/2}\left(1 - \tfrac{3}{2}\Pi '(\mb{p},-\mb{p})|_{p^2=0}\right)\Gamma^{(3)}(\mb{0},\mb{0},\mb{0})
\;, \label{eq:matching_correlator_3}\\
\Gamma_3^{(4)}(\mb{0},\mb{0},\mb{0},\mb{0}) &= T\left(1 - 2\Pi '(\mb{p},-\mb{p})|_{p^2=0}\right)\Gamma^{(4)}(\mb{0},\mb{0},\mb{0},\mb{0})
\;. \label{eq:matching_correlator_4}
\end{align}
From Eqs.~\eqref{eq:matching_field_1}-\eqref{eq:matching_correlator_4}, we can read off the effective couplings.
Using the explicit expressions for the correlation functions in Appendix~\ref{appendix:correlation_functions}, and expanding up to $O(\lambda^2)$ and we find the effective parameters.

Before writing the matching relations, we first make a couple of judicious modifications following Refs.~\cite{Braaten:1995cm,Kajantie:1995dw}.
We run the \MSbar\ parameters of the 4d theory from the matching scale $\Lambda$, introduced by dimensional regularisation, to some new renormalisation scale $\mu$.
The beta functions are collected in Appendix~\ref{appendix:zero_temperature}.
This running may be essential to minimise large logarithms in perturbation theory, but can also be used to investigate any scale dependence of our result.
After this, the tadpole and mass parameters retain some $\Lambda$-dependence, which will eventually be cancelled by two-loop diagrams in the 3d EFT.
To extend this cancellation, which occurs at $O(\lambda^2)$, to all-loop orders, we rewrite the coefficients of the $\Lambda$-dependent terms and the 3d counterterms in terms of the 3d effective parameters.
This possibility is a consequence of the superrenormalisability of the 3d theory.

After these modifications, and assuming $g\sim \lambda T$, we arrive at our final result for the matching relations to the 3d EFT up to $O(\lambda^2)$:
\begin{align}
\phi_3^2 &= \frac{\varphi_0^2}{T}
\;,\label{eq:dr_Z} \\
\sigma_3 &= \frac{1}{\sqrt{T}}\bar{\sigma}
+ \frac{\bar{g}T^{3/2}}{24}
+\frac{g_3 \lambda_3}{6(4\pi)^2} \left[ \log \left(\frac{\Lambda }{3 T}\right) - c\right]
\;, \label{eq:dr_1}\\
m^2_3 &= \bar{m}^2 + \frac{\bar{\lambda}T^2}{24}
+\frac{\lambda_3^2}{6(4\pi)^2}\left[ \log \left(\frac{\Lambda }{3 T}\right) - c\right]
\;, \label{eq:dr_2}\\
g_3 &= \sqrt{T}\bar{g}
\;, \label{eq:dr_3}\\
\lambda_3 &= T\bar{\lambda}
\;, \label{eq:dr_4}
\end{align}
where, in order to simplify the formulae, and following Ref.~\cite{Kajantie:1995dw}, we have introduced the constant
\begin{equation}
c \equiv -\log \left(\frac{3e^{\gE/2}A^6}{4 \pi} \right) = -0.348723\dots
\; .
\end{equation}
To make clear the renormalisation scale dependence of the result, we have defined barred couplings $\bar{\kappa}$, with $\kappa\in \{\sigma, m^2, g, \lambda \}$, which are renormalisation scale invariant at $O(\lambda^2)$,
\begin{equation}
\bar{\kappa} \equiv \kappa - \frac{1}{2}\beta_\kappa  L_b(\mu)
\;, \label{eq:rg_invariant_lambda}
\end{equation}
where $\beta_{\kappa}\equiv d\kappa/d\log\mu$ denotes the one-loop beta functions, given in Appendix~\ref{appendix:zero_temperature}. 
The parameters on the right hand side of Eqs.~\eqref{eq:dr_Z} to \eqref{eq:rg_invariant_lambda} have been run to the renormalisation scale $\mu$, e.g.\ $\lambda=\lambda(\mu)$, whereas the 3d effective parameters are defined at $\Lambda$.
The leading corrections to Eqs.~\eqref{eq:dr_Z} to \eqref{eq:dr_4} arise at $O(\lambda^3)$; the contributions of hard modes yield an expansion in integer powers of $\lambda$.

There also remain temperature dependent $1/\epsilon$ poles from the computation in the full theory, which must cancel against identical poles in the EFT.
Matching these, we obtain
\begin{align}
\delta\sigma_3 &= \frac{g_3 \lambda_3}{24 (4\pi)^2\epsilon }
\;, \eqendlab{dr_1ct}
\delta m^2_3 &= \frac{\lambda_3^2}{24 (4\pi)^2 \epsilon }
\;. \label{eq:dr_2ct}
\end{align}

The matching relations, Eqs.~\eqref{eq:dr_1} to \eqref{eq:dr_2ct}, pass several nontrivial checks.
First, the counterterms, which have been derived from loop computations in the full 4d theory at finite temperature, are just the right counterterms required to cancel UV divergences within the 3d EFT;
see Sec.~\ref{sec:effective_potential}.
Second, all dependence on the renormalisation scale $\Lambda$ cancels up to the order that we have calculated.
In particular, the matching relations for the field, the 3-point and the 4-point couplings are independent of the renormalisation scale $\Lambda$,
\begin{align}
\frac{d g_3(\Lambda)}{d\log\Lambda} &= 0
\;, \eqendlab{betafns_g3}
\frac{d \lambda_3(\Lambda)}{d\log\Lambda} &= 0
\;. \label{eq:betafns_lambda3}
\end{align}
For the tadpole and mass, $\sigma_3$ and $m_3^2$, instead we find,
\begin{align}
\frac{d \sigma_3(\Lambda)}{d\log\Lambda} &= \frac{g_3 \lambda _3}{6 (4 \pi) ^2}
\;, \eqendlab{betafns_sigma3}
\label{eq:betafns_m3}
\frac{d m_3^2 (\Lambda)}{d\log\Lambda} &= \frac{\lambda _3^2}{6 (4 \pi) ^2}
\;.
\end{align}
Eqs.~\eqref{eq:betafns_g3} to \eqref{eq:betafns_m3} are in fact the exact beta functions of the 3d EFT;
see Sec.~\ref{sec:perturbation_theory}.
As a consequence, the $\Lambda$-dependence arising from loop calculations within the 3d EFT cancels the $\Lambda$-dependence of the matching relations.
This allows us to exchange $\Lambda$ for some new scale $\mu_3$, chosen for example to minimise large logarithms within perturbative calculations in the 3d theory.
Thus the original scale $\Lambda$ (described as the matching scale in Ref.~\cite{Braaten:1995cm}) all but disappears, to be replaced by two renormalisation scales, $\mu$ and $\mu_3$ which may be chosen independently.

We have also performed some checks against the literature.
We find agreement with Refs.~\cite{Arnold:1992rz,Farakos:1994kx,Braaten:1995cm,Andersen:1997zx} (excluding what is presumably a sign error in Eq.~(60) of Ref.~\cite{Farakos:1994kx}), which provide checks in the $Z_2$-symmetric limit.
Further, we find agreement with Ref.~\cite{Brauner:2016fla}, which provides a check of all our correlation functions at one-loop order.
Finally, we are grateful to the authors of Refs.~\cite{Schicho:2021gca,Niemi:2021qvp}, with whom we have cross-checked the full matching relations.

\subsection{Higgs interactions} \label{sec:dr_higgs}

A singlet scalar may couple to the Standard Model Higgs, $H$, via the following portal couplings,
\begin{equation}
\ms{L}_{\rm portal} = -\frac{1}{2}a_1 \phi H^\dagger H - \frac{1}{2}a_2 \phi^2 H^\dagger H
\;,
\end{equation}
where we have followed the notation of ref.~\cite{Profumo:2007wc} for the interaction terms.
In fact, generically such terms should be included as they are renormalisable and do not explicitly break any symmetries of the theory.
Additionally, if the field $\phi$ is interpreted as the inflaton, there must be nonzero couplings to Standard Model particles in order to reheat the universe after inflation; see for example Ref.~\cite{Bezrukov:2009yw}.

If the temperature $T$ is above, or around, the electroweak symmetry breaking scale, then the zero Matsubara modes of the Higgs will also enter the 3d EFT.
So too will the gauge bosons of the SM, though these do not couple directly to the singlet scalar.
Further, such Higgs-portal couplings will give corrections to Eqs.~\eqref{eq:dr_1} to \eqref{eq:dr_2ct} arising from the nonzero Matsubara modes of the Higgs, which at leading order amount to
\begin{align} \label{eq:dr_higgs1}
\Delta \sigma_3 & = \frac{a_1 T^{3/2}}{12}
\;, &
\Delta m_3^2 & = \frac{a_2 T^{2}}{6}
\;. 
\end{align}
In calculating this correction, we have used the one-loop thermal correlation functions from Ref.~\cite{Brauner:2016fla}.
However, a complete calculation of the dimensional reduction of the SM plus singlet is beyond the scope of this article; for which see Refs.~\cite{Schicho:2021gca,Niemi:2021qvp}.

If the singlet scalar appears around the electroweak scale, the dynamics of the coupled system will be a complicated interplay of the fields, perhaps involving a two-step transition~\cite{Hammerschmitt:1994fn,Laine:1998qk,Patel:2012pi,Vaskonen:2016yiu,Niemi:2020hto}.
This possibility has been studied in Refs.~\cite{Espinosa:1993bs,Ham:2004cf,Ashoorioon:2009nf,Espinosa:2011ax,Vaskonen:2016yiu,Beniwal:2017eik,Alves:2018jsw,Beniwal:2018hyi,Gould:2019qek,Ghorbani:2020xqv}.
While the full two-step transition goes beyond the scope of this article,
our analysis is in principle applicable to a first step in the singlet scalar direction.

In Eq.~\eqref{eq:dr_higgs1} we have assumed the temperature is above, or around, the electroweak symmetry breaking scale.
Below this temperature, the Higgs takes a nonzero vev, modifying the effective couplings of the singlet at tree-level.
As the temperature lowers further, perturbative excitations of the Higgs become exponentially suppressed, so that temperature-dependent Higgs corrections to the singlet EFT can be neglected,
leaving only the tree-level effects of the Higgs vev.

\subsection{Other possible interactions} \label{sec:dr_other}

Due to the lack of gauge charges, a real scalar field cannot couple to gauge fields or to the charged chiral fermions of the Standard Model.
However, a variety of other interactions are possible.

Yukawa interactions with Dirac or Majorana fermions are possible, of the form,
\begin{equation}
\ms{L}_{\rm Yukawa} = -\sum_A y_A \phi \bar{\psi}_A \psi_A
\;,
\end{equation}
where $A$ is a flavour index, $y_A$ are the Yukawa couplings and $\psi_A$ are the fermion fields.
Such a Yukawa term may arise in models with sterile neutrinos~\cite{Dodelson:1993je}, constructed to explain the observed small neutrino masses and to provide a dark matter candidate~\cite{Kusenko:2006rh,Shaposhnikov:2006xi,Petraki:2007gq,Petraki:2008ef,Merle:2013wta,Adulpravitchai:2014xna,Merle:2015vzu,Merle:2015oja,Konig:2016dzg,Manso:2018cba,Baker:2019ndr,DiBari:2020bvn,Kelly:2020aks}.
In the $\nu$MSM, for example, the scalar field $\phi$ plays the role of the inflaton~\cite{Shaposhnikov:2006xi}.
Such terms may also arise in simplified dark matter models involving a fermionic dark matter candidate and real scalar portal~\cite{Abdallah:2015ter}.
In the following, for simplicity, we consider the $\psi_A$ to be Dirac fermions.

Fermions do not have zero Matsubara modes, due to the different boundary conditions implied by Fermi-Dirac statistics.
As a consequence, they cannot enter the 3d effective theory.
However, the Yukawa couplings will give corrections to the matching relations of the low-energy effective theory, which at LO amount simply to
\begin{align} \label{eq:yukawa_dr}
\Delta \sigma_3 &= \sum_A \frac{y_A m_A T^{3/2}}{6}
\;, &
\Delta m_3^2 &= \sum_A \frac{y_A^2 T^2}{6}
\;,
\end{align}where $m_A$ are the tree-level fermion masses.
Here we have assumed that the tree-level masses of the fermions are small compared with the thermal mass scale $\pi T$.
If, on the other hand, the fermions are much heavier than the thermal mass scale, their contributions to $m_3^2$ will be Boltzmann suppressed and can be neglected.
In the intermediate case where the fermion masses are of the same order as the thermal mass scale, the full temperature dependence of the fermionic thermal functions must be retained.

Finally, we should mention that $\phi$ must inevitably couple to gravity, modifying the singlet Lagrangian as,
\begin{equation}
\ms{L}_{\rm singlet} \to \sqrt{-g}\left(\ms{L}_{\rm singlet} - \xi_1\phi R - \frac{1}{2}\xi_2\phi^2 R\right)
\;,
\end{equation}
where $g$ is the metric determinant (not to be confused with the scalar cubic coupling), $R$ is the Ricci scalar and $\xi_1$ and $\xi_2$ are coupling constants.
Treating the metric as a classical, slowly-varying background field, we can see that the Ricci scalar corrects the tadpole and mass terms of the scalar $\phi$ already at tree-level,%
\footnote{
Additionally, there are loop-level gravitational corrections to the running of matter Lagrangian parameters; see for example Refs.~\cite{Buchbinder:1992rb,Herranen:2014cua}.
}
$\Delta\sigma_3=\xi_1 R/\sqrt{T}$ and $\Delta m_3^2=\xi_2 R$,
and hence can potentially drive the scalar field through a phase transition.
This has been studied in the context of a rapid quench, causing a spinodal, or tachyonic, instability for the scalar field in the $Z_2$-symmetric version of this model~\cite{Dimopoulos:2018wfg,Bettoni:2019dcw}.
On the other hand, a sufficiently slow time variation of the Ricci scalar can be incorporated in the effective parameters of the 3d EFT.

We note that the full non-$Z_2$-symmetric scalar field theory, with both Yukawa interactions and nonminimal couplings to gravity has recently been studied in Ref.~\cite{Martins:2020oxv}, in which the scalar $\phi$ acts as the inflaton field.
Under the assumption that the $Z_2$-symmetry breaking terms are small, this model was shown to approximately reproduce the spectral index and tensor-to-scalar ratio of the $Z_2$-symmetric Starobinsky inflation model.

In the following, we will focus on the thermodynamics of the 3d EFT containing only the real singlet scalar.
We will remain agnostic about the full UV theory, and as a consequence will treat the 3d effective parameters, $\{\sigma_3,m_3^2,g_3,\lambda_3\}$, as unknown functions of temperature $T$.

\section{Phase diagram} \label{sec:phase_diagram}

As the temperature changes, the couplings of the low-energy effective theory change accordingly.
Cosmological history therefore traces out a path through the space of effective couplings, a path parameterised by the temperature.
Observables, such as the free-energy density, or the expectation value of the field operator, will also change with temperature, following the variation of the effective couplings.
In this way, the hard, nonzero Matsubara modes, which dictate the temperature dependence of the effective couplings, can drive the soft, zero Matsubara mode through a phase transition.

In this section we learn what we can about the phase diagram of the 3d EFT using only very general arguments.
In Sec.~\ref{sec:order_transition} we determine the structure of the phase diagram from symmetry and the known properties of the $Z_2$-symmetric theory.
In Sec.~\ref{sec:latent_heat} we derive a simple expression for the latent heat, Eq.~\eqref{eq:latent_heat_simple}, which factorises into a product of infrared and ultraviolet parts.

\subsection{Order of the transition} \label{sec:order_transition}

In this article, we are particularly interested in the case where the thermal evolution leads to a first-order phase transition.
For there to be a first-order phase transition, there must be a coexistence of phases in some temperature range. Homogeneous phases of the real scalar theory may be distinguished by the value of the field condensate,
\begin{equation}
\cond{} = \frac{1}{Z_3 \Vol}\int \mc{D}\phi_3\ \int d^3x\ \phi_3(x)\ \mr{e}^{-\int d^3x \ms{L}_3(\phi_3)}
\;,
\end{equation}
where $\Vol$ denotes the volume of space, and $Z_3$ denotes the partition function of the 3d EFT.
The field condensate acts just as the density does in a liquid~$\leftrightarrow$~gas transition, or the gauge-invariant Higgs condensate $\langle H^\dagger H \rangle$ in the electroweak phase transition.

At the critical temperature two different phases are equally likely to occur, their free energies being equal. In the case where the 3d effective field theory consists only of the real scalar field, there is a freedom to shift the field by a (temperature dependent) constant. Because of this freedom, one can find a basis in which $g_3(T)=0$ for all $T$, which can be achieved by shifting
\begin{equation} \label{eq:g3_zero_basis}
\phi_3 \to -\frac{g_3}{\lambda_3} + \phi_3
\;,
\end{equation}
after which the bare potential takes the form,
\begin{equation} \label{eq:V_g30}
V = \underbrace{\bigg(\sigma_3 + \frac{g_3^3}{3 \lambda _3^2} - \frac{g_3 m_3^2}{\lambda _3}\bigg)}_{\textstyle \tilde{\sigma}_3}\phi_3
+ \frac{1}{2} \bigg(\underbrace{m_3^2-\frac{g_3^2}{2 \lambda _3}}_{\textstyle \rr} + \delta m^2_3\bigg)\phi_3^2
+ \frac{1}{4!}\lambda _3 \phi_3^4
\;,
\end{equation}
where we have used that $\lambda_3\delta \sigma_3 = g_3 \delta m_3^2$, and have dropped the constant additive term.
This shift reduces the general theory to the Ising-like $Z_2$-symmetric theory (i.e.~$\phi^4$ theory) in the presence of a finite external field $\tilde{\sigma}_3(T)$.
Thus, if there is a phase transition, the critical temperature, $T_c$, occurs at
\begin{equation} \label{eq:critical_mass}
\tilde{\sigma}_3(T) = 0\;,
\end{equation}
at which point the whole partition function possesses a $Z_2$ symmetry.
Thus, at this point, either there are two identical phases related by the $Z_2$ symmetry, in which case there is a first-order phase transition, or there is only a single phase, in which case there is either a higher-order phase transition, or a crossover.
Due to the symmetry, this is an exact equation determining the critical temperature: it is not corrected at any order in the loop expansion of the 3d EFT.
By using the beta functions of the 3d effective parameters, Eqs.~\eqref{eq:betafns_g3} to \eqref{eq:betafns_m3}, one can see that this equality holds independently of the renormalisation scale.

\begin{figure}
	\centering
	\includegraphics[width=\textwidth]{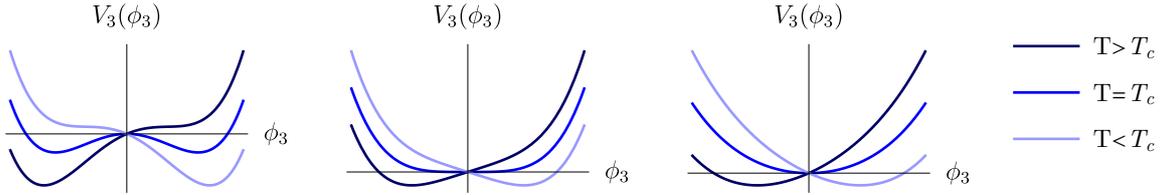}
	\caption{From left to right, the evolution of the tree-level potential in the 3d effective theory for first-order ($\rr<\rs$), second-order ($\rr = \rs$) and crossover ($\rr > \rs$) transitions.
	Here we adopt the basis in which the cubic term is zero, i.e. $\phi_3\to-g_3/\lambda_3+\phi_3$, and choose $\mu_3 = 0.07860 (18) \lambda_3$, such that $\rs(\mu_3)=0$.
	In all cases the system starts at $\cond{}<0$ at high temperatures and transitions to $\cond{}>0$ as the system cools.}
	\label{fig:phase-transition-orders}
\end{figure}

\begin{figure}[t]
	\centering
	\includegraphics[width=0.45\textwidth]{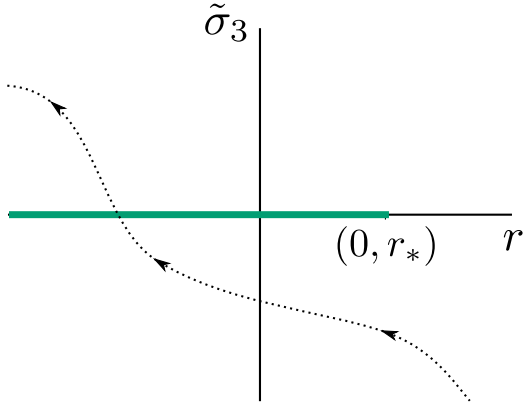}
	\caption{The phase diagram of the 3d effective theory, on the plane of $(\tilde{\sigma}_3,\rr)$, which are the linear and quadratic coefficients of the tree-level potential, in the basis shifted by Eq.~\eqref{eq:g3_zero_basis}.
	The green thick line shows the line of first-order phase transitions which ends at  the second-order critical point at $(0,\rs)$.
	For a given 4d theory, the matching relations of dimensional reduction describe a (curved) line on this plot parameterised by temperature, an example of which is shown here as the dotted line with the arrows denoting the direction of decreasing temperature.
	The order of the phase transition depends on the value of $\rr$ as this line crosses the critical surface $\tilde{\sigma}_3=0$.
	In the example trajectory shown, the transition is first order.
	}
	\label{fig:phase-diagram}
\end{figure}

The nature of the transition, at this critical point, can be found from the value of the coefficient of $\phi_3^2$ in Eq.~\eqref{eq:V_g30} (minus the \MSbar\ counterterm) which we have denoted by $\rr$, following the literature in statistical mechanics~\cite{Wilson:1973jj}.
At tree-level the transition goes from first-order, through second order, and then crossover as the value of $\rr$ goes respectively from negative, through zero, to positive.
This is illustrated in Fig.~\ref{fig:phase-transition-orders}.
Beyond tree-level, the value of $\rr$ for which the transition is second order, $\rs$, shifts away from zero and is renormalisation scale dependent.
In Ref.~\cite{Sun:2002cc}, a lattice Monte-Carlo study of the $Z_2$-symmetric theory, it was found that,
\begin{equation} \label{eq:r_second_order}
\rs(\mu_3) = \left[0.0015249 (48) + \frac{1}{6(4\pi)^2}\log\left(\frac{3\mu_3}{\lambda_3}\right)\right]\lambda_3^2
\;,
\end{equation}
where the number in parenthesis is the statistical uncertainty in the last two digits.
Combining this result with our previous arguments for the generic non-$Z_2$-symmetric theory, in summary we find that
\begin{equation} \label{eq:phase_transition_order}
    \text{phase transition order} = 
\begin{cases}
    \text{first order}, & \text{if } \rr < \rs \\
    \text{second order}, & \text{if } \rr = \rs \\
    \text{crossover}, & \rr > \rs
\end{cases}
\end{equation}
where $\rr$ and $\rs$ are to be evaluated at the critical temperature, i.e. where Eq.~\eqref{eq:critical_mass} holds.
The phase diagram of the theory is shown in Fig.~\ref{fig:phase-diagram}.
By using the beta functions of the 3d EFT, Eqs.~\eqref{eq:betafns_g3} to \eqref{eq:betafns_m3}, in the definition of $\rr$, one can see that these conditions giving the order of the phase transition hold independently of the renormalisation scale.
Eq.~\eqref{eq:phase_transition_order} is exact, up to the statistical uncertainty in the determination of $\rs$.

As one considers weaker and weaker first-order phase transitions, i.e. as $\rr$ tends towards $\rs$ from below, both the jump in the field expectation value and the screening mass tend towards zero.
For asymptotically weak first-order transitions, the approach to the second order point is determined by universality, with the universality class being that of the 3d Ising model.%
\footnote{
Interestingly, this is the same universality class as the 3d EFT of the electroweak theory for its second-order phase transition~\cite{Rummukainen:1998as}, which for the Standard Model field content occurs for a Higgs mass of somewhere in the vicinity of 70-80~GeV.
}
For example, the difference in the field condensate between the two phases, $\Delta \langle \bar{\phi}_3 \rangle$, and the screening mass, $m_s$, are given by
\begin{align}
\Delta \langle \bar{\phi}_3 \rangle &\propto \left( -\rr + \rs\right)^{\beta}
\;, \label{eq:dphi_r}\\
m_s &\propto \left( -\rr + \rs\right)^\nu
\;, \label{eq:ms_r}
\end{align}
with critical exponents $\beta=0.3258(14)$ and $\nu = 0.6304(13)$~\cite{ZinnJustin:2002ru}.

\subsection{Latent heat} \label{sec:latent_heat}

The latent heat, $L$, of a first-order phase transition can be determined by the following thermodynamic relation, evaluated at the critical temperature,
\begin{align}
L &= -\frac{\partial \Delta f}{\partial \log T}
\;, \label{eq:latent_heat_evac}
\end{align}
where $f$ is the free energy density of the full 4d theory,
and $\Delta$ denotes the difference between the two phases.
This in turn can be expressed in terms of $\eVac$, the vacuum energy density of the 3d EFT,
using $\Delta f = T \Delta \eVac$.
By definition $\eVac =-\log Z_3/\Vol$, where $Z_3$ denotes the partition function of the 3d EFT and $\Vol$ denotes the volume of space.

The vacuum energy density of the 3d EFT depends on temperature only through its four effective parameters.
Thus Eq.~\eqref{eq:latent_heat_evac} can be expanded out using the chain rule, in terms of derivatives with respect to these parameters.
From the definition of the vacuum energy, one can see that such derivatives give the condensates of the corresponding operators which appear in the action.
For example, the linear field condensate is given by
\begin{equation} \label{eq:condensate1_definition}
\cond{} = -\frac{\partial \eVac}{\partial \sigma_3}
\;.
\end{equation}
Taking derivatives of $\eVac$ with respect to the bare couplings yields UV divergent bare condensates.
On the other hand, taking derivatives with respect to the renormalised couplings introduces counterterm corrections producing finite renormalised condensates~\cite{Farakos:1994xh}.

In terms of these field condensates, the latent heat is then
\begin{equation}
\frac{L}{T} =  \frac{\partial \sigma_3}{\partial \log T}\Delta\cond{}
+ \frac{1}{2}\frac{\partial m_3^2}{\partial \log T}\Delta\cond{2}
+ \frac{1}{3!}\frac{\partial g_3}{\partial \log T}\Delta\cond{3}
+ \frac{1}{4!}\frac{\partial \lambda_3}{\partial \log T}\Delta\cond{4}
\;,
\end{equation}
all evaluated at the critical temperature.
This expression for the latent heat can be simplified significantly by shifting the origin of the field following Eq.~\eqref{eq:g3_zero_basis}.
In this basis, the coefficient of the cubic term vanishes by construction.
Further, due to the $Z_2$ symmetry at the critical temperature, condensates of even powers of the field are equal in both phases and hence
\begin{equation}
\Delta \left\langle \left(\bar{\phi} + \frac{g_3}{\lambda_3}\right)^2 \right\rangle =
\Delta \left\langle \left(\bar{\phi} + \frac{g_3}{\lambda_3}\right)^4 \right\rangle =
... = 0
\;, \label{eq:condensates_even}
\end{equation}
where, for clarity, we have shown the shift of Eq.~\eqref{eq:g3_zero_basis} explicitly.

Thus, in this basis, the latent heat simplifies to,
\begin{equation} \label{eq:latent_heat_simple}
\frac{L}{T} =  \frac{\partial \tilde{\sigma}_3}{\partial \log T}\Delta\cond{}
\;,
\end{equation}
where $\tilde{\sigma}_3$ is the coefficient of $\phi$ about $-g_3/\lambda_3$ in the potential, given explicitly in Eq.~\eqref{eq:V_g30}.
Eq.~\eqref{eq:latent_heat_simple} shows the factorisation of IR ($\Delta\cond{}$) and UB ($\partial \tilde{\sigma}_3/\partial \log T$) contributions.
While the IR part is universal, the UV part receives contributions from all the modes which were integrated out in the construction of the 3d EFT.

\section{Phase transition in perturbation theory} \label{sec:perturbation_theory}

Perturbation theory is applicable rather generically to the nonzero Matsubara modes, as long as the theory is perturbative at $T=0$.
However the infrared physics of the zero Matsubara mode can become nonperturbative at high temperatures.
In this section, we will investigate the applicability of perturbation theory to the infrared EFT, and apply it to the computation of various equilibrium thermodynamic properties.
In particular, in Sec.~\ref{sec:loop_expansion} we will consider the general form of the loop expansion within the 3d EFT.
Then in Sec.~\ref{sec:effective_potential} we will compute the effective potential to three-loop order.
Using this result, and performing an explicit $\hbar$-expansion, we will compute the discontinuity of the order parameter in Sec.~\ref{sec:condensates_hbar}, and discuss its renormalisation group improvement in Sec.~\ref{sec:rgi}.

\subsection{The loop expansion near \texorpdfstring{$T_c$}{Tc}} \label{sec:loop_expansion}

For a consistent perturbative expansion, one should expand around a background field which is a minimum of the tree-level action.
For homogeneous phases, this can be achieved by shifting the origin of the field to set $\sigma_3$ to zero.

The couplings in the 3d effective theory both have positive mass dimension. As such, each successive loop order comes with either a factor of $\lambda_3$ or $g_3^2$, as well as compensating powers of the tree-level mass parameter, making the dimensionless combinations
\begin{equation}
\frac{\lambda_3}{m_3}\;,\ \frac{g_3^2}{m^3_3}
\;.
\end{equation}
We are interested in the case $m_3 \sim m_{3,c}$. Using Eq.~\eqref{eq:critical_mass} we find that both loop-expansion parameters are of the same order, and the combination of parameters which dictates the convergence of the 3d loop expansion near the critical temperature is
\begin{equation} \label{eq:expansion_parameter}
\ep \equiv \frac{\hbar}{4\pi}\frac{\lambda_3^{3/2}}{|g_3|}\;, \quad \text{for } \sigma_3=0
\;.
\end{equation}
Shifting the field back to a generic point with nonzero $\sigma_3$, the loop-expansion parameter becomes complicated by the introduction of cubic roots, but as this is simply a change of basis, it does not change the underlying physics.
For a generic basis, but at the critical temperature, i.e. at $\tilde{\sigma}_3 = 0$, we are also able to find a simple form for the loop-expansion parameter,
\begin{align} \label{eq:critical_expansion_parameter}
\ep &\equiv \frac{\hbar}{4\pi}\frac{\lambda_3^{1/2}}{|v_{0}|}\;, \quad \text{for } \tilde{\sigma}_3 = 0
\;, \\
v_0^2 &= \frac{g_3^2}{\lambda_3^2}-\frac{6 \sigma _3}{g_3}
\;. \label{eq:v0}
\end{align}
where $v_0$ is the field-value of the tree-level minima about the $Z_2$-symmetric origin.

From the above, it would appear that the loop expansion breaks down in the $Z_2$-symmetric limit, $g_3,\sigma_3 \to 0$.
This is related to the fact that the phase transition is of second order in the $Z_2$-symmetric limit, and hence the scalar mass and the jump in the order parameter both go to zero.
As these quantities enter the loop-expansion parameter inversely,
the loop expansion breaks down in this limit.
Thus, one must either find some other expansion parameter or resort to nonperturbative methods.

At the critical temperature in this model, the two phases are identical and hence the expansion parameter is the same in both phases. This is unlike the case of the Standard Model Higgs field, for which the presence of perturbatively massless gauge bosons causes the loop expansion to fail around the symmetric minimum, though it may converge well around the broken minimum~\cite{Farakos:1994kx,Farakos:1994xh,Laine:1994zq}.

Assuming the scaling relations of Sec.~\ref{sec:dr_minimal}, it follows that $\ep \sim \sqrt{\lambda}$.
This should be contrasted with the matching relations of dimensional reduction, in which only integer powers of $\lambda$ arise.
To see the consequences of this, consider Eq.~\eqref{eq:latent_heat_simple} for the latent heat.
The UV factor in Eq.~\eqref{eq:latent_heat_simple} yields an expansion in powers of $\lambda$, starting at $O(\lambda)$, with the dimensions made up by powers of $T$.
The IR factor instead yields an expansion in powers of $\sqrt{\lambda}$ starting at $O(\lambda^0)$.
As a consequence, to achieve a given accuracy for the latent heat, one must work to higher loop orders for the infrared physics.

\subsection{The effective potential} \label{sec:effective_potential}

In calculating the effective potential of the 3d EFT, we start with the bare Lagrangian, given in Eqs.~\eqref{eq:lagrangian_3d} and \eqref{eq:potential_3d}.
Unlike in Sec.~\ref{sec:dr}, where we were only interested in UV physics at the scale $\sim 2\pi T$, in studying the IR physics of the phase transition we must make a different split between free and interacting terms.
In particular, the tadpole and mass terms now enter the free Lagrangian.

\begin{figure}
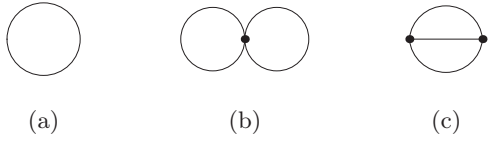

\begin{feynartspicture}(432,75)(3,1)

\FADiagram{(a)}
\FAProp(6.,10.)(6.,10.)(14.,10.){Straight}{0}

\FADiagram{(b)}
\FAProp(10.,10.)(10.,10.)(17.,10.){Straight}{0}
\FAProp(10.,10.)(10.,10.)(3.,10.){Straight}{0}
\FAVert(10.,10.){0}

\FADiagram{(c)}
\FAProp(6.,10.)(14.,10.)(-0.917,){Straight}{0}
\FAProp(6.,10.)(14.,10.)(0.,){Straight}{0}
\FAProp(6.,10.)(14.,10.)(0.917,){Straight}{0}
\FAVert(6.,10.){0}
\FAVert(14.,10.){0}

\end{feynartspicture}
\caption{Feynman loop diagrams contributing to the effective potential up to two-loop order.
They are shown here in the same order that they appear in Eqs.~\eqref{eq:Veff_3d_integrals} and \eqref{eq:Veff_3d}.}
\label{fig:effective-potential}
\end{figure}

In the study of first-order phase transitions, the convex effective potential defined by a Legendre transform~\cite{Jackiw:1974cv} is not particularly relevant.
Instead it is appropriate to define the effective potential as the result of integrating over all non-zero momentum modes, following Ref.~\cite{Fukuda:1974ey} (see also Ref.~\cite{Laine:2016hma}).
Only the 1-particle-irreducible diagrams contribute~\cite{Fukuda:1974ey},
of which there is
one at one-loop,
two at two-loop
and six at three-loop order.
The one- and two-loop diagrams are shown in Fig.~\ref{fig:effective-potential}, and the three-loop diagrams can be found in Ref.~\cite{Rajantie:1996np}.

For the effective potential up to two-loop order, we find
\begin{align}
V_{\rm 3, eff} &=
\sigma_3 \phi_3
+ \frac{1}{2}m_3^2\phi_3^2
+ \frac{1}{3!}g_3\phi_3^3
+ \frac{1}{4!}\lambda_3\phi_3^4
-\frac{\hbar}{2}\int_p \log\left(p^2+M_3^2\right)
 + \frac{\hbar^2}{8}\lambda_3 \int_{pq}\frac{1}{(p^2+M_3^2)(q^2+M_3^2)}\nonumber \\
&\quad
+ \frac{\hbar^2}{12}G_3^2\int_{pq} \frac{1}{(p^2+M_3^2)(q^2+M_3^2)((\mb{p}+\mb{q})^2+M_3^2)}
+ \delta V_3
+ \delta \sigma_3 \phi_3 
+ \frac{1}{2}\delta m_3^2 \phi_3^2
+ O(\hbar^3)
\;, \label{eq:Veff_3d_integrals}
\end{align}
where, for the second and third derivatives of the tree-level potential, we have defined
\begin{align}
M_3^2 &= m_3^2 + g_3 \phi_3 + \frac{1}{2}\lambda_3 \phi_3^2
\;, &
G_3 &= g_3+\lambda_3  \phi_3
\;.
\end{align}
The required loop integrals are collected in Appendix~\ref{appendix:loop_integrals}.
At two-loop order, there is one logarithmically divergent diagram: the sunset.
The corresponding counterterms are,
\begin{align}
\delta V_{3} &= \frac{\hbar^2g_3^2}{48 (4\pi) ^2 \epsilon }\;, &
\delta \sigma_{3} &=  \frac{\hbar^2 g_3 \lambda_3  }{24 (4\pi) ^2 \epsilon }\;, &
\delta m_{3}^2 &= \frac{\hbar^2 \lambda_3 ^2}{24 (4\pi) ^2 \epsilon }
\;.
\end{align}
Due to the superrenormalisability of the theory, the $\delta \sigma_3$ and $\delta m_3^2$ counterterms are in fact exact, to all orders in $\hbar$.
However, the constant $\delta V_3$ counterterm will receive further contributions at four-loop order.

Note that these counterterms match the temperature-dependent divergences that we found in dimensional reduction, Eqs.~\eqref{eq:dr_1ct} and \eqref{eq:dr_2ct}.
Further, by demanding that the bare parameters are independent of the scale $\Lambda$, we recover the beta functions of the 3d parameters that we found in dimensional reduction, Eqs.~\eqref{eq:betafns_g3} to \eqref{eq:betafns_m3}.

After inserting the results for the loop integrals from Appendix~\ref{appendix:loop_integrals}, Eq.~\eqref{eq:Veff_3d_integrals} becomes
\begin{align}
V_{\rm 3, eff} &= \sigma_3 \phi_3 +  \frac{1}{2}m_3^2\phi_3^2 + \frac{1}{3!}g_3\phi_3^3+ \frac{1}{4!}\lambda_3\phi_3^4 -\frac{\hbar}{3(4\pi)}M_3^3 \nonumber \\
&\quad + \frac{\hbar^2}{(4\pi) ^2} \left(\frac{1}{8}\lambda_3 M_3^2 + \frac{1}{12}G_3^2\left(\log\left(\frac{3 M_3}{\mu_3 }\right)-\frac{1}{2}\right) \right) + O(\hbar^3)
\;, \label{eq:Veff_3d}
\end{align}
where $\mu_3$ is the \MSbar\ renormalisation scale of the 3d EFT.
The $Z_2$-symmetric limit of Eq.~\eqref{eq:Veff_3d} agrees with that in Refs.~\cite{Farakos:1994kx,Braaten:1995cm}.

The three-loop contribution can be constructed using the results of Ref.~\cite{Rajantie:1996np}, and reads,
\begin{align}
V_{\rm 3, eff}^{\rmi{3-loop}} &= \frac{\hbar^3}{(4 \pi) ^3}\Bigg(
\frac{\lambda _3^2 M_3}{36} 
   \left(3 \log \left(\frac{\mu_3 }{4 M_3}\right)+\frac{27}{8}\right)
+\frac{G_3^2 \lambda _3}{216 M_3} \left(-27 \text{Li}_2\left(\frac{1}{4}\right)-\frac{9}{2}+\frac{9 \pi
   ^2}{4}-\frac{27}{2} \log ^2\left(\frac{4}{3}\right)\right)
\nonumber \\
&\quad
+\frac{G_3^4}{1296 M_3^3} \left(\frac{81}{2} \text{Li}_2\left(\frac{1}{4}\right)-\frac{27
   \pi ^2}{8}+\frac{81}{4} \log ^2\left(\frac{4}{3}\right)+54 \log \left(\frac{4}{3}\right) 
-27 \sqrt{2} \xi   
   \right)
\Bigg)
\;, \label{eq:V_3loop}
\end{align}
where $\xi=0.03074157526289594\dots$ is the result of performing a 1d integral numerically.
For the special case $G_3 = 0$, this result reproduces Eq.~(34) of Ref.~\cite{Braaten:1995cm}.
Note that for the power counting relations we assumed in Sec.~\ref{sec:dr}, this three-loop contribution is of order $O(\lambda^{5/2})$.
Hence it is of lower order than the terms neglected in the dimensional reduction of Sec.~\ref{sec:dr_minimal}, which are of $O(\lambda^{3})$.

\subsection{Condensates in the \texorpdfstring{$\hbar$}{hbar}-expansion} \label{sec:condensates_hbar}

The vacuum energy of the EFT determines the free energy, latent heat and field condensates.
The vacuum energy is the sum of all connected diagrams, whereas the effective potential is the sum of the connected, 1PI diagrams.
In the vicinity of a minimum of the tree-level potential, 
the 1-particle-reducible (1PR) diagrams, which are missing from the effective potential, can be generated from the 1PI diagrams by performing a strict $\hbar$ expansion~\cite{Fukuda:1975di} (see also Refs.~\cite{Farakos:1994xh,Laine:1994zq,Laine:1995np,Patel:2011th}).
In this, one expands the effective potential and the vev in powers of $\hbar$,
\begin{align}
	V_{\rm 3, eff}(v_3) &= \sum_{n=0}^N \hbar^n V_{(n)}(v_3)
    \;, &
    v_3 &= \sum_{n=0}^N \hbar^n v_{(n)}
    \;,
\end{align}
and solves for the vev order-by-order in $\hbar$.
This approach avoids the spurious imaginary parts which arise when one simply numerically minimises the effective potential~\cite{Weinberg:1987vp,Delaunay:2007wb}.
In gauge theories, it also gives rise to gauge-invariant results order-by-order in $\hbar$~\cite{Nielsen:1975fs,Fukuda:1975di}.
Note, however, that the $\hbar$-expansion is not applicable to transitions that are radiatively-induced within the 3d EFT.
For instance, in the SU(2)-Higgs model at two-loop order in the $\hbar$-expansion, the critical temperature is IR divergent and the latent heat receives an unphysical imaginary part~\cite{Laine:1994zq}.

In calculating the condensates, it is convenient to shift the origin of the field according to Eq.~\eqref{eq:g3_zero_basis}, so that the third derivative of the tree-level potential vanishes,
About this origin, and at the critical temperature the tree-level minima are located at $v_{(0)}=\pm v_0$, where $v_0$ is given by Eq.~\eqref{eq:v0}.
Expanding around this tree-level result to $O(\hbar^3)$, one finds
\begin{align}
v_{(1)} &= \frac{-V'_{(1)}}{V''_{(0)}}
\;, \label{eq:v_hbar} \\
v_{(2)} &= - \frac{V_{(1)}'{}^{2}V'''_{(0)}}{2V''_{(0)}{}^3}
+ \frac{V_{(1)}''V_{(1)}' }{V''_{(0)}{}^2}
-\frac{V_{(2)}'}{V''_{(0)}}
\;,  \eqend
v_{(3)} &= -\frac{V'_{(1)}{}^3 V_{(0)}'''{}^2}{2 V_{(0)}''{}^5}
+\frac{3V'_{(1)}{}^2 V_{(1)}'' V_{(0)}'''}{2V_{(0)}''{}^4}
+\frac{V_{(1)}'{}^3 V_{(0)}'''' }{6V_{(0)}''{}^4}
-\frac{V'_{(1)}{}^2  V_{(1)}'''}{2V_{(0)}''{}^3}
\nonumber \\
&\qquad
-\frac{V'_{(2)} V'_{(1)} V_{(0)}'''}{V_{(0)}''{}^3}
-\frac{V_{(1)}''{}^2 V'_{(1)}}{V_{(0)}''{}^3}
+\frac{V_{(2)}'' V'_{(1)} }{V_{(0)}''{}^2}
+\frac{V'_{(2)} V_{(1)}''}{V_{(0)}''{}^2}
-\frac{V'_{(3)}}{V_{(0)}''}
\;,
\end{align}
where all the functions on the right-hand sides of the equations are evaluated at the tree-level minimum.
Evaluating these, we find
\begin{align}
\Delta \cond{} &= 
2 v_0
+\frac{\hbar\sqrt{3\lambda _3}}{4 \pi }
+\frac{\hbar^2\lambda _3}{2(4\pi)^2 v_0}
\left[1 + \log \left(\frac{\mu_3^2}{3\lambda_3 v_0^2}\right)\right]  \nonumber \\
&\quad + \frac{\hbar^3\sqrt{3}\ \lambda_3^{3/2}}{(4\pi)^3 v_0^2}\left[-\frac{3}{8\sqrt{2}} \xi +\frac{21}{32} 
   \text{Li}_2\left(\frac{1}{4}\right)
   -\frac{7  \pi^2}{128}
   -\frac{1}{2}
   +\frac{21}{64} \log^2\left(\frac{4}{3}\right)
   +\frac{5}{8} \log
   \left(\frac{4}{3}\right)\right]
   \;. \label{eq:condensate1}
\end{align}
The explicit $\log\mu_3$ term at two-loop order cancels the running of $v_0$ at order $\hbar^2$, and further the absence of a $\log\mu_3$ term at three-loop order is a result of the absence of running of the one-loop correction to $\Delta \cond{}$, which only depends on $\lambda_3$.
Numerically Eq.~\eqref{eq:condensate1} reads
\begin{equation}
\frac{1}{v_0}\Delta \cond{} =
2
+ 1.73205\; \ep
+ \frac{1}{2}\left[1 + \log \left(\tilde{\mu}_3^2\right)\right]\ep^2
- 1.15232\; \ep^3
+ O\left(\ep^4\right)
\;, \label{eq:condensate1_numeric}
\end{equation}
where we have introduced $\tilde{\mu}_3^2 = \mu_3^2/(3\lambda_3 v_0^2)$, and have indicated the size of the four-loop corrections.
The loop-expansion parameter, $\ep$, is given in Eq.~\eqref{eq:critical_expansion_parameter}.
As can be seen, the expansion coefficients are all $O(1)$, suggesting that the magnitude of $\ep$ should give a reliable estimate of how well the series converges.
As $\ep$ is inversely proportional to the jump in the order parameter, the series converges more quickly for stronger transitions.

\subsection{Renormalisation group improvement} \label{sec:rgi}

The effective potential, or rather its $\phi_3$-derivative to avoid the cosmological constant, satisfies the renormalisation group (or Callan-Symanzik) equation,
\begin{equation} \label{eq:Veff_rge}
\left(\frac{\partial}{\partial \log\mu_3} + \frac{\hbar^2g_3\lambda_3}{6(4\pi)^2}\frac{\partial}{\partial \sigma_3} + \frac{\hbar^2\lambda_3^2}{6(4\pi)^2}\frac{\partial}{\partial m_3^2}\right) \frac{dV_{\rm 3, eff}}{d\phi_3} = 0
\;,
\end{equation}
where the nonzero beta functions are taken from Eqs.~\eqref{eq:betafns_sigma3} and \eqref{eq:betafns_m3}.
Due to the superrenormalisability of the theory, this equation is exact.
Expanding it in $\hbar$, leads to an infinite set of relations linking the running of couplings at $O(\hbar^{n})$ to explicit logarithms at $O(\hbar^{n+2})$.
Using this, one can deduce the explicit $\mu_3$ dependence present at four- and five-loop order from the running of the couplings at two- and three-loop order,
\begin{align}
\frac{\partial V_{\rm 3, eff}^{\rmi{4-loop}} } {\partial\log \mu_3} &= - \frac{\hbar^2\lambda_3^2}{6(4\pi)^2}\frac{\partial V_{\rm 3, eff}^{\rmi{2-loop}}}{\partial m_3^2}
\;, &
\frac{\partial V_{\rm 3, eff}^{\rmi{5-loop}}}{\partial \log \mu_3} &= - \frac{\hbar^2\lambda_3^2}{6(4\pi)^2}\frac{\partial V_{\rm 3, eff}^{\rmi{3-loop}}}{\partial m_3^2}
\;.
\end{align}
However, Eq.~\eqref{eq:Veff_rge} is an exact equation, and hence it seems reasonable to solve it exactly, rather than order-by-order in $\hbar$.
Doing so resums the most ultraviolet sensitive higher order contributions, giving the renormalisation group improved (RGI) effective potential.
By incorporating this nonperturbative information, one would hope to improve the accuracy or convergence of perturbative results.
The construction of the RGI effective potential was presented in this context in Ref.~\cite{Farakos:1994kx}, though we will adopt an alternative approach, as follows.

Correlation functions, such as the linear condensate $\Delta\cond{}$, satisfy an identical renormalisation group equation.
This can be solved order-by-order in $\hbar$, to find the scale dependent parts of the 4- and 5-loop contributions,
\begin{align}
\frac{1}{v_0}\frac{\partial}{\partial \log \mu_3}\Delta\cond{}^{\rmi{4-loop}} &= \left(\frac{3}{4}+\frac{1}{2}\log\tilde{\mu}_3 \right) \ep^4
\;, \\
\frac{1}{v_0}\frac{\partial}{\partial \log \mu_3}\Delta\cond{}^{\rmi{5-loop}} &= - 1.15232\ \ep^5
\;, \label{eq:cond_5_loop}
\end{align}
where the numerical constant in Eq.~\eqref{eq:cond_5_loop} is the same as the 3-loop coefficient in Eqs.~\eqref{eq:condensate1} and \eqref{eq:condensate1_numeric}.
Alternatively, one can solve the renormalisation group equation using the method of characteristics, which amounts to replacing the coupling constants by the corresponding running solutions.
For Eq.~\eqref{eq:condensate1_numeric}, this amounts simply to
\begin{align}
v_0 &\to v_0\sqrt{1 - \ep^2 \log\left(\frac{\mu_3}{\mu_{3,0}}\right)}\;, &
\ep &\to \frac{\ep}{\sqrt{1 - \ep^2 \log\left(\frac{\mu_3}{\mu_{3,0}}\right)}}
\;,
\end{align}
where $\mu_{3,0}$ is some initial scale, at which $v_0$ and $\ep$ are given.
This defines our RGI perturbative calculation.
Note that this improvement is naturally incorporated upon performing dimensional reduction, if the exact running of the parameters is used to cancel the dependence on the matching scale $\Lambda$, replacing it with a new renormalisation scale $\mu_3$; see the end of Sec.~\ref{sec:dr_minimal}.

Interestingly, while both $g_3$ and $\lambda_3$ are independent of $\mu_3$, the expansion parameter $\ep$ is not.
While this is clear in the form given in Eq.~\eqref{eq:critical_expansion_parameter}, to see it in the form given in Eq.~\eqref{eq:expansion_parameter} one should note that changing $\mu_3$ shifts the minimum of the potential, and the basis transformation required to shift back to remove the tadpole induces a $\mu_3$-dependence in the cubic coupling.

This RGI perturbative calculation still retains some $\mu_3$-dependence.
For the jump in the linear condensate, Eq.~\eqref{eq:condensate1_numeric} shows that one must choose $\mu_3\sim \sqrt{3\lambda_3}v_0$ in order to avoid large logarithms.
An optimal choice of renormalisation scale, $\mu_3^{\text{opt}}$, can be found according to the {\em principle of minimal sensitivity}~\cite{Stevenson:1981vj}, whereby, for some approximation to a physical quantity, one solves for the point of minimal sensitivity to the renormalisation scale.
A natural choice in this context is the linear condensate,
\begin{equation} \label{eq:minimal_sensitivity}
\frac{\partial}{\partial\log\mu_3} \Delta\cond{}^{\rmi{RGI}}= 0
\; .
\end{equation}
This equation can be used to determine $\mu_3^{\text{opt}}$, the optimal choice for the renormalisation scale for a given perturbative approximation.
Numerically, we find solutions to Eq.~\eqref{eq:minimal_sensitivity} with $\mu_3^{\text{opt}}\sim \sqrt{3\lambda_3}v_0$ for the two- and three-loop order approximations to the linear condensate.
However, at lower loop order we find no such solutions and hence simply take $\mu_3^{\text{opt}} = \sqrt{3\lambda_3}v_0$, a conclusion which might have been expected because two-loop order is the lowest order at which the couplings run in this EFT.
An estimate of the magnitude of missing higher order terms can be found by varying $\mu_3$ about $\mu_3^{\text{opt}}$ by some multiplicative $O(1)$ factor, which we take to be 10 to be conservative.

\section{Phase transition on the lattice} \label{sec:lattice}

Perturbation theory can only take us so far.
As discussed in Sec.~\ref{sec:loop_expansion}, at the critical temperature the loop-expansion parameter for the 3d EFT scales as $\ep\propto\lambda_3^{3/2}/|g_3|$.
So, as one approaches the $Z_2$-symmetric limit, the loop-expansion parameter grows without bound, signalling the complete breakdown of perturbation theory.

To reliably study the phase transition for both small and large $\ep$, we resort to lattice Monte-Carlo simulations; for relevant overviews see Refs.~\cite{Berg:1992qua,Montvay:1994cy,Kajantie:1995kf,KariLattice}, and for recent studies in other models see Refs.~\cite{Kainulainen:2019kyp,Niemi:2020hto} and \cite{Huang:2020mso,Cossu:2020yeg}.
The first step, which we carry out in Sec.~\ref{sec:lattice-continuum}, is to find explicit relations between the bare parameters of the lattice Lagrangian, and those of the continuum theory in the \MSbar\ renormalisation scheme.
Once this is done, measurements from Monte-Carlo simulations can be directly interpreted in terms of \MSbar\ observables.

In Sec.~\ref{sec:lattice-continuum} we derive the lattice-continuum relations, applicable in the limit $a\to 0$.
Following this we discuss the Monte-Carlo simulations in Sec.~\ref{sec:monte-carlo}, with details of the algorithms given in Appendix~\ref{appendix:algorithms}.
Results for the latent heat, extrapolated to the continuum limit, are presented in Sec.~\ref{sec:lattice_latent_heat}, with additional plots of the continuum extrapolations given in Appendix~\ref{appendix:numerical-results}.

\subsection{Lattice-continuum relations} \label{sec:lattice-continuum}

The simplest lattice Lagrangian which converges to Eq.~\eqref{eq:lagrangian_3d} in the continuum is
\begin{align}
\ms{L}_{\rm 3,L} &= \frac{1}{2a^2}\sum_i[\phi_3(x+i)-\phi_3(x)]^2 + V_{\rm 3, L}(\phi_3)
\;, \\
V_{\rm 3, L}(\phi_3) &=  \left(\sigma_{\rm 3} + \delta \sigma_{\rm 3, L} \right)\phi_3(x) + \frac{1}{2}\left(m_{\rm 3}^2+\delta m_{\rm 3,L}^2\right)\phi_3(x)^2 + \frac{1}{3!}g_{3}\phi_3(x)^3 + \frac{1}{4!}\lambda_{3}\phi_3(x)^4
\;,
\end{align}
where $x$ labels lattice sites, $i$ denotes the link from the site $x$ to its nearest neighbour in one Cartesian direction, and the sum over $i$ denotes a sum over Cartesian directions.%
\footnote{
Improved convergence can be obtained by using a more complicated finite-difference approximation to the kinetic term involving both nearest and next-to-nearest neighbouring sites.
However, to achieve this improved convergence also requires computing the lattice-continuum relations to higher order in the lattice spacing, $a$;
see Refs.~\cite{Moore:1996bf,Moore:1997np,Moore:2001vf,Arnold:2001ir}.
}
We have added the subscript $L$ to the tadpole and mass counterterms because, due to the different regularisation of divergent integrals, these quantities differ from their \MSbar\ counterparts.
Conversely, due to the absence of momentum-dependent divergences, or divergences of diagrams with three or four external legs, the field renormalisation and three- and four-point couplings may be chosen to be equal to their \MSbar\ counterparts.

Due to the superrenormalisability of the theory, all divergences of the theory turn up at finite loop order and hence can be calculated analytically, in lattice perturbation theory.
Thus, it is possible to derive the exact relationship between the parameters in the lattice and continuum theories, in the limit $a\to 0$.
To do so, we follow Refs.~\cite{Laine:1995np,Laine:1997dy} in computing the effective potential in lattice perturbation theory, and equating the result to the effective potential in the continuum, in the \MSbar\ scheme.

The computation of the effective potential in lattice perturbation theory, mirrors almost exactly that in Sec.~\eqref{sec:effective_potential}.
The diagrams and combinatorics are the same; only the values of the loop integrals are different.
A notable consequence of this difference is that with lattice regularisation, unlike with dimensional regularisation, there are linear divergences at one-loop, which demand compensating one-loop counterterms.

The one-loop order contribution to the effective potential is
\begin{align}
V_{\rm 3, eff }^{\rmi{1-loop}} &=  
\hbar J\left(M_3(\phi_3)\right)
+ \delta V^{\rmi{1-loop}}_{\rm 3, L}
+ \delta \sigma^{\rmi{1-loop}}_{\rm 3, L} \phi_3
+ \frac{1}{2}\left(\delta m^2_{\rm 3, L}\right)^{\rmi{1-loop}} \phi_3^2
\;,
\end{align}
where the lattice loop-integral $J$ is defined in Eq.~(171) of Ref.~\cite{Farakos:1994xh}.
In the infinite volume limit, this lattice loop integral may be carried out as a Fourier integral over the first Brillouin zone, the result of which can also be found in Ref.~\cite{Farakos:1994xh}.

Expanding for small $a$, we find the following one-loop counterterms,
\begin{align}
\delta V_{\rm 3,L}^{\rmi{1-loop}} &= -\frac{\hbar M_3^2 \Sigma }{2 (4\pi)  a}\;, &
\delta \sigma_{\rm 3,L}^{\rmi{1-loop}} &= -\frac{\hbar g_3 \Sigma }{2 (4\pi)  a}\;, &
\left(\delta m^2_{\rm 3,L}\right)^{\rmi{1-loop}} &= -\frac{\hbar \lambda_3  \Sigma }{2 (4\pi)  a}
\;,
\end{align}
were $\Sigma$ is a numerical constant, the result of a dimensionless integral.
It is given analytically in Eq.~(170) of Ref.~\cite{Farakos:1994xh}, and its numerical value is $\Sigma=3.17591153562522...$.

Progressing to two-loop order involves nothing qualitatively new.
The lattice-regularised, two-loop effective potential is equal to its \MSbar\ counterpart in the $a\to 0$ limit if we choose the following counterterms (including their finite parts),
\begin{align}
\delta V_{\rm 3,L}^{\rmi{2-loop}} &= \frac{\hbar^2\lambda_3  \Sigma ^2}{8 (4\pi) ^2 a^2} + \frac{\hbar^2 g_3^2 }{12(4\pi) ^2} \left[\log \left(\frac{6}{a \mu_3 }\right)+ \zeta \right]
\;, \\
\delta \sigma_{\rm 3,L}^{\rmi{2-loop}} &= \frac{\hbar^2 g_3 \lambda_3}{6 (4\pi) ^2} \left[\log \left(\frac{6}{a \mu_3 }\right)+\zeta \right]
\;, \\
\left(\delta m^2_{\rm 3,L}\right)^{\rmi{2-loop}} &= \frac{\hbar^2 \lambda_3 ^2}{6 (4\pi) ^2}\left[\log \left(\frac{6}{a\mu_3 }\right)+ \zeta \right]
\;.
\end{align}
Here $\zeta$ is another numerical constant, the value of which is approximately $\zeta\approx 0.0884801$ (called $C_{\rm 3,U}$ in Ref.~\cite{Moore:2001vf}, see also Ref.~\cite{Laine:1997dy}).
There are also divergent additive contributions to the (cosmological) constant counterterm at three- and four-loop order, but there are no divergent, field-dependent contributions at higher-loop orders.
Thus, the two-loop results for the tadpole and mass lattice counterterms are exact in the $a\to 0$ limit.

Finally, we would like to comment on the lattice-continuum relations for the four special condensates, defined as in Eq.~\eqref{eq:condensate1_definition}.
Now that the lattice-continuum relations have been found for the bare lagrangians, those for the condensates follow simply by differentiation with respect to the renormalised \MSbar\ parameters~\cite{Farakos:1994xh}.
Due to the absence of $\sigma_3$-dependence in the lattice-continuum relations above, the result for the one-point condensate is simple,
\begin{equation}
\Delta\cond{}_L = \Delta\cond{}(1 + O(a))
\;, \label{eq:lattice_condensate_1}
\end{equation}
as $a\to 0$.

\subsection{Monte-Carlo simulations} \label{sec:monte-carlo}

With the lattice-continuum relations in hand, one can carry out Markov-chain Monte-Carlo simulations.
For this we wrote a simulation code in {\tt C}.
Starting from an arbitrary initial lattice field configuration, we perform successive updates on the field such as to sample the Boltzmann-weighted probability distribution of the theory.
The list of configurations forms the Markov chain.
After a large number of updates of the whole lattice (typically a few million), the Markov chain converges to a relatively accurate representation of the Boltzmann probability distribution.
From this measurements of physical quantities, such as the field condensates, can be made.
In order to ensure the field configuration does not get stuck in one phase, we utilised multicanonical methods~\cite{Berg:1991cf,Berg:1992qua}.
Details of the algorithms we adopted are discussed in Appendix~\ref{appendix:algorithms}.

\begin{figure}[t]
\begin{subfigure}{.5\textwidth}
  \centering
  \includegraphics[width=\linewidth]{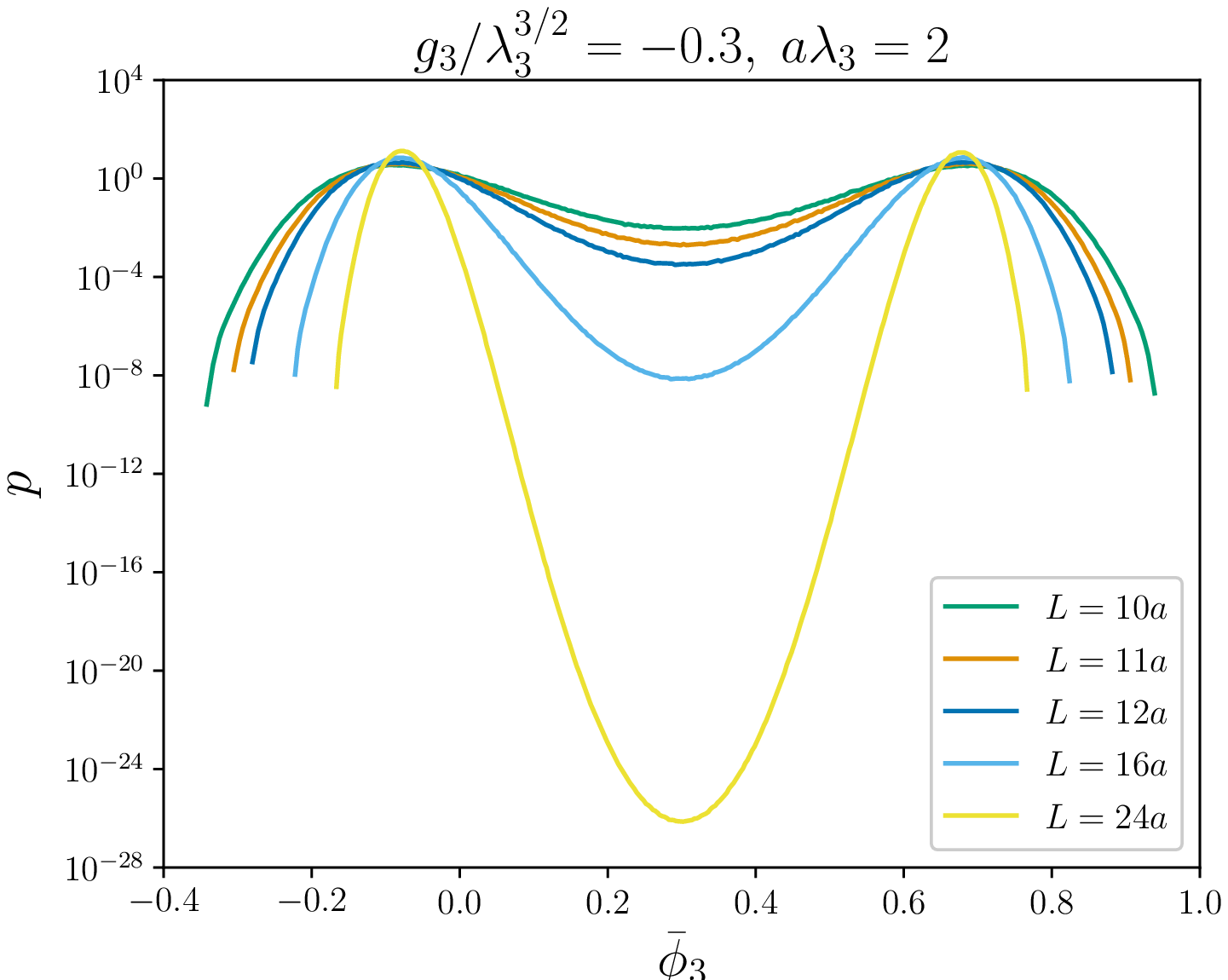}  
  \caption{}
  \label{fig:hists_N}
\end{subfigure}
\begin{subfigure}{.5\textwidth}
  \centering
  \includegraphics[width=\linewidth]{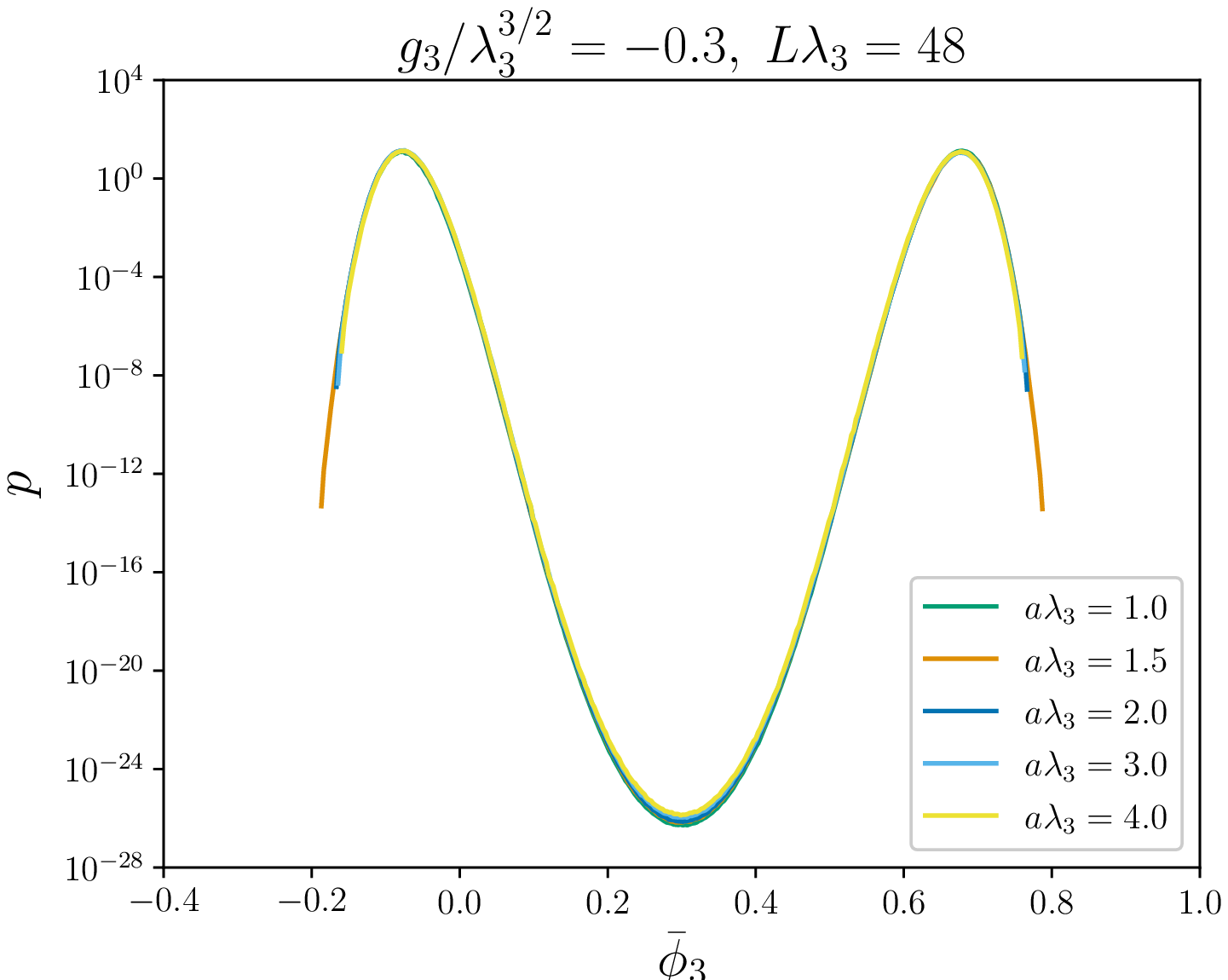}  
  \caption{}
  \label{fig:hists_a}
\end{subfigure}
  \caption{Probability distributions, $p$, of lattice Monte-Carlo measurements of $\bar{\phi}_3$ at the critical temperature.
  Fig.~(a) shows the dependence on the lattice volume at fixed lattice spacing, $a$.
  Fig.~(b) shows the dependence on the lattice spacing, $a$, at fixed lattice volume.
  The coexistence of clearly separated phases shows the first-order nature of the transition.}
  \label{fig:hists}
\end{figure}
Binning measurements of $\bar{\phi}_3$ to form a histogram gives an approximation to the probability distribution for this observable; see Fig.~\ref{fig:hists}.
At the critical temperature this probability distribution shows a two-peaked structure, characteristic of a first-order phase transition.
To extract physical results relevant for continuum physics requires taking first the infinite volume limit, and then the zero lattice spacing limit.
In Fig.~\ref{fig:hists_N} we show the effect on the probability distribution of increasing the lattice volume, while keeping the lattice spacing fixed, and in Fig.~\ref{fig:hists_a} we show the effect of decreasing the lattice spacing while keeping the lattice volume fixed.

At the critical temperature there is only one dimensionless parameter which characterises the 3d EFT, and this can be chosen to be $g_3/\lambda_3^{3/2}$ (alternative choices are $(r-r_*)/\lambda_3^2$ and $\alpha_3$).
This is because, of the four initial Lagrangian parameters, one parameter can be fixed by shifting the field origin, $\sigma_3=0$ say, a second parameter can be fixed by the condition of being at the critical temperature, $m_3^2$ say, and a third parameter can be fixed by a choice of units, $\lambda_3=1$ say.
This is analogous to what happens in the SU(2)-Higgs theory at the critical temperature, where the dimensionless parameter which controls the character of the phase transition is $x\equiv \lambda_{\rmi{Higgs}, 3}/g_{\rmi{SU(2)}, 3}^2$~\cite{Kajantie:1995kf}.
In our case the most closely corresponding dimensionless combination would be $x_{\rmi{singlet}}\equiv \lambda_3/|g_3|^{2/3}$.

We chose six parameter points to simulate on the lattice, starting at $g_3/\lambda_3^{3/2}=-1.2$ and decreasing by powers of 2 until $g_3/\lambda_3^{3/2}=-0.0375$.
In all cases, the \MSbar\ renormalisation scale was taken to be $\mu_{\rm 3,L} = 1.066496\lambda_3$.%
\footnote{
This odd value of $\mu_3$ arose due to an earlier error in the parameter $\zeta$ in the lattice-continuum relations, which fortunately could be rectified by a shift in $\mu_3$.
}
To extrapolate to the continuum limit for each such parameter point, we simulated between five and eight different lattice spacings, and for each lattice spacing we simulated between five and eight different lattice volumes.
Thus, all in all, our data set consists of more than 200 different simulations.
The continuum-extrapolated results are independent of the renormalisation scale.

The different lattice spacings, $a$, and volumes, $L^3$, should all be chosen to satisfy
\begin{equation}
a \ll \xi \ll L
\;,
\end{equation}
where $\xi$ is the correlation length of the system, or the inverse screening mass.
For parameter values where the EFT is perturbative, the screening mass will be close to the tree-level mass, in which case $\xi\approx 1/m_3$.
Including also the one-loop corrections, for $\sigma_3=0$, we find
\begin{equation} \label{eq:screening_mass_one_loop}
\frac{1}{\xi^2} = m_3^2\left(1-\frac{\hbar}{2(4\pi)}\left[\frac{\lambda _3}{m_3}+\frac{g_3^2}{2m_3^3}(\log (3)-2)\right] + O(\hbar^2)\right)
\;.
\end{equation}
In principle the two-loop corrections can be constructed from the results of Ref.~\cite{Rajantie:1996np}, though we have not done so.
In the opposite limit, near the second-order phase transition where perturbation theory does not work, Eq.~\eqref{eq:ms_r} should provide a better estimate.
In choosing appropriate lattice spacings and volumes, we have used a combination of these two estimates.
However, these approximations are not perfect and, especially for small values of $-g_3/\lambda_3^{3/2}$, additional lattice spacings and volumes were necessary to attain reasonable continuum limits.
A more robust alternative, which we did not attempt, would be to directly measure $\xi$ on the lattice, by the exponential decay of correlation functions with distance; see for example Ref.~\cite{Kajantie:1995kf}.

Our methods for analysing the simulation data are fairly standard, and generally follow Refs.~\cite{KariLattice,Kajantie:1995kf}.
Error bars for simulation data points show statistical errors, calculated using jackknife resampling on blocked measurements, with each block much larger than one autocorrelation time.

\subsection{Latent heat on the lattice} \label{sec:lattice_latent_heat}

The latent heat is proportional to the change in the linear field condensate, $\Delta\cond{}$, with the proportionality constant being dependent on the details of the full 4d theory, but not on the dynamics of the 3d EFT; see Eq.~\eqref{eq:latent_heat_simple}.
Thus, for a theory which is weakly coupled at $T=0$, this proportionality constant can be calculated perturbatively.
The possibly nonperturbative dynamics of the zero Matsubara mode only enters the latent heat through $\Delta\cond{}$.

In a finite volume, the linear condensate can be defined as
\begin{equation}
\Delta\cond{} = 
\frac{1}{2}\int\displaylimits_{ > \bar{\phi}_3^\rmi{min}}d\bar{\phi}_3\ \bar{\phi}_3 P(\bar{\phi}_3)
- \frac{1}{2}\int\displaylimits_{< \bar{\phi}_3^\rmi{min}}d\bar{\phi}_3\ \bar{\phi}_3 P(\bar{\phi}_3)
\;,
\end{equation}
where $P(\bar{\phi}_3)$ denotes the probability density of being in a state $\bar{\phi}_3$ and $\bar{\phi}_{3}^{\rmi{min}}=-g_3/\lambda_3$ denotes the position of the minimum of the probability density between the two phases.
In a small enough volume, states between the two phases are not uncommon, however as the lattice volume grows such states become exponentially rarer; see Fig.~\ref{fig:hists_N}.

\begin{figure}[t]
\begin{subfigure}{.5\textwidth}
  \centering
  \includegraphics[width=\linewidth]{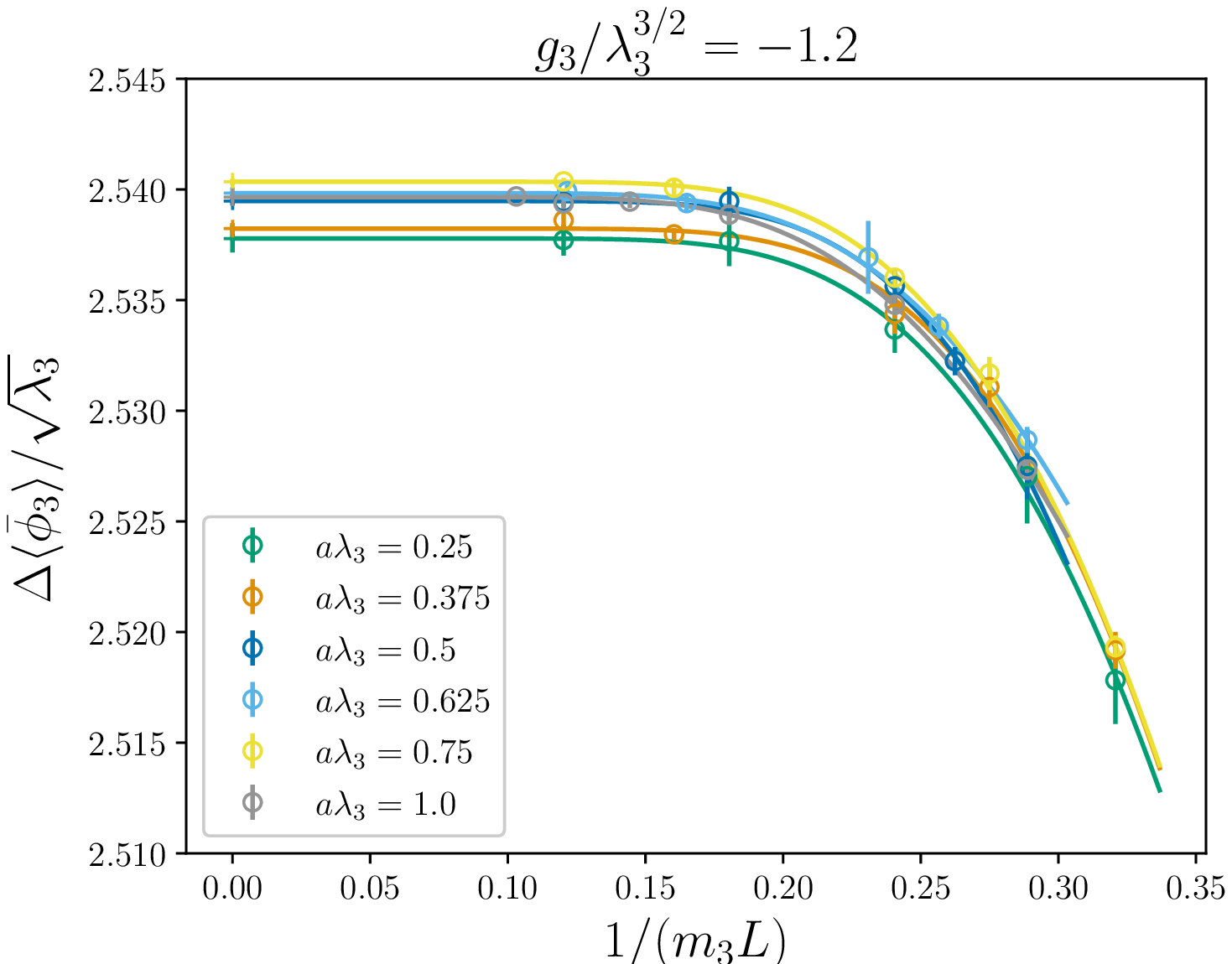}  
  \caption{}
  \label{fig:dphi_vol_x_1.2}
\end{subfigure}
\begin{subfigure}{.5\textwidth}
  \centering
  \includegraphics[width=\linewidth]{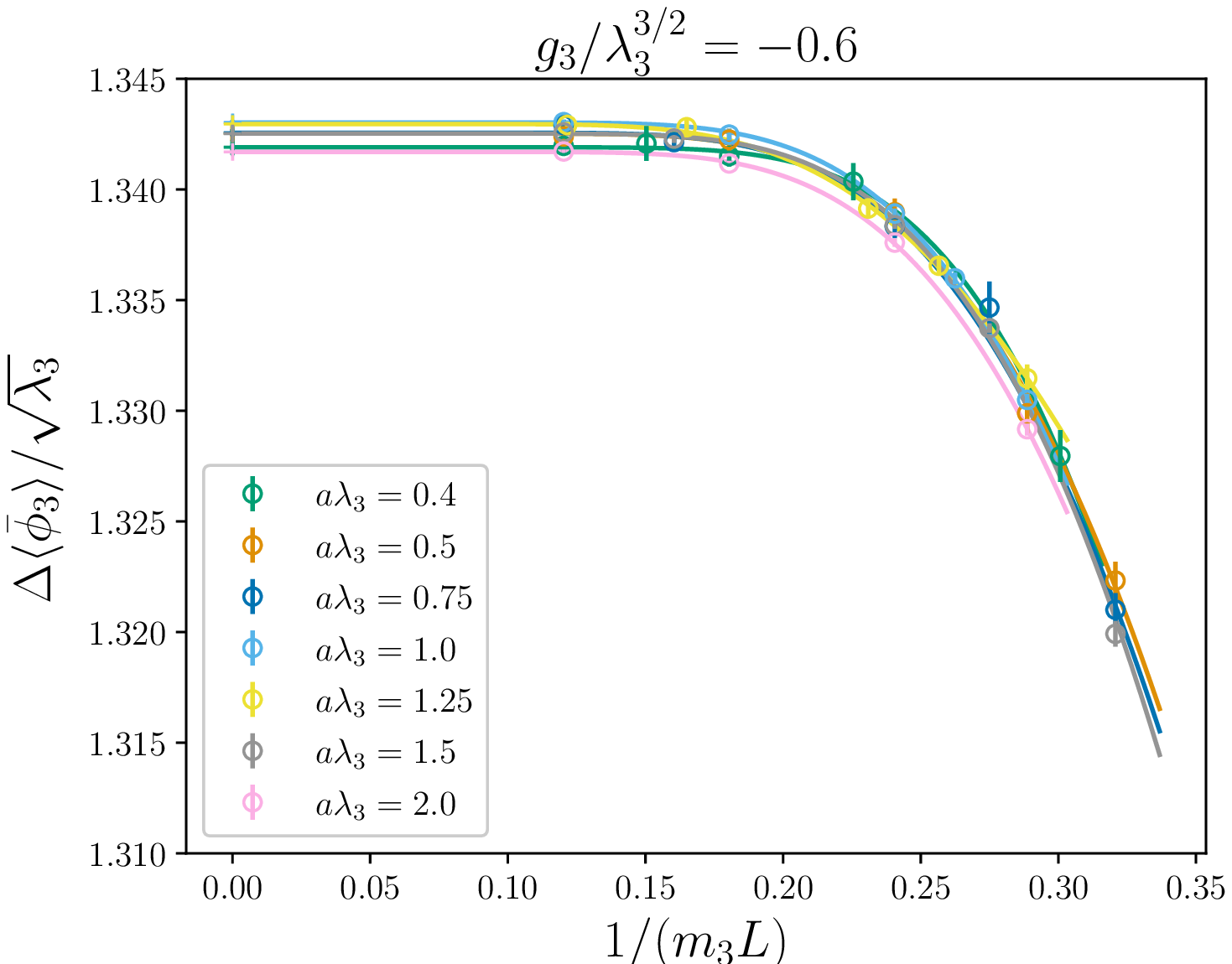}  
  \caption{}
  \label{fig:dphi_vol_x_0.6}
\end{subfigure}

\begin{subfigure}{.5\textwidth}
  \centering
  \includegraphics[width=\linewidth]{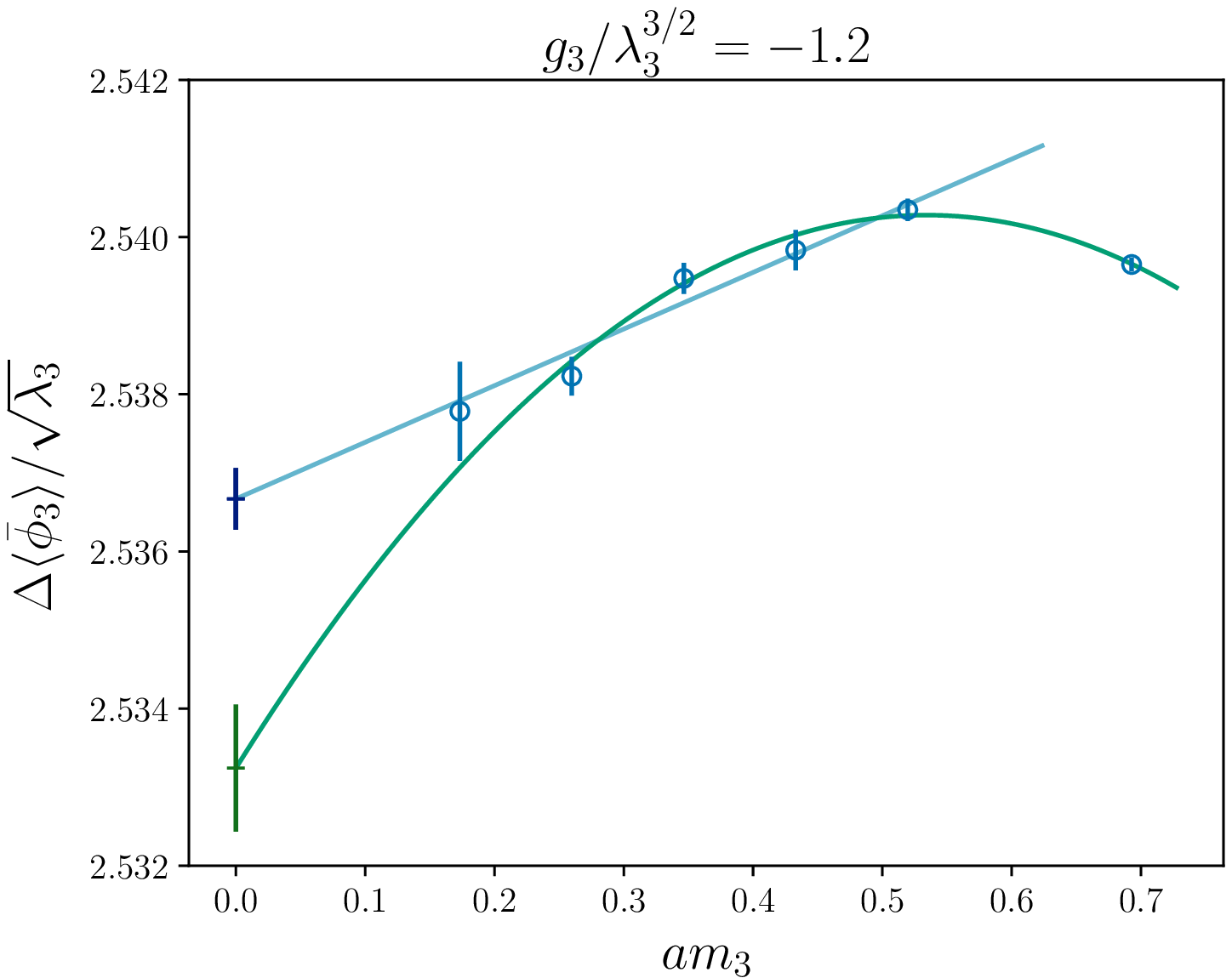}  
  \caption{}
  \label{fig:dphi_a_x_1.2}
\end{subfigure}
\begin{subfigure}{.5\textwidth}
  \centering
  \includegraphics[width=\linewidth]{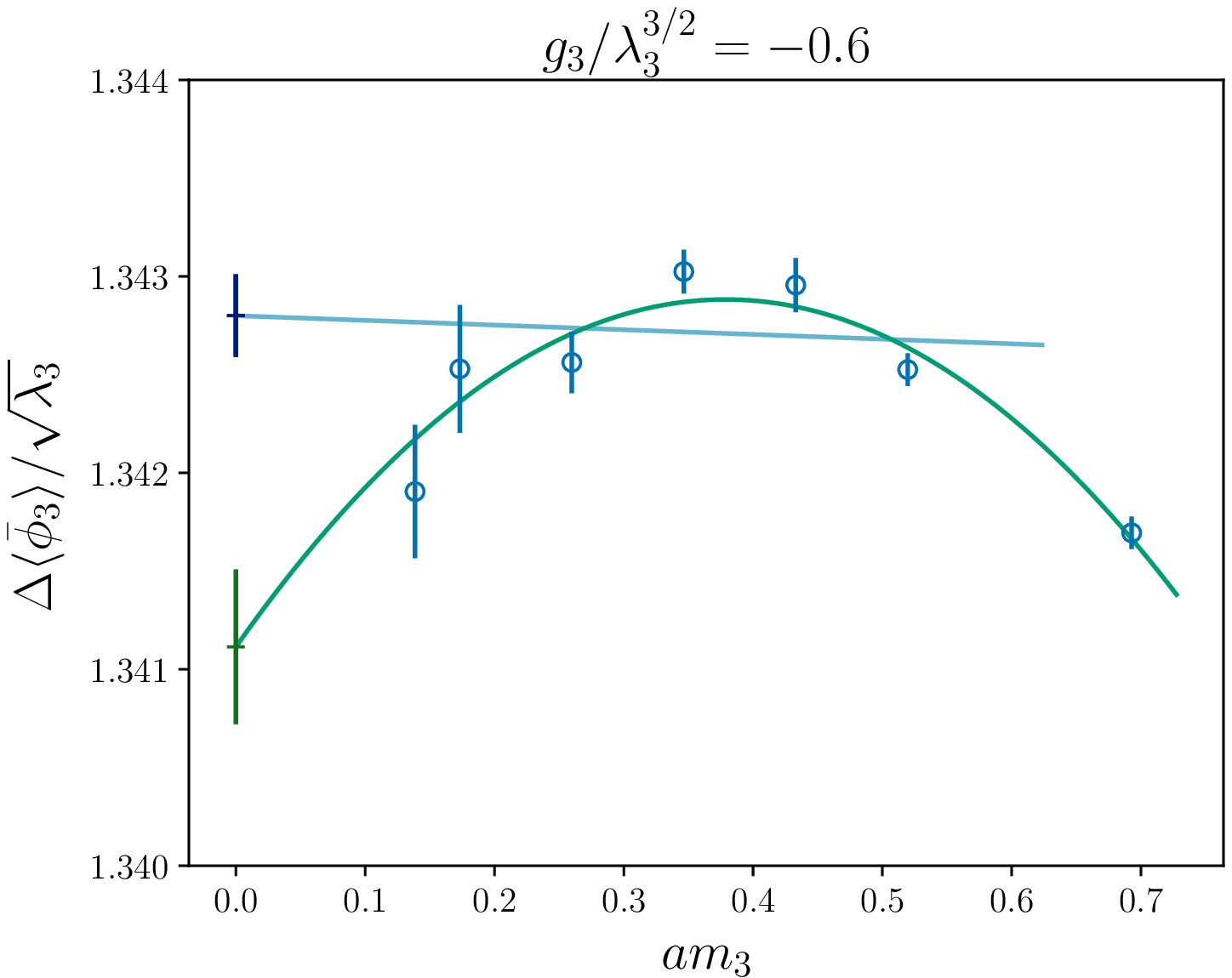}  
  \caption{}
  \label{fig:dphi_a_x_0.6}
\end{subfigure}
\caption{Taking the infinite volume, followed by the zero lattice spacing limits of lattice data for $\Delta\langle\phi_3\rangle$ at two parameters points.
For the fits to the lattice spacing dependence here, we show both a linear fit excluding the data point with largest $a$, and a quadratic fit to all the data.
In Fig.~\ref{fig:dphi_a_x_1.2} the linear and quadratic fits give reduced $\chi^2\approx 1.5$ and 0.9 respectively.
In Fig.~\ref{fig:dphi_a_x_0.6} these are instead 5.6 and 1.5.
We take the difference between these fits as a measure of the systematic error introduced in the extrapolation.
}
\label{fig:dphi_continuum_x_1.2_0.6}
\end{figure}

In Fig.~\ref{fig:dphi_continuum_x_1.2_0.6} we show our results for $g_3/\lambda_3^{3/2}=-1.2$ and $-0.6$.
Plots for the other four parameter points are collected in Appendix~\ref{appendix:numerical-results}.
The continuum-extrapolated results are collected in Table~\ref{table:results_lattice}.

\begin{table}[t]
	\centering
    \begin{tabular}{llll}
      \hline
      $g_3/\lambda_3^{3/2}$ & $(\rr-\rs)/\lambda_3^2$ & $\Delta \langle \bar{\phi_3} \rangle_c/\lambda_3^{1/2}$ & $\Delta \langle \left(\bar{\phi}_3 + g_3/\lambda_3\right)^3 \rangle_c/\lambda_3^{3/2}$ \\
      \hline
      -1.2 & -0.242752(5) & 2.5332(8)(39) & 3.640(15)(10) \\
	  -0.6 & -0.062752(5) & 1.3411(4)(20) & 0.4830(16)(13) \\ 
      -0.3 & -0.017752(5) & 0.75533(15)(16) & 0.06797(18)(2) \\ 
      -0.15 & -0.006502(5) & 0.48211(8)(8) & 0.010847(25)(0) \\
      -0.075 & -0.003690(5) & 0.3757(5)(12) & 0.002127(23)(35) \\ 
      -0.0375 & -0.002987(5) & 0.3431(5)(25) & 0.000488(13)(13)
    \end{tabular}
  \caption{
  Continuum-extrapolated lattice results for the discontinuities in the linear and cubic condensates.
  The UV divergences of the cubic condensate has been removed, following Ref.~\cite{Farakos:1994xh}.
  Errors quoted are statistical followed by systematic.
  \label{table:results_lattice}}
\end{table}

Figs.~\ref{fig:dphi_vol_x_1.2} and \ref{fig:dphi_vol_x_0.6} show the infinite volume extrapolations at fixed lattice spacing.
As the theory is gapped, at large volumes the infinite volume limit is approached exponentially fast, with corrections $\sim \exp(-L/\xi)$.
The lines shown in these figures show least-squares fits to the data of the form $c_1 + c_2 \exp(c_3 L)$, where the $c_i$ are fit parameters.
In all cases the reduced $\chi^2$ values for the fits are in the range $\sim [0.2,5]$.
As a measure of the systematic uncertainty associated with the infinite volume extrapolations, we repeat the succeeding analysis using only the largest volume simulations at each value of $a$, rather than the extrapolation of the fit.

The results of the infinite volume extrapolations are the data points in Figs.~\ref{fig:dphi_a_x_1.2} and \ref{fig:dphi_a_x_0.6}, with the error bars resulting from the least-squares fit.
The lines in these figures in turn show the extrapolations to zero lattice spacing.
As a consequence of the lattice-continuum relations of Sec.~\ref{sec:lattice-continuum}, for sufficiently small $a m_3$, the lattice results should approach the continuum limit linearly,
\begin{equation} \label{eq:a_dependence}
\Delta\cond{}(\kappa,a) = \Delta\cond{}(\kappa,0) + C_1(\kappa) a m_3 + C_2(\kappa) (a m_3)^2 + \dots
\;,
\end{equation}
where we have denoted the \MSbar\ parameters of the theory by $\kappa = \{\sigma_3,m_3^2,g_3,\lambda_3 , \mu_3 \}$.
For each parameter point we attempt to extrapolate to $a=0$ linearly in $a$, but consider also higher powers in a Taylor expansion in $a$ if the linear fit is poor.
In such cases we truncate at the lowest power which gives a reduced $\chi^2$ of order one.

Some, but not all, of the lattice spacing fits reveal the presence of significant nonlinear terms in $am_3$.
By comparing the six different parameter points, we suggest an explanation in terms of the coefficients $C_i(\kappa)$ in Eq.~\eqref{eq:a_dependence}.
For the strongest transition, $g_3/\lambda_3^{3/2}=-1.2$, the fit in Fig.~\ref{fig:dphi_a_x_1.2} shows $C_1>0$.
Progressing to weaker transitions, shown in Figs.~\ref{fig:dphi_a_x_0.6}, \ref{fig:dphi_a_x_0.3}, \ref{fig:dphi_a_x_0.15}, \ref{fig:dphi_a_x_0.075} and \ref{fig:dphi_a_x_0.0375}, one can see that $C_1$ changes sign around $g_3/\lambda_3^{3/2}\approx -0.6$, remaining negative all the way to $g_3/\lambda_3^{3/2}\approx -0.0375$.
Regarding $C_2$, Fig.~\ref{fig:dphi_a_x_1.2} suggests the curvature is negative, $C_2<0$, for $g_3/\lambda_3^{3/2}=-1.2$.
Likewise progressing to weaker and weaker transitions, one can see that $C_2$ changes sign somewhere between $g_3/\lambda_3^{3/2}=-0.6$ and $g_3/\lambda_3^{3/2}=-0.075$.
These considerations can explain, for example, the comparatively poor linear fit in Fig.~\ref{fig:dphi_a_x_0.6} as due to the smallness of $C_1$ at this parameter point, and the comparatively good linear fits in Figs.~\ref{fig:dphi_a_x_0.3} and \ref{fig:dphi_a_x_0.15} as due to the smallness of $C_2$.
For the parameter points where a nonlinear fit is shown, we use the difference between this and the linear fit as a measure of the systematic uncertainty in the extrapolation.
The relatively large systematic errors introduced by the extrapolation suggest that for future studies, it would be worthwhile to use the $O(a)$ improved lattice-continuum relations~\cite{Moore:1996bf,Moore:1997np,Moore:2001vf,Arnold:2001ir}.

\section{Discussion} \label{sec:discussion}

In this work, we have studied the phase transitions of a generic real scalar field.
We have been agnostic about couplings to other fields, such as to the Higgs, to sterile neutrinos and to gravity.
However, we have focused on the case where the infrared dynamics of the phase transition is dominated by the real scalar;
the contributions of all other particles to the transition being simply to modify the effective couplings of the infrared EFT.
With this restriction, we have characterised the order and strength of the phase transition nonperturbatively.

Our results can be applied to a variety of 4d particle physics and cosmological models.
As long as the phase transition is dominated by a real scalar field, no new simulations need be performed to nonperturbatively determine the phase diagram and latent heat of the 4d model in question.
This is a key advantage of the EFT approach.
One needs only to compute the matching relations to the 3d EFT, discussed with examples in Secs.~\ref{sec:dr_higgs} and \ref{sec:dr_other}.
In a similar way the lattice results of the SU(2)-Higgs 3d EFT~\cite{Kajantie:1995kf,Kajantie:1996qd,Moore:2000jw}, have been repurposed to determine nonperturbatively parts of the phase diagram (and other quantities) for the xSM~\cite{Gould:2019qek}, the two higgs doublet model (2HDM)~\cite{Andersen:2017ika} and the triplet scalar extended SM ($\Sigma$SM)~\cite{Niemi:2018asa}.

\begin{figure}[p]
\centering
\begin{subfigure}{.7\textwidth}
  \centering
    \includegraphics[width=\linewidth]{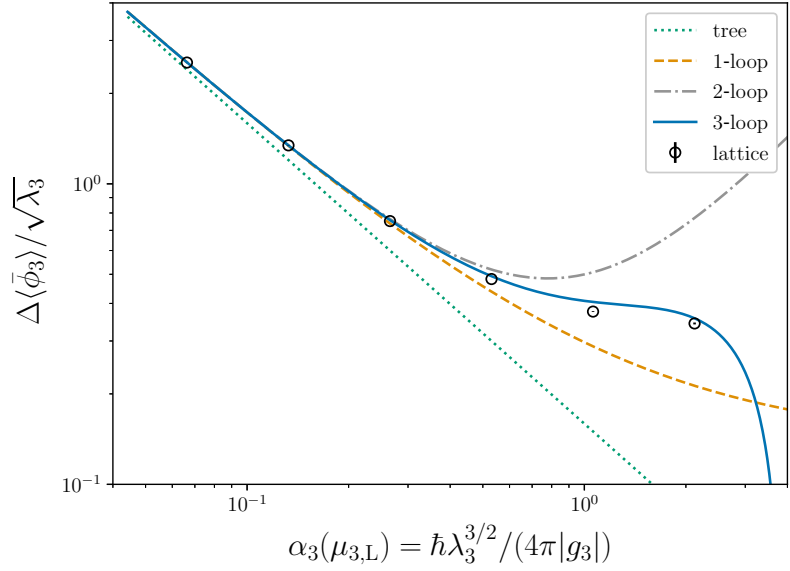}  
  \caption{Lattice versus unimproved perturbation theory.}
  \label{fig:linear_condensate_naive}
\end{subfigure}
\\
\centering
\begin{subfigure}{.7\textwidth}
  \centering
  \includegraphics[width=\linewidth]{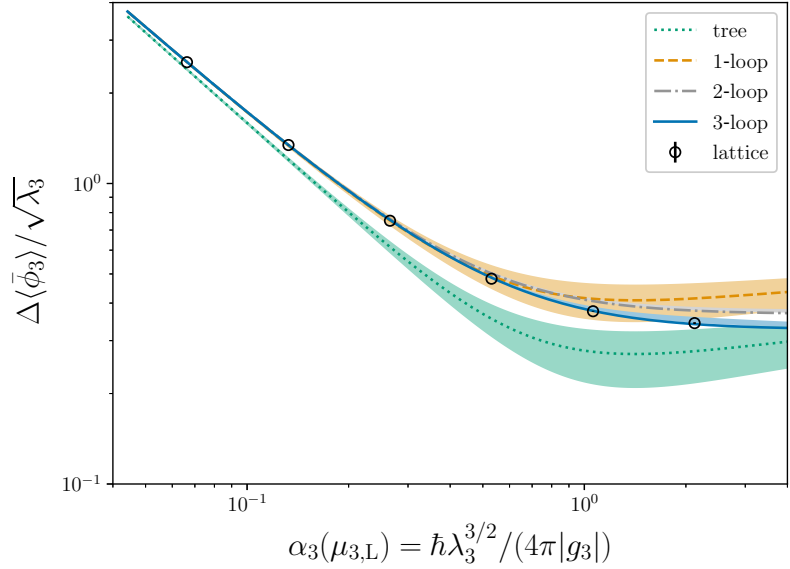}  
  \caption{Lattice versus renormalisation group improved perturbation theory.}
  \label{fig:linear_condensate_rge}
\end{subfigure}

\caption{The change in the linear condensate (proportional to the latent heat) versus the loop expansion parameter within the high temperature EFT.
Black circles are lattice results (continuum extrapolations) with error bars shown within (but barely visible on this scale).
The coloured lines show perturbative results in various approximations.
In Fig.~\ref{fig:linear_condensate_naive} the calculations are performed at a single renormalisation scale $\mu_{\rm 3, L}\approx \lambda_3$.
In Fig.~\ref{fig:linear_condensate_rge}, the renormalisation scale for each perturbative approximation is run from $\mu_{\rm 3, L}$ to some optimal scale $\mu_3^{\text{opt}}$ (discussed in Sec.~\ref{sec:rgi}), and for each approximation the bands reflect the renormalisation scale dependence for $\mu_3/\mu_3^{\text{opt}} \in \{1/\sqrt{10},1,\sqrt{10}\}$.
While both unimproved and RGI perturbation theory agree well with the lattice results at small couplings, the unimproved approach breaks down badly above $\ep \approx 1$, whereas the RGI approach continues to work remarkably well.
}
\label{fig:linear_condensate}
\end{figure}

We leave for future work the computation of the bubble nucleation rate within the 3d EFT, from which one can determine the nucleation temperature, the duration of the phase transition and the change in the trace-anomaly through the transition.
These in turn are crucial ingredients in determining the gravitational wave spectrum resulting from a first-order phase transition.

\begin{figure}[t]
 \centering
  \includegraphics[width=0.75\textwidth]{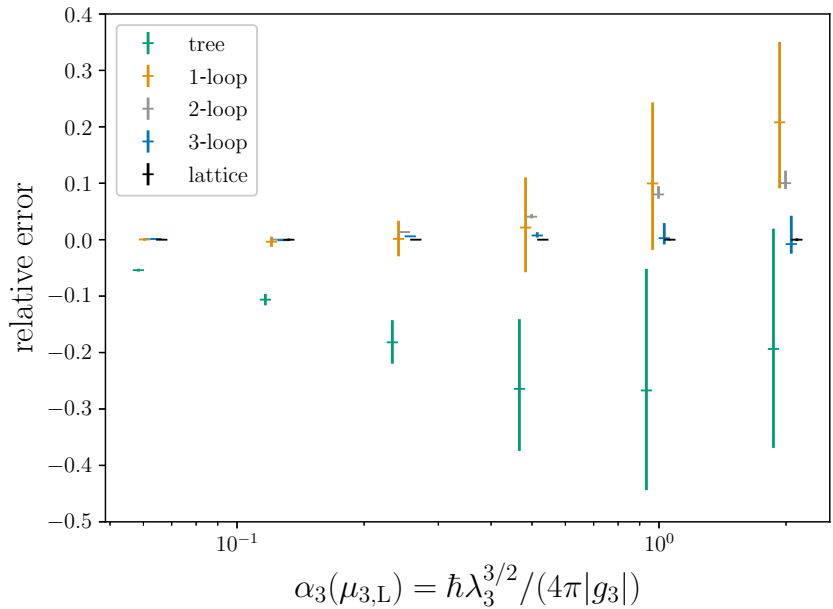}
  \caption{Convergence of the RGI loop expansion towards the lattice results, as a function of the loop expansion parameter.
  The relative error plotted is defined as $(n\text{-loop}-\text{lattice})/\text{lattice}$, $n\in \{0,1,2,3\}$, for the change in the linear condensate.
  The bands shown for each of the perturbative approximations reflect the renormalisation scale dependence for $\mu_3/\mu_3^{\text{opt}} \in \{1/\sqrt{10},1,\sqrt{10}\}$, where $\mu_3^{\text{opt}}$ is defined in Sec.~\ref{sec:rgi}.
  For clarity, the perturbative results have been offset horizontally.
  }
  \label{fig:convergence}
\end{figure}

Theoretically, an important result of this work is a quantitative answer to the question: how well does perturbation theory fare in this model, in comparison to nonperturbative lattice results?
We can answer this question as a function of the loop-expansion parameter, which is also the only dimensionless parameter on which the strength of the phase transition depends nontrivially.
Our answer is shown in Figs.~\ref{fig:linear_condensate} and \ref{fig:convergence}, which also summarise our results.

Fig.~\ref{fig:linear_condensate} shows the lattice and perturbative results across the full range of parameter space that we have studied.
The horizontal axis is the loop expansion parameter of the 3d EFT, $\ep$, as defined in Eq.~\eqref{eq:expansion_parameter}, and evaluated at the renormalisation scale $\mu_3=\mu_{\rm 3, L}\approx\lambda_3$.
It is approximately equal to $\ep \approx \lambda^{3/2}T/(4\pi |g|)$.%
\footnote{
In this rough expression for $\ep$, we have dropped the tadpole coupling in both the full and the effective theory.
More complete expressions are given in the text.
}
In Fig.~\ref{fig:linear_condensate_rge} the perturbative results plotted use the renormalisation group improvement of Sec.~\ref{sec:rgi}, whereas those in Fig.~\ref{fig:linear_condensate_naive} use unimproved perturbation theory.
For the former, results at three different renormalisation scales are plotted in order to estimate the uncertainty due to missing higher-loop orders.

For small $\ep$, there is very good agreement between lattice and perturbation theory, at least when the one-loop term is included.
For the smallest two values of $\ep$, the 2- and 3-loop perturbative results lie entirely within the lattice error bands.
By $\ep \approx 0.25$, though the agreement is still good, the scale dependence has grown significantly, and there are small discrepancies with the lattice result, even at 3-loop order.
The unimproved and RGI perturbative calculations differ little at these smaller values of $\ep$.

At larger $\ep$, the unimproved perturbative results deviate increasingly from the lattice, and at around $\ep \approx 1$ this perturbative calculation breaks down altogether, with successively higher-loop approximations diverging wildly.
By contrast, the RGI perturbative results remain remarkably under control all the way up to the largest expansion parameter we study, $\ep \approx 2$.
The scale dependence of the RGI results is quite small,%
\footnote{
In fact this is true even if we vary the renormalisation scale over a factor of 10 larger range (i.e.\ a factor of 100).
As a consequence there is only a minor additional improvement from solving Eq.~\eqref{eq:minimal_sensitivity}, compared to simply choosing $\mu_3^{\text{opt}}=\sqrt{3\lambda_3}v_0$.
}
the agreement with the lattice results is much better, and each additional loop order improves the agreement.
Fig.~\ref{fig:convergence} shows this more clearly: it shows that the 3-loop RGI calculation has an accuracy of a few percent even at $\ep \approx 2$.

The apparently miraculous convergence of RGI perturbation theory at expansion parameters as large as $\ep\approx 2$ deserves explanation.
One perhaps natural explanation is that this is due to the superrenormalisability of the 3d EFT, and the consequent exactness of the renormalisation group equations.
It should also be noted that $\ep$, being the loop expansion parameter, depends on the split between the free and interacting Lagrangians, and hence on the renormalisation scale.
Consequently, the limit $\ep\to\infty$ corresponds to the tree-level mass going to zero, rather than the screening mass going to zero.
Pushing the RGI perturbative calculation even closer to the second-order phase transition at $\rr=\rs$, indeed we find that it eventually breaks down.
Nevertheless, the efficacy of RGI perturbation theory deserves further study in other superrenormalisable 3d theories.

Overall, the discrepancies found here between perturbation theory and the lattice are smaller than those that have been found in studies of non-Abelian gauge theories~\cite{Kajantie:1995kf,Kajantie:1996qd,Moore:2000jw,Kainulainen:2019kyp,Niemi:2020hto}.
Two possible explanations for this are:
(i) that there are no perturbatively massless particles in first-order phase transitions in this model, and hence no true infrared divergences, unlike in non-Abelian gauge theories~\cite{Linde:1980ts},
(ii) that there is a tree-level barrier between the phases in this model, unlike in gauge-Higgs theories, and hence an $\hbar$-expansion is amenable.
Although both explanations surely play a role, an understanding of their relative importance could inform theoretical studies attempting to reduce uncertainties in calculations of first-order phase transitions.
Such an understanding could be achieved, for example, by performing a similar lattice study around the electroweak scale in the xSM, including couplings to the electroweak sector in the 3d EFT.

\section*{Acknowledgements}

The author would like to thank Joonas Hirvonen, Anna Kormu, Johan L\"{o}fgren, Lauri Niemi, Hiren Patel, Kari Rummukainen, Philipp Schicho, Tuomas Tenkanen, David Weir and Juuso {\" O}sterman for enlightening discussions and guidance on various aspects of this work.
The author was supported by the Research Funds of the University of Helsinki, and U.K.~Science and Technology Facilities Council (STFC) Consolidated grant ST/T000732/1.
The author wishes to thank the Finnish Grid and Cloud Infrastructure (FGCI) for supporting this project with computational and data storage resources. 

\appendix

\section{Loop integrals} \label{appendix:loop_integrals}

Our notation for momenta and loop integration follows Ref.~\cite{Braaten:1995cm,Braaten:1995jr}. In particular, thermal four-momenta are denoted by uppercase letters, $P=(p_0,\mb{p})$, their components being the Matsubara frequencies, $p_0=2\pi T n$, and the spatial momenta, $\mb{p}$. Their norms squared are $P^2=p_0^2 + p^2$, where $p^2\equiv\mb{p}.\mb{p}$. The loop integration measure in $d=3-2\epsilon$ dimensions is defined as,
\begin{equation}
\int_p \equiv \left(\frac{e^\gamma  \Lambda^2 }{ 4 \pi }\right)^\epsilon \int \frac{d^{3-2\epsilon}p}{(2\pi)^{3-2\epsilon}}
\;,
\end{equation}
which includes powers of the \MSbar\ renormalisation scale, $\Lambda$, to make the measure up to mass dimension 3. Sum-integrals over loop momenta are then defined as,
\begin{equation}
\sumint{P} \equiv T \sum_{n=-\infty}^{\infty} \int_p
\;.
\end{equation}

The necessary one-loop integrals in the effective theory are,
\begin{align}
\int_p \log\left(p^2+m^2\right) &= -\frac{m^{3}}{6 \pi }\left(1+O\left(\epsilon\right)\right)
\;, \\
\int_p \frac{1}{(p^2+m^2)^{\alpha}} &= 
\left(\frac{e^\gamma  \Lambda^2 }{ 4 \pi }\right)^\epsilon \frac{1}{(4\pi)^{d/2}}\frac{ \Gamma \left(\alpha - \frac{d}{2}\right) }{\Gamma (\alpha )}m^{d/2-\alpha}
\;.
\end{align}
The only two-loop integral in the effective theory that we need is the sunset integral~\cite{Farakos:1994kx},
\begin{equation}
\int_{pq} \frac{1}{(p^2+m^2)(q^2+m^2)((\mb{p}+\mb{q})^2+m^2)} = \frac{1}{(4\pi)^2}\left(\frac{1}{4 \epsilon }+\log \left(\frac{\Lambda}{3 m}\right)+\frac{1}{2} + O(\epsilon)\right)
\;.
\end{equation}

In the full theory, at one-loop we have the following master sum-integral,
\begin{align} 
\sumint{P} \frac{(P_0^2)^{\beta}(\mb{p}^2)^{\gamma}}{(P^2)^{\alpha}} &=\left(\frac{e^\gamma  \Lambda^2 }{ 4 \pi }\right)^\epsilon
\frac{2T (2\pi T)^{d-2 \alpha+2 \beta+2 \gamma}}{(4\pi)^{d/2}} \nonumber \\
&\qquad\qquad \frac{\Gamma \left(\frac{d}{2}+\gamma\right) \Gamma \left(-\frac{d}{2}+\alpha-\gamma\right)}{\Gamma \left(\frac{d}{2}\right) \Gamma \left(\alpha\right)}\zeta \left(-d+2 \alpha-2 \beta-2 \gamma\right)
\;.
\end{align}
We also note the result for the logarithmic integrand,
\begin{equation}
\sumint{P} \log\left(P^2\right) = -\frac{\pi ^2}{45} T^4 \left( 1 + O(\epsilon) \right)
\;.
\end{equation}

Massless two-loop sum-integrals in the full theory can be most simply calculated using integration-by-parts techniques, which reduces them to products of the one-loop sum-integrals above~\cite{Ghisoiu:2012yk,Schicho:2020ml}. The relevant sum-integrals for us are,%
\footnote{We thank Juuso {\" O}sterman and Philipp Schicho for bringing this to our attention, and for verifying the results.} 
\begin{align}
\sumint{PQ} \frac{1}{P^2 Q^2 (P+Q)^2} &= 0
\;, \\
\sumint{PQ} \frac{1}{P^4 Q^2 (P+Q)^2} &= -\frac{1}{(d-2)(d-5)}\sumint{PQ} \frac{1}{P^4 Q^4}
\;, \\
\sumint{PQ} \frac{1}{P^6 Q^2 (P+Q)^2} &= -\frac{4}{(d-2)(d-7)}\sumint{PQ} \frac{1}{P^4 Q^6}
\;, \\
\sumint{PQ} \frac{1}{P^4 Q^4 (P+Q)^2} &= 0
\;.
\end{align}
These sum-integrals have also been calculated directly in Refs.~\cite{Braaten:1995cm,Gynther:2007bw,OstermanEvaluation}.

\section{Correlation functions} \label{appendix:correlation_functions}

In this appendix, we give the results for the connected, 1PI correlation functions in the minimal real singlet scalar model.
The one- and two-point functions are calculated to two-loop order, and the three-and four-point functions are calculated to one-loop order, which are input to the dimensional reduction matching relations discussed in Sec.~\ref{sec:dr}.
Results for all the sum-integrals that appear are given in Appendix~\ref{appendix:loop_integrals}.
As we are interested only in zero Matsubara mode external legs, we will explicitly show only the dependence on the spatial momenta.

\begin{figure}
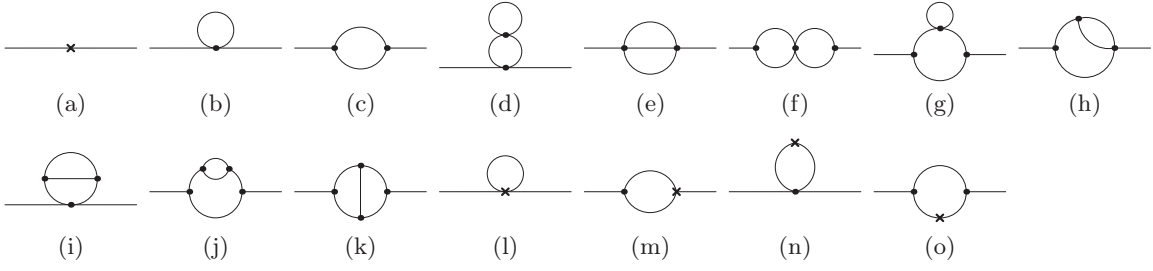

\begin{feynartspicture}(432,150)(8,2)

\FADiagram{(a)}
\FAProp(0.,10.)(10.,10.)(0.,){Straight}{0}
\FAProp(20,10.)(10.,10.)(0.,){Straight}{0}
\FAVert(10.,10.){1}

\FADiagram{(b)}
\FAProp(0.,10.)(10.,10.)(0.,){Straight}{0}
\FAProp(20,10.)(10.,10.)(0.,){Straight}{0}
\FAProp(10.,10.)(10.,10.)(10.,15.5){Straight}{0}
\FAVert(10.,10.){0}

\FADiagram{(c)}
\FAProp(0.,10.)(6.,10.)(0.,){Straight}{0}
\FAProp(20,10.)(14.,10.)(0.,){Straight}{0}
\FAProp(6.,10.)(14.,10.)(0.8,){Straight}{0}
\FAProp(6.,10.)(14.,10.)(-0.8,){Straight}{0}
\FAVert(6.,10.){0}
\FAVert(14.,10.){0}

\FADiagram{(d)}
\FAProp(0.,7.)(10.,7.)(0.,){Straight}{0}
\FAProp(20,7.)(10.,7.)(0.,){Straight}{0}
\FAProp(10.,7.)(10.,12.)(-1.,){Straight}{0}
\FAProp(10.,7.)(10.,12.)(1.,){Straight}{0}
\FAProp(10.,12.)(10.,12.)(10.,17.){Straight}{0}
\FAVert(10.,7.){0}
\FAVert(10.,12.){0}

\FADiagram{(e)}
\FAProp(0.,10.)(6.,10.)(0.,){Straight}{0}
\FAProp(20,10.)(14.,10.)(0.,){Straight}{0}
\FAProp(14.,10.)(6.,10.)(-1.,){Straight}{0}
\FAProp(14.,10.)(6.,10.)(0.,){Straight}{0}
\FAProp(14.,10.)(6.,10.)(1.,){Straight}{0}
\FAVert(14.,10.){0}
\FAVert(6.,10.){0}

\FADiagram{(f)}
\FAProp(0.,10.)(4.,10.)(0.,){Straight}{0}
\FAProp(20,10.)(16.,10.)(0.,){Straight}{0}
\FAProp(4.,10.)(10.,10.)(1.,){Straight}{0}
\FAProp(4.,10.)(10.,10.)(-1.,){Straight}{0}
\FAProp(16.,10.)(10.,10.)(1.,){Straight}{0}
\FAProp(16.,10.)(10.,10.)(-1.,){Straight}{0}
\FAVert(4.,10.){0}
\FAVert(16.,10.){0}
\FAVert(10.,10.){0}

\FADiagram{(g)}
\FAProp(0.,9.)(6.,9.)(0.,){Straight}{0}
\FAProp(20,9.)(14.,9.)(0.,){Straight}{0}
\FAProp(6.,9.)(14.,9.)(1.,){Straight}{0}
\FAProp(6.,9.)(10.,13.)(-0.441,){Straight}{0}
\FAProp(14.,9.)(10.,13.)(0.426,){Straight}{0}
\FAProp(10.,13.)(10.,13.)(10.,17.){Straight}{0}
\FAVert(6.,9.){0}
\FAVert(14.,9.){0}
\FAVert(10.,13.){0}

\FADiagram{(h)}
\FAProp(0.,10.)(5.5,10.)(0.,){Straight}{0}
\FAProp(20,10.)(14.5,10.)(0.,){Straight}{0}
\FAProp(9.,14.5)(5.5,10.)(0.378,){Straight}{0}
\FAProp(9.,14.5)(14.5,10.)(0.465,){Straight}{0}
\FAProp(9.,14.5)(14.5,10.)(-0.532,){Straight}{0}
\FAProp(5.5,10.)(14.5,10.)(1.,){Straight}{0}
\FAVert(9.,14.5){0}
\FAVert(5.5,10.){0}
\FAVert(14.5,10.){0}

\FADiagram{(i)}
\FAProp(0.,8.)(10.,8.)(0.,){Straight}{0}
\FAProp(20,8.)(10.,8.)(0.,){Straight}{0}
\FAProp(6.,12.)(14.,12.)(-1.,){Straight}{0}
\FAProp(6.,12.)(14.,12.)(0.,){Straight}{0}
\FAProp(6.,12.)(10.,8.)(0.432,){Straight}{0}
\FAProp(14.,12.)(10.,8.)(-0.451,){Straight}{0}
\FAVert(6.,12.){0}
\FAVert(14.,12.){0}
\FAVert(10.,8.){0}

\FADiagram{(j)}
\FAProp(0.,10.)(6.,10.)(0.,){Straight}{0}
\FAProp(20,10.)(14.,10.)(0.,){Straight}{0}
\FAProp(6.,10.)(14.,10.)(1.,){Straight}{0}
\FAProp(8.,13.5)(6.,10.)(0.316,){Straight}{0}
\FAProp(12.,13.5)(8.,13.5)(0.8,){Straight}{0}
\FAProp(12.,13.5)(8.,13.5)(-0.8,){Straight}{0}
\FAProp(12.,13.5)(14.,10.)(-0.31,){Straight}{0}
\FAVert(12.,13.5){0}
\FAVert(8.,13.5){0}
\FAVert(6.,10.){0}
\FAVert(14.,10.){0}

\FADiagram{(k)}
\FAProp(0.,10.)(6.,10.)(0.,){Straight}{0}
\FAProp(20,10.)(14.,10.)(0.,){Straight}{0}
\FAProp(10.,6.)(6.,10.)(-0.435,){Straight}{0}
\FAProp(10.,6.)(14.,10.)(0.413,){Straight}{0}
\FAProp(10.,14.)(10.,6.)(0.,){Straight}{0}
\FAProp(10.,14.)(6.,10.)(0.426,){Straight}{0}
\FAProp(10.,14.)(14.,10.)(-0.413,){Straight}{0}
\FAVert(10.,14.){0}
\FAVert(10.,6.){0}
\FAVert(6.,10.){0}
\FAVert(14.,10.){0}

\FADiagram{(l)}
\FAProp(0.,10.)(10.,10.)(0.,){Straight}{0}
\FAProp(20,10.)(10.,10.)(0.,){Straight}{0}
\FAProp(10.,10.)(10.,10.)(10.,15.5){Straight}{0}
\FAVert(10.,10.){1}

\FADiagram{(m)}
\FAProp(0.,10.)(6.,10.)(0.,){Straight}{0}
\FAProp(20,10.)(14.,10.)(0.,){Straight}{0}
\FAProp(6.,10.)(14.,10.)(0.8,){Straight}{0}
\FAProp(6.,10.)(14.,10.)(-0.8,){Straight}{0}
\FAVert(6.,10.){0}
\FAVert(14.,10.){1}

\FADiagram{(n)}
\FAProp(0.,10.)(10.,10.)(0.,){Straight}{0}
\FAProp(20,10.)(10.,10.)(0.,){Straight}{0}
\FAProp(10.,17.5)(10.,10.)(-0.8,){Straight}{0}
\FAProp(10.,17.5)(10.,10.)(0.8,){Straight}{0}
\FAVert(10.,10.){0}
\FAVert(10.,17.5){1}

\FADiagram{(o)}
\FAProp(0.,10.)(6.,10.)(0.,){Straight}{0}
\FAProp(20,10.)(14.,10.)(0.,){Straight}{0}
\FAProp(10.,6.)(6.,10.)(-0.431,){Straight}{0}
\FAProp(10.,6.)(14.,10.)(0.431,){Straight}{0}
\FAProp(6.,10.)(14.,10.)(-1.,){Straight}{0}
\FAVert(6.,10.){0}
\FAVert(14.,10.){0}
\FAVert(10.,6.){1}

\FADiagram{}
\end{feynartspicture}

\caption{Feynman diagrams contributing to the two-point function up to two-loop order. They are shown here in the same order that they appear in Eq.~\eqref{eq:two_point}.}
\label{fig:two_point}
\end{figure}

The two-point function is given, up to two-loop order, by the Feynman diagrams in Fig.~\ref{fig:two_point}. In the full theory, these are
\begin{align}
\Gamma^{(2)}(\mb{p},-\mb{p}) &\approx  p^2 + m^2 + \delta m^2 
+ \frac{\lambda}{2}\ \sumint{Q}\frac{1}{Q^2} 
- \frac{g^2}{2}\ \sumint{Q} \frac{1}{Q^4} \left( 1 - \frac{p^2}{Q^2} + \frac{4\left(\mb{p}.\mb{q}\right)^2}{Q^4}\right) \nonumber \\
&\qquad
- \frac{\lambda^2}{4}\ \sumint{QR} \frac{1}{Q^4 R^2}
- \frac{\lambda^2}{6}\ \sumint{QR} \frac{1}{Q^2 R^2 (Q+R)^2} 
+ \frac{g^2\lambda}{4}\ \sumint{QR} \frac{1}{Q^4 R^4}\nonumber \\
&\qquad
+ \frac{g^2\lambda}{2}\ \sumint{QR} \frac{1}{Q^6 R^2} 
+ g^2\lambda\ \sumint{QR} \frac{1}{Q^4 R^2 (Q+R)^2}
+ \frac{g^2\lambda}{4}\ \sumint{QR} \frac{1}{Q^4 R^2 (Q+R)^2}\nonumber \\
&\qquad
- \frac{g^4}{2}\ \sumint{QR} \frac{1}{Q^6 R^2 (Q+R)^2}
- \frac{g^4}{2}\ \sumint{QR} \frac{1}{Q^4 R^4 (Q+R)^2}
+\frac{\delta \lambda}{2}\ \sumint{Q} \frac{1}{Q^2}
- g\delta g\ \sumint{Q} \frac{1}{Q^4}\nonumber \\
&\qquad
- \frac{1}{2}\lambda\left(m^2+\delta m^2\right)\ \sumint{Q} \frac{1}{Q^4}
+ g^2\left(m^2+\delta m^2\right)\ \sumint{Q} \frac{1}{Q^6}
\;, \label{eq:two_point} \\
&\approx p^2 + m^2
+\frac{\lambda  T^2}{24} 
- \frac{g^2 L_b(\Lambda)}{2 (4\pi) ^2}
+ \left[\frac{g^2 \zeta (3)}{3 (4\pi)^4 T^2}\right] p^2
\nonumber \\
&\qquad
+ \frac{1}{(4\pi)^2}\bigg[
\frac{\lambda^2T^2}{12}\left(\frac{1}{2\epsilon} + \frac{1}{4}L_b(\Lambda) - \gE + 12\log(A)\right)
\nonumber \\
&\qquad
-\frac{1}{2} \lambda  m^2 L_b(\Lambda)
+\frac{g^2\lambda}{(4\pi)^2}\left(
\frac{15}{8}
+\frac{\zeta (3)}{12} 
+\frac{5}{4}L_b(\Lambda)
+\frac{7}{8}L_b^2(\Lambda)
\right)\nonumber \\
&\qquad
+\frac{2g^2 m^2 \zeta (3)}{(4 \pi) ^2 T^2}
+\frac{g^4\zeta (3)}{(4\pi)^4 T^2}\left(
-\frac{3}{2} - L_b(\Lambda)
\right)
\bigg]
\;,
\end{align}
where we have expanded assuming $p \sim \sqrt{\lambda} T$.
As regards the momentum dependence, this justifies the following expansion,
\begin{equation}
\Gamma^{(2)}(\mb{p},-\mb{p}) \approx \Gamma^{(2)}(\mb{0},\mb{0}) + p^2 \frac{\partial \Gamma^{(2)}(\mb{q},-\mb{q})}{\partial q^2}\bigg|_{q^2=0}
\;. \label{eq:two_point_expansion}
\end{equation}
In accordance with the philosophy of the strict perturbative expansion, this expansion projects out any non-analytic IR contributions. In our scheme in which the mass term is treated as a perturbation, the self-energy, $\Pi$, is equal to the two-point function up to the $p^2$ term,
\begin{equation}
\Gamma^{(2)}(\mb{p},-\mb{p}) = p^2 + \Pi(\mb{p},-\mb{p})
\;.
\end{equation}

In the effective theory the corresponding calculation is trivial, due to the vanishing of scaleless integrals in dimensional regularisation,
\begin{equation}
\Gamma_3^{(2)}(\mb{p},-\mb{p}) \approx p^2 + m_3^2 + \delta m_3^2
\;. \label{eq:two_point_3d}
\end{equation}
The vanishing of momentum-dependent loop contributions follows an expansion identical to Eq.~\eqref{eq:two_point_expansion} for $\Gamma_3^{(2)}$.
Just as for the one-point function, the scaleless integrations in the 3d effective theory exactly match scaleless integrals for the zero Matsubara modes in Eq.~\eqref{eq:two_point}.

\begin{figure}
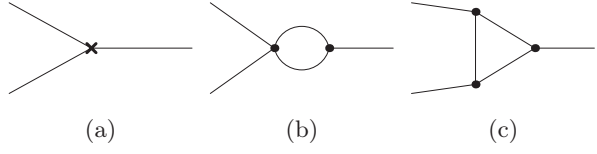

\begin{feynartspicture}(432,75)(3,1)

\FADiagram{(a)}
\FAProp(0.,15.)(9.,10.)(0.,){Straight}{0}
\FAProp(0.,5.)(9.,10.)(0.,){Straight}{0}
\FAProp(20,10.)(9.,10.)(0.,){Straight}{0}
\FAVert(9.,10.){1}

\FADiagram{(b)}
\FAProp(0.,15.)(7.,10.)(0.,){Straight}{0}
\FAProp(0.,5.)(7.,10.)(0.,){Straight}{0}
\FAProp(20,10.)(13.,10.)(0.,){Straight}{0}
\FAProp(13.,10.)(7.,10.)(0.8,){Straight}{0}
\FAProp(13.,10.)(7.,10.)(-0.8,){Straight}{0}
\FAVert(13.,10.){0}
\FAVert(7.,10.){0}

\FADiagram{(c)}
\FAProp(0.,15.)(7.,14.)(0.,){Straight}{0}
\FAProp(0.,5.)(7.,6.)(0.,){Straight}{0}
\FAProp(20,10.)(13.5,10.)(0.,){Straight}{0}
\FAProp(7.,14.)(7.,6.)(0.,){Straight}{0}
\FAProp(7.,14.)(13.5,10.)(0.,){Straight}{0}
\FAProp(7.,6.)(13.5,10.)(0.,){Straight}{0}
\FAVert(7.,14.){0}
\FAVert(7.,6.){0}
\FAVert(13.5,10.){0}

\end{feynartspicture}
\caption{Feynman diagrams contributing to the three-point function up to one-loop order. They are shown here in the same order that they appear in Eq.~\eqref{eq:three_point}.}
\label{fig:three_point}
\end{figure}

The three-point function is given, up to one-loop order, by the Feynman diagrams in Fig.~\ref{fig:three_point}. In the full theory, these are
\begin{align}
\Gamma^{(3)}(\mb{p},\mb{q},-\mb{p}-\mb{q}) &\approx g + \delta g 
- \frac{3}{2} g \lambda\ \sumint{Q} \frac{1}{Q^4} 
+ g^3\ \sumint{Q} \frac{1}{Q^6}
\;, \label{eq:three_point} \\
&=   g
+ \frac{1}{(4\pi)^2}\left(
-\frac{3}{2} g \lambda  L_b(\Lambda)
+ \frac{2g^3 \zeta (3)}{(4 \pi) ^2 T^2}
\right)
\;.
\end{align}
Corrections related to the soft external momenta arise at higher order than we consider. In the effective theory the corresponding calculation is again trivial,
\begin{equation}
\Gamma_3^{(3)}(\mb{p},\mb{q},-\mb{p}-\mb{q}) \approx g_3 + \delta g_3
\;. \label{eq:three_point_3d}
\end{equation}

\begin{figure}
\begin{feynartspicture}(432,75)(4,1)

\FADiagram{(a)}
\FAProp(0.,15.)(10.,10.)(0.,){Straight}{0}
\FAProp(0.,5.)(10.,10.)(0.,){Straight}{0}
\FAProp(20,15.)(10.,10.)(0.,){Straight}{0}
\FAProp(20,5.)(10.,10.)(0.,){Straight}{0}
\FAVert(10.,10.){1}

\FADiagram{(b)}
\FAProp(0.,15.)(6.,10.)(0.,){Straight}{0}
\FAProp(0.,5.)(6.,10.)(0.,){Straight}{0}
\FAProp(20,15.)(14.,10.)(0.,){Straight}{0}
\FAProp(20,5.)(14.,10.)(0.,){Straight}{0}
\FAProp(6.,10.)(14.,10.)(0.8,){Straight}{0}
\FAProp(6.,10.)(14.,10.)(-0.8,){Straight}{0}
\FAVert(6.,10.){0}
\FAVert(14.,10.){0}

\FADiagram{(c)}
\FAProp(0.,15.)(6.,13.5)(0.,){Straight}{0}
\FAProp(0.,5.)(6.,6.5)(0.,){Straight}{0}
\FAProp(20,15.)(12.,10.)(0.,){Straight}{0}
\FAProp(20,5.)(12.,10.)(0.,){Straight}{0}
\FAProp(6.,13.5)(6.,6.5)(0.,){Straight}{0}
\FAProp(6.,13.5)(12.,10.)(0.,){Straight}{0}
\FAProp(6.,6.5)(12.,10.)(0.,){Straight}{0}
\FAVert(6.,13.5){0}
\FAVert(6.,6.5){0}
\FAVert(12.,10.){0}

\FADiagram{(d)}
\FAProp(0.,15.)(6.5,13.5)(0.,){Straight}{0}
\FAProp(0.,5.)(6.5,6.5)(0.,){Straight}{0}
\FAProp(20,15.)(13.5,13.5)(0.,){Straight}{0}
\FAProp(20,5.)(13.5,6.5)(0.,){Straight}{0}
\FAProp(6.5,13.5)(6.5,6.5)(0.,){Straight}{0}
\FAProp(6.5,13.5)(13.5,13.5)(0.,){Straight}{0}
\FAProp(6.5,6.5)(13.5,6.5)(0.,){Straight}{0}
\FAProp(13.5,13.5)(13.5,6.5)(0.,){Straight}{0}
\FAVert(6.5,13.5){0}
\FAVert(6.5,6.5){0}
\FAVert(13.5,13.5){0}
\FAVert(13.5,6.5){0}

\end{feynartspicture}
\caption{Feynman diagrams contributing to the four-point function up to one-loop order. They are shown here in the same order that they appear in Eq.~\eqref{eq:four_point}.}
\label{fig:four_point}
\end{figure}

The four-point function is given, up to one-loop order, by the Feynman diagrams in Fig.~\ref{fig:four_point}. In the full theory, these are
\begin{align}
\Gamma^{(4)}(\mb{p},\mb{q},\mb{r},-\mb{p}-\mb{q}-\mb{r}) &\approx \lambda + \delta \lambda 
- \frac{3}{2} \lambda^2\ \sumint{Q} \frac{1}{Q^4} 
+ 6 g^2\lambda\ \sumint{Q} \frac{1}{Q^6} 
- 3 g^4\ \sumint{Q} \frac{1}{Q^8}
\;, \label{eq:four_point} \\
&\approx \lambda
+\frac{1}{(4\pi)^2}\left(
-\frac{3}{2} \lambda ^2 L_b(\Lambda)
+\frac{12 g^2 \lambda  \zeta (3)}{(4 \pi) ^2 T^2}
-\frac{12 g^4 \zeta (5)}{(4 \pi) ^4 T^4}
\right)
\;.
\end{align}
Corrections related to the soft external momenta arise at higher order than we consider. In the effective theory the corresponding calculation is again trivial,
\begin{equation}
\Gamma_3^{(4)}(\mb{p},\mb{q},\mb{r},-\mb{p}-\mb{q}-\mb{r}) \approx \lambda_3 + \delta \lambda_3
\;. \label{eq:four_point_3d}
\end{equation}

\section{Some relations at zero temperature} \label{appendix:zero_temperature}

To preserve the accuracy achieved in our dimensional reduction throughout the calculation, it is necessary to:
(i) relate the \MSbar\ parameters to physical quantities at some input scale at one-loop order~\cite{Fleischer:1980ub,Kajantie:1995dw}
and (ii) run the \MSbar\ couplings from the input scale up to some temperature-dependent scale chosen to minimise large logarithms.
The details follow, for the minimal pure real scalar theory.

The matching of physical quantities to \MSbar\ parameters is carried out at zero temperature at some input renormalisation scale which we take to be equal to the physical (pole) mass of the particle $\mu_{\rmi{input}}=m_{\rmi{phys}}$.
The tadpole and three- and four-point \MSbar\ couplings can be fixed by requiring
\begin{align}
V'(0) &= 0
\;, \eqendlab{tadpole_condition}
V'''(0) &= g_{\rmi{phys}}
\;, \eqendlab{g_condition}
V''''(0) &= \lambda_{\rmi{phys}}
\;, \label{eq:lambda_condition}
\end{align}
at one-loop order.
At zero temperature, the one-loop contribution to the \MSbar\;-renormalised effective potential is given by
\begin{equation}
V^{\rmi{1-loop}} = \frac{\hbar M^4(\phi)}{4(4\pi)^2} \left(\log \left(\frac{ M^2(\phi)}{\mu
   ^2}\right)-\frac{3}{2}\right)
\;,
\end{equation}
where $M^2(\phi) = m^2+g \phi +\frac{1}{2} \lambda  \phi ^2$.
In principle one should in fact match the three- and four-point couplings on-shell, rather than at zero external momentum, but in lieu of a measurement of these couplings Eqs.~\eqref{eq:g_condition} and \eqref{eq:lambda_condition} are reasonable renormalisation conditions.

For the \MSbar\ mass parameter, one should match the pole mass to the physical mass,
\begin{equation}
m^2 + \Pi(m_{\rmi{phys}}^2) = m_{\rmi{phys}}^2
\;,
\end{equation}
where, approximating $m^2\approx m_{\rmi{phys}}^2$ within the one-loop self-energy, we have
\begin{equation}
\Pi(m^2) =  
\frac{\hbar}{2(4\pi)^2}\left(
\lambda  m^2\left(1 + \log \left(\frac{\mu ^2}{m^2}\right)\right)
+ g^2\left(2 - \frac{\pi}{\sqrt{3}} + \log \left(\frac{\mu ^2}{m^2}\right)\right)
\right)
\;.
\end{equation}
Here we have included only the 1PI part of the self-energy as the 1PR part cancels by virtue of Eq.~\eqref{eq:tadpole_condition}.
We have used {\tt Package-X}~\cite{Patel:2016fam} to evaluate the one-loop integrals.

The necessary (zero temperature) counterterms in dimensional regularisation are
\begin{align}
\delta \sigma &= \frac{\hbar g m^2}{2(4\pi)^2 \epsilon } + \frac{\hbar^2 g^3}{(4\pi)^4}\left(- \frac{1}{8\epsilon } + \frac{1}{8 \epsilon ^2}\right)
\;, &
\delta g &= \frac{3\hbar g \lambda }{2(4\pi)^2 \epsilon }
\;,\\
\delta m^2 &= \frac{\hbar(g^2 + \lambda  m^2)}{2(4\pi)^2\epsilon} + \frac{\hbar^2 g^2 \lambda}{(4\pi)^4}\left(- \frac{5 }{8 \epsilon } + \frac{7 }{8 \epsilon ^2}\right)
\;, &
\delta \lambda &= \frac{ 3 \hbar \lambda ^2}{2(4\pi)^2 \epsilon }
\;.
\end{align}
By demanding that the bare coefficients are independent of the cut-off scale, we derive the beta functions, $\beta_{\mc{X}}\equiv d\mc{X}/d\log\Lambda$, up to the same order:
\begin{align}
\beta_\sigma &= \frac{\hbar g m^2}{(4\pi)^2} - \frac{\hbar ^2 g^3}{2(4\pi)^4}
\;, &
\beta_{m^2} &= \frac{\hbar (g^2 + \lambda  m^2)}{(4\pi)^2} - \frac{5 \hbar ^2g^2 \lambda}{2 (4\pi)^4}
\;, \\
\beta_g &= \frac{3 \hbar g \lambda }{(4\pi)^2}
\;, &
\beta_\lambda &= \frac{3 \hbar \lambda ^2}{(4\pi)^2}
\;.
\end{align}
If we assume the parametric scaling relation $\lambda\sim g/m$, then the two-loop $O(\hbar^2)$ parts of the tadpole and mass beta functions do not contribute at the order we work.

\section{Monte-Carlo update algorithm} \label{appendix:algorithms}

A vanilla Monte-Carlo update strategy is not well suited to the study of first-order phase transitions as it is unable to efficiently sample both phases.
For strong transitions, the significant barrier between the two phases will mean that any practical run will get stuck in one of the two phases.
A standard solution to this problem is the multicanonical method~\cite{Berg:1991cf,Berg:1992qua}, which takes advantage of the freedom to change the probability measure thuswise,
\begin{align}
\langle \mc{X} \rangle &\propto \int \mc{D}\phi_3 \left(\mc{X}\right) \mr{e}^{-S[\phi_3]}
\;, \nonumber \\
& = \int \mc{D}\phi_3 \left(\mc{X}\mr{e}^{-W[\phi_3]}\right) \mr{e}^{-S[\phi_3]+W[\phi_3]}
\;.
\end{align}
Here $W$ is a weight function, chosen in order to increase the probability of field configurations between the two phases.
The weight function is taken to depend on a single variable, $\mc{O}[\phi_3]$, and itself is calculated in a Monte-Carlo simulation.
For the choice of $\mc{O}[\phi_3]$, the key requisites are that it should
distinguish between the two phases,
be relatively quick to calculate,
and its use should enhance as much as possible the rate at which the simulation tunnels between the two phases.
We investigated a variety of field polynomials, and settled with the spatial average of $(\phi_3+g_3/\lambda_3)^3$.
The specific iterative multicanonical algorithm that we adopted follows Ref.~\cite{Laine:1998qk}.

\begin{figure}[t]
 \centering
 \includegraphics[width=0.7\textwidth]{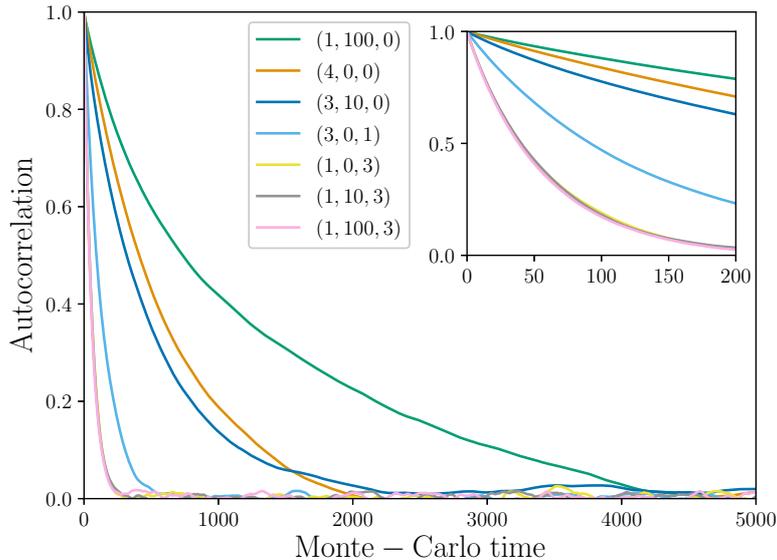}
  \caption{Autocorrelation function of $\langle \phi^2 \rangle$ for Monte-Carlo simulations using different combinations of update algorithms. The legend gives the number of each of the algorithms (i)-(ii) performed per sweep, so that e.g. $(1,100,0)$ refers to 1 local Metropolis, 100 global, and 0 overrelaxation updates. The inset shows the first 200 steps of the Markov chain. The specific parameter point chosen here has $\sigma_3 = 0$, $g_3 = -0.6\lambda_3^{3/2}$, $m_3^2 = 0.12 \lambda_3^2$, $a=1/\lambda_3$ and $V=(16/\lambda_3)^3$.}
  \label{fig:autocorrelation}
\end{figure}

\begin{table}[t]
	\centering
    \begin{tabular}{lll}
      \hline
      $|g_3|/\lambda_3^{3/2}$  & $\Delta \langle \left(\bar{\phi}_3 + g_3/\lambda_3\right)^2 \rangle_c/\lambda_3$ & $\Delta \langle \left(\bar{\phi}_3 + g_3/\lambda_3\right)^4 \rangle_c/\lambda_3^2$ \\
      \hline
      1.2  & \ 0.0001(35)(10) & -0.000(64)(5) \\ 
	  0.6  & \ 0.0001(8)(7) & \ 0.0000(35)(5) \\ 
      0.3  & \ 0.00007(17)(9) & -0.00001(20)(2) \\ 
      0.15  & -0.00006(5)(5) & \ 0.000003(14)(3) \\
      0.075  & -0.00026(10)(8) & \ 0.000005(7)(8) \\ 
      0.0375  & -0.00017(7)(15) & \ 0.000004(4)(4)
    \end{tabular}
  \caption{
  The continuum-extrapolated changes in the quadratic and quartic condensates across the transition, with statistical and systematic uncertainties in the final digits following in parentheses.
    The UV divergences of the quartic condensate has been removed, following Ref.~\cite{Farakos:1994xh}.
  The results should be exactly zero, by symmetry, thus providing a consistency check on the simulation and analysis pipeline, including the error estimates.
  \label{table:lattice_z2}}
\end{table}

As regards the specific Monte-Carlo update algorithm, we wrote and tested three different algorithms: (i) a local Metropolis algorithm, (ii) a global, radial, Metropolis algorithm and (iii) a local overrelaxation algorithm.
The inspiration for these algorithms came from Ref.~\cite{Kajantie:1995kf}.

Algorithm (i) is standard and (ii) follows Ref.~\cite{Kajantie:1995kf} closely. We used the checkerboard update order for (i) and (iii), with a random choice of whether the first sweep was over the odd or the even sites. The overrelaxation algorithm, (iii), for the update $\phi_i \to \phi'_i$ starts by solving the equation,%
\footnote{
Here we drop the subscript 3 on the field, to avoid notational clutter.
}
\begin{equation}\label{eq:overrelaxation}
S_i(\phi'_i) = S_i(\phi_i)
\;,
\end{equation}
where $S_i$ is the part of the action which depends on the field at lattice site $i$. This quartic equation for $\phi'_i$ can be reduced to a cubic equation by factorising the solution $\phi'_i = \phi_i$. The first step in the overrelaxation algorithm is to choose for $\phi'_i$ one of the real solutions of the cubic equation. In the case where there are three real solutions to Eq.~\eqref{eq:overrelaxation}, we choose each solution with equal probability (though in practice this case is vanishingly rare). For this step we thus have that,
\begin{equation}
\frac{p(\phi_i\to\phi')d\phi'_i}{p(\phi_i'\to \phi_i)d\phi_i} = \bigg| \frac{d\phi'_i}{d\phi_i}\bigg|
\;,
\end{equation}
where we have used that for this step the probability densities are equal, $p(\phi_i\to\phi'_i) = p(\phi'_i\to\phi_i)$. To ensure detailed balance, we must then take a second step, which we perform as a Metropolis, accept/reject on the measure,
\begin{equation}
\bigg| \frac{d\phi_i}{d\phi'_i}\bigg| = \bigg| \frac{dS_i(\phi'_i)}{d\phi'_i}\bigg/\frac{dS_i(\phi_i)}{d\phi_i}\bigg|
\;.
\end{equation}
The two-step algorithm then has equal probability of going forward or backwards and hence, due to Eq.~\eqref{eq:overrelaxation}, satisfies detailed balance. 

For one specific parameter point, we measured the autocorrelation times of Markov chains using combinations of these three algorithms. The autocorrelation functions are shown in Fig.~\ref{fig:autocorrelation}. Note that neither the overrelaxation nor the global, radial update are ergodic, and hence they must be combined with the local Metropolis update. When combined with the overrelaxation and Metropolis, the global, radial update does not further diminish the autocorrelation time, and hence in the final runs we did not use it. Taking into account also the computational cost of the algorithms, our final update algorithm consisted of 1 local Metropolis update followed by 4 overrelaxation updates.

To validate the simulation code, we performed a range of tests.
The implementations of the action and its various terms were shown to converge towards exact analytic results for specific smooth field configurations.
Our final results at small loop expansion parameters $\ep\ll 1$ agree well with perturbation theory; see Fig.~\ref{fig:linear_condensate}.
Our results for the condensates of even powers of the field are close to zero, as required by Eq.~\eqref{eq:condensates_even}, with all discrepancies being consistent with the estimated statistical and systematic errors.
This is shown in Table~\ref{table:lattice_z2}.

At a small subset of parameter points we performed additional checks.
The results of different combinations of algorithms, both multicanonical and canonical, were shown to agree within errors.
Our chosen random number generator, 
the Tausworthe generator of L’Ecuyer \cite{l1996maximally,l1999tables} (implemented in {\tt GSL}~\cite{GSL} as {\tt gsl\_rng\_taus2}), was shown to produce results in agreement with two others:
the Mersenne Twister algorithm~\cite{matsumoto1998mersenne}
and the RANLUX algorithm of Luscher~\cite{Luscher:1993dy,James:1993np}.
Finally, we would like to thank Kari Rummukainen, who shared his Monte-Carlo simulation code for the cubic anisotropy model~\cite{Moore:2001vf}, and with which we found agreement using a version of our code which implements that model.

\section{Additional continuum extrapolations} \label{appendix:numerical-results}

\begin{figure}
\begin{subfigure}{\figwidthfudge\textwidth}
  \centering
  \includegraphics[width=\linewidth]{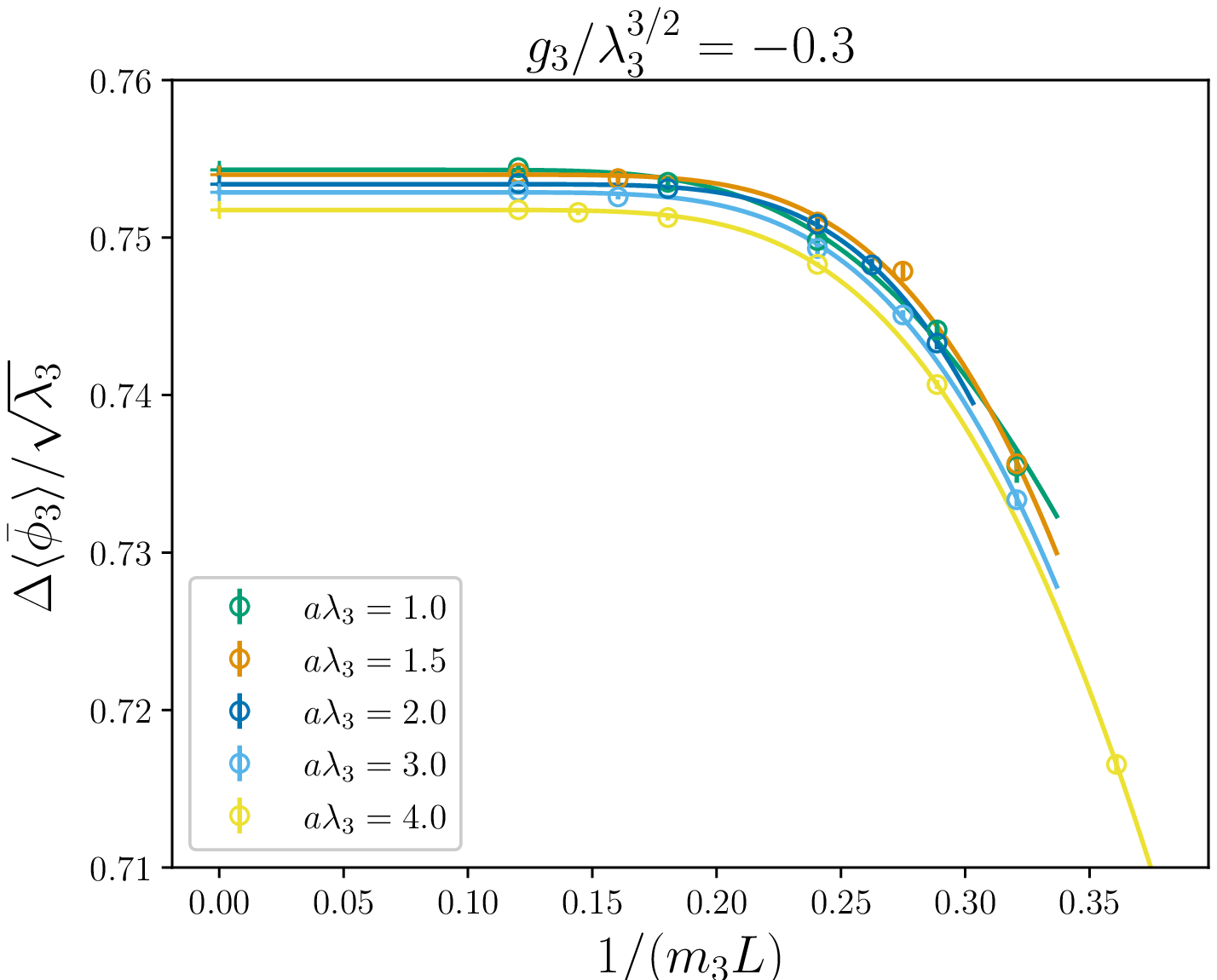}  
  \caption{}
  \label{fig:dphi_vol_x_0.3}
\end{subfigure}
\begin{subfigure}{\figwidthfudge\textwidth}
  \centering
    \includegraphics[width=\linewidth]{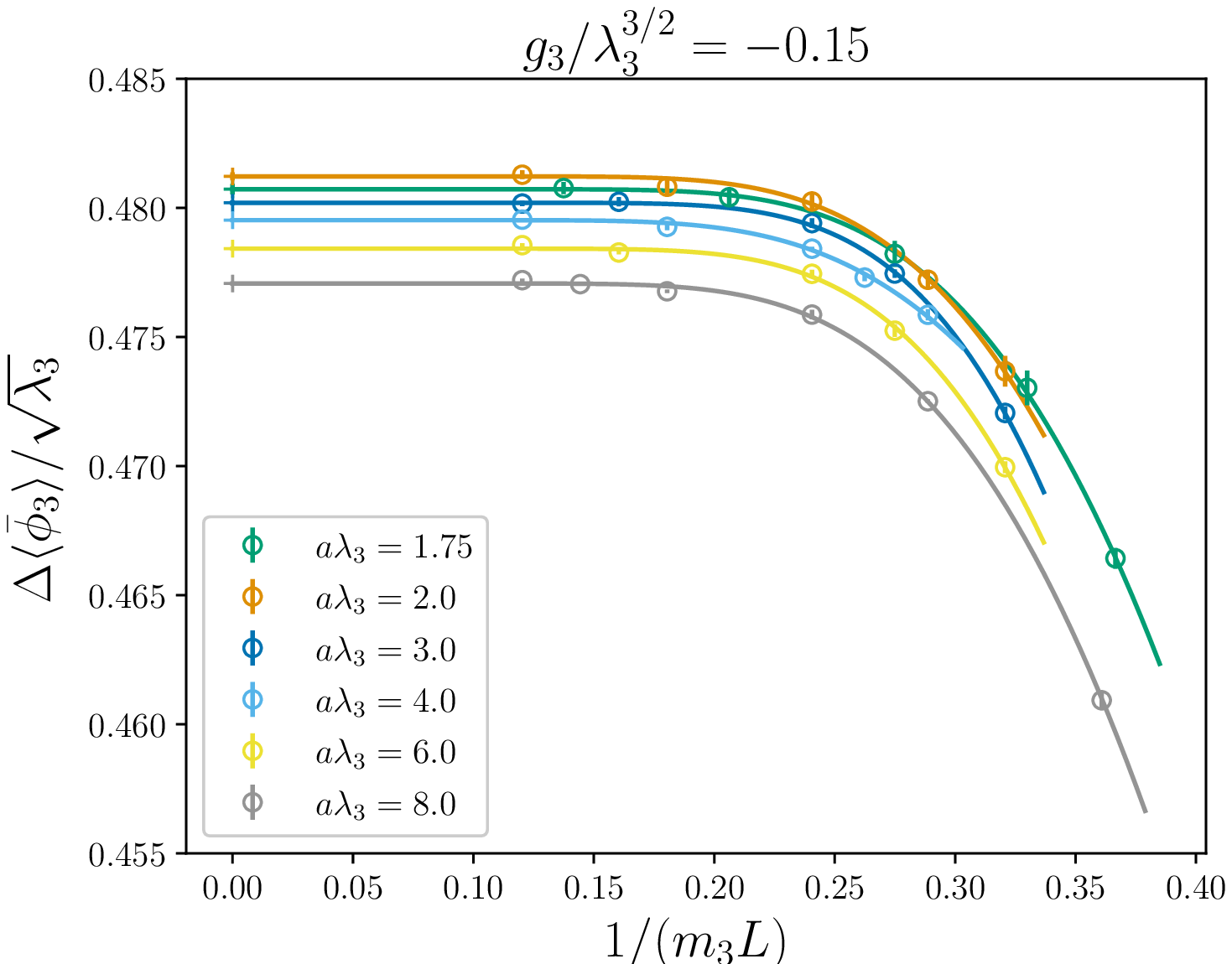}  
  \caption{}
  \label{fig:dphi_vol_x_0.15}
\end{subfigure}

\begin{subfigure}{\figwidthfudge\textwidth}
  \centering
  \includegraphics[width=\linewidth]{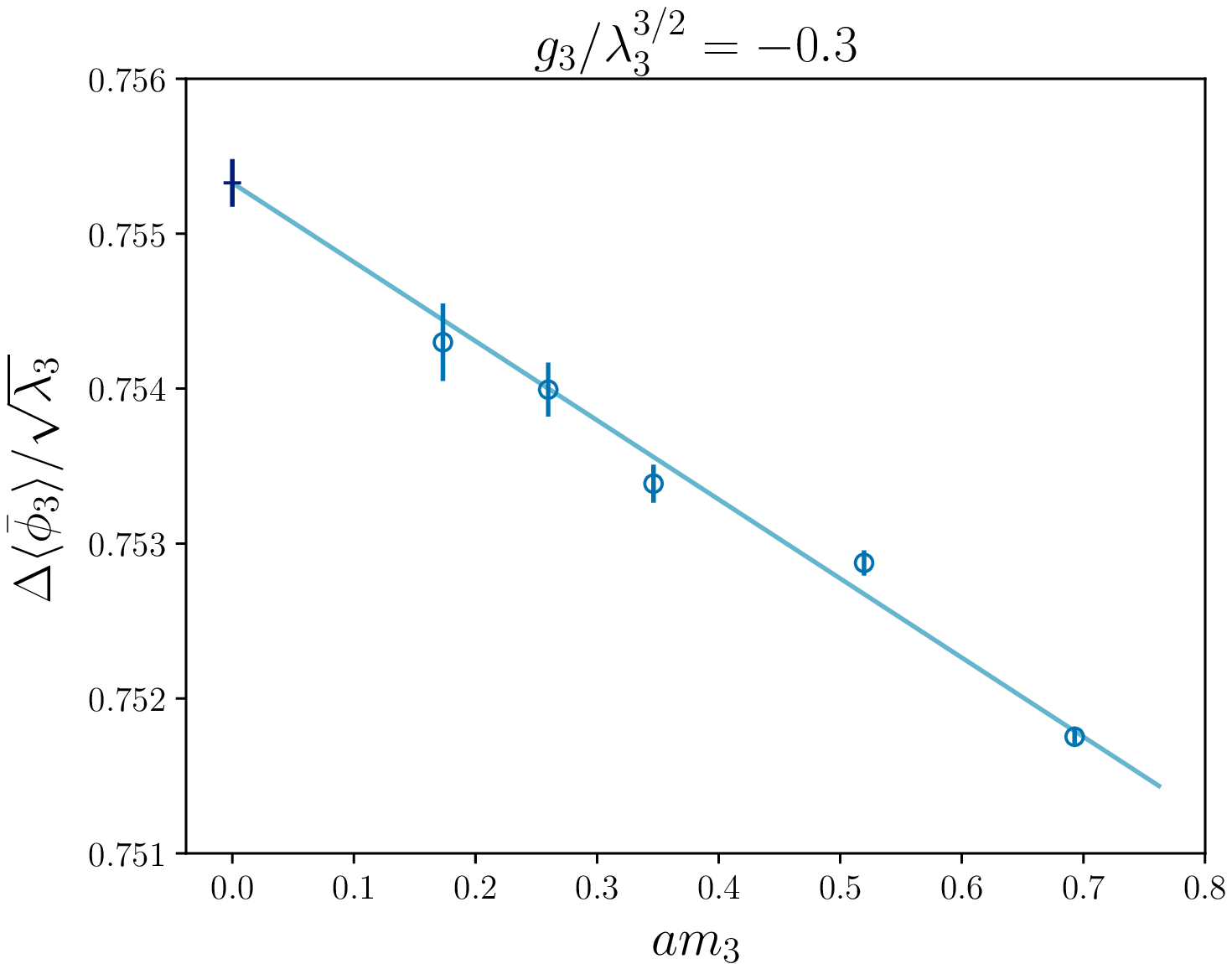}  
  \caption{}
  \label{fig:dphi_a_x_0.3}
\end{subfigure}
\begin{subfigure}{\figwidthfudge\textwidth}
  \centering
  \includegraphics[width=\linewidth]{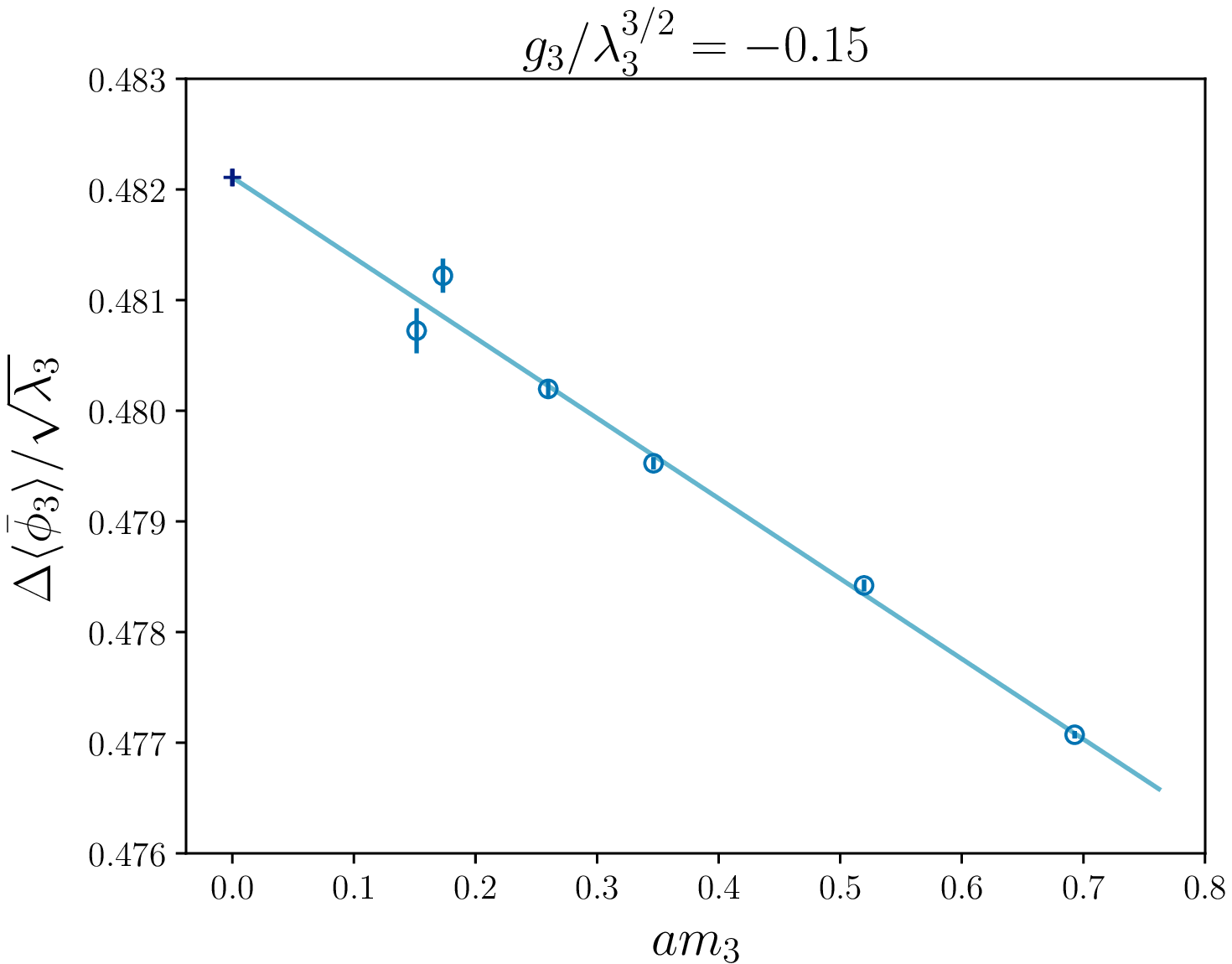}  
  \caption{}
  \label{fig:dphi_a_x_0.15}
\end{subfigure}
\caption{%
The infinite volume, followed by the zero lattice spacing limits of lattice data for $\Delta\langle\phi_3\rangle$ at two parameters points.
For the fits to the lattice spacing dependence in Figs.~\ref{fig:dphi_a_x_0.3}   and \ref{fig:dphi_a_x_0.15}, we perform linear fits to all the data.
This gives reduced $\chi^2\approx 3$ in both cases.
}
\label{fig:dphi_continuum_x_0.3_0.15}
\end{figure}

\begin{figure}

\begin{subfigure}{\figwidthfudge\textwidth}
  \centering
  \includegraphics[width=\linewidth]{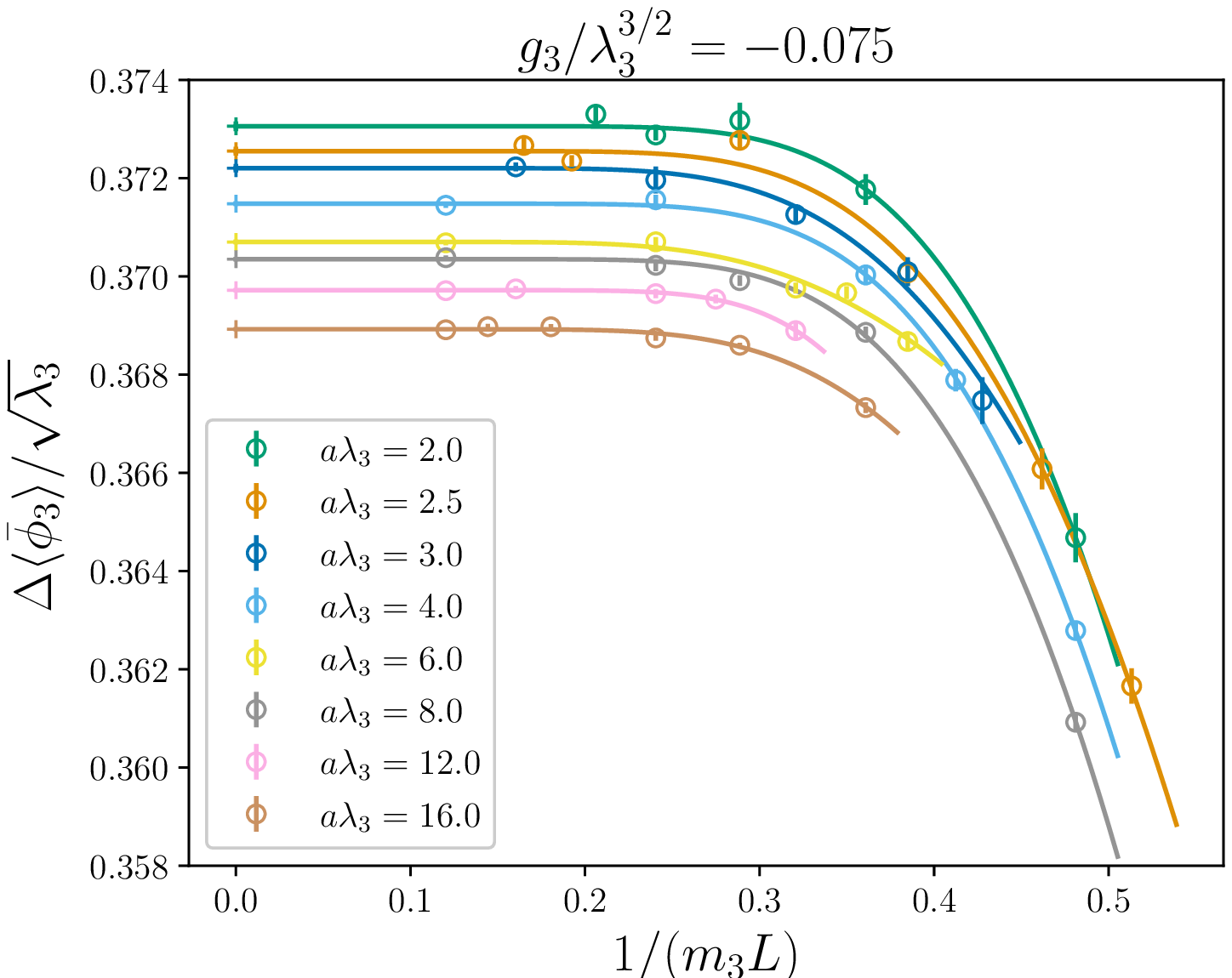}  
  \caption{}
  \label{fig:dphi_vol_x_0.075}
\end{subfigure}
\begin{subfigure}{\figwidthfudge\textwidth}
  \centering
    \includegraphics[width=\linewidth]{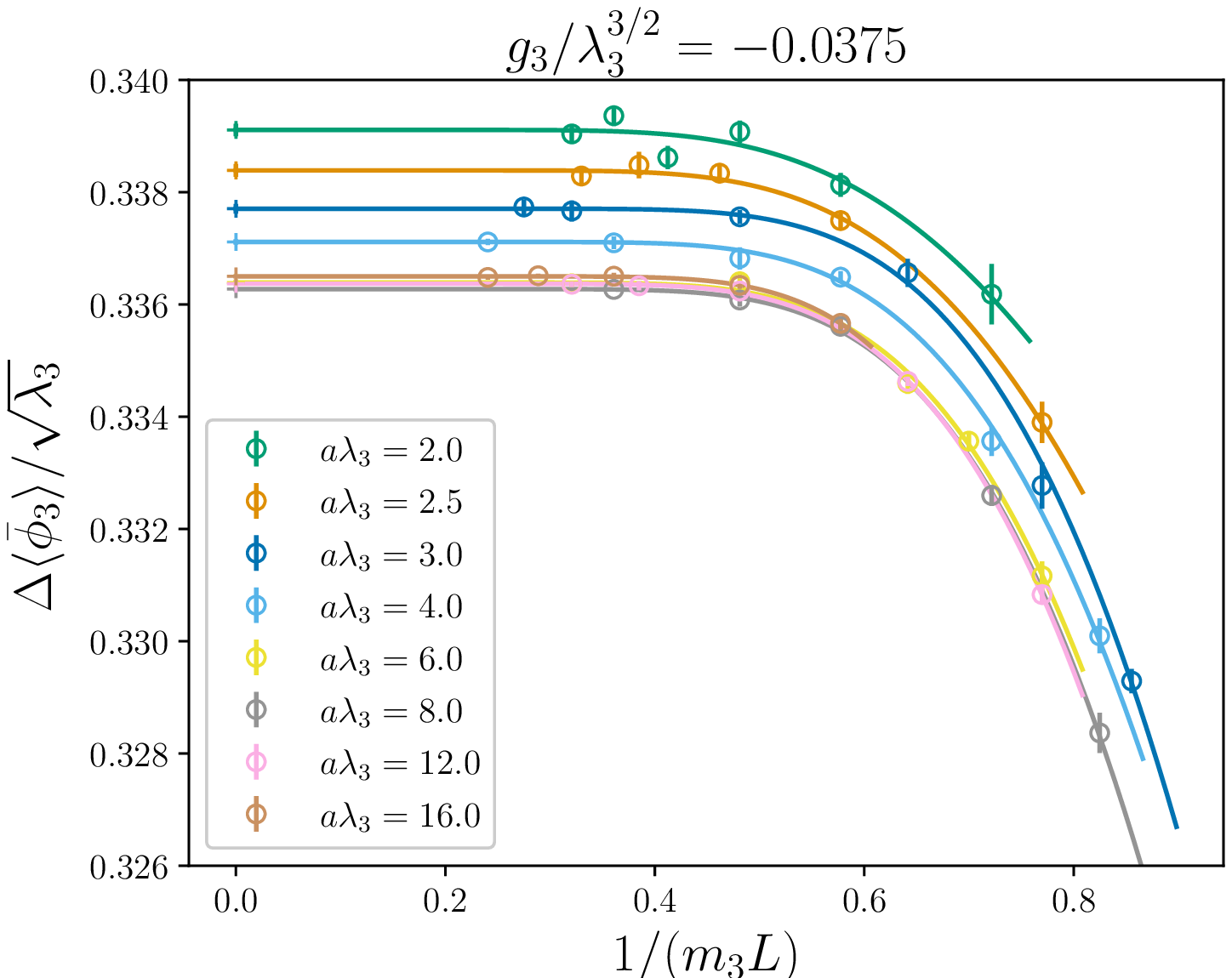}  
  \caption{}
  \label{fig:dphi_vol_x_0.0375}
\end{subfigure}

\begin{subfigure}{\figwidthfudge\textwidth}
  \centering
  \includegraphics[width=\linewidth]{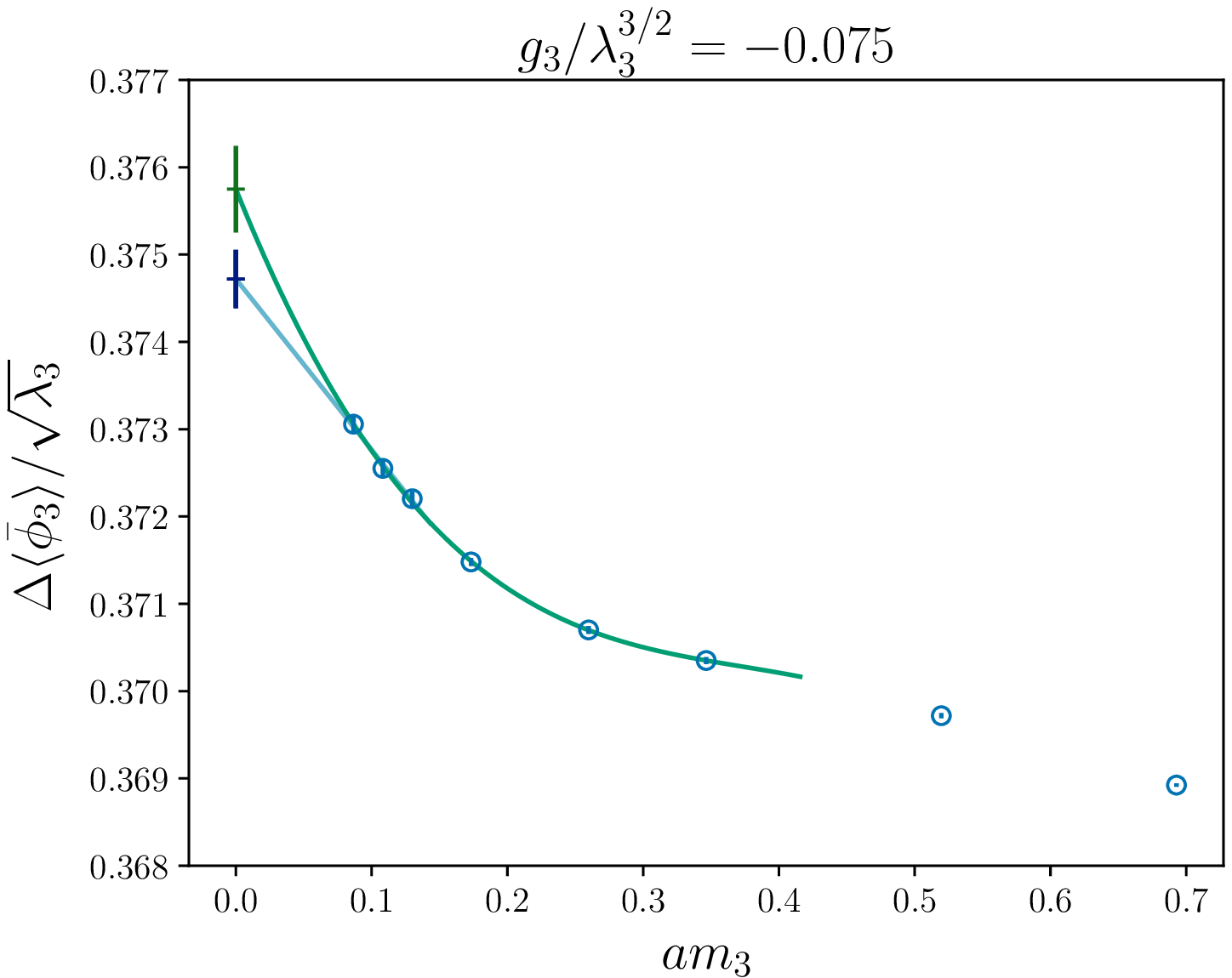}  
  \caption{}
  \label{fig:dphi_a_x_0.075}
\end{subfigure}
\begin{subfigure}{\figwidthfudge\textwidth}
  \centering
  \includegraphics[width=\linewidth]{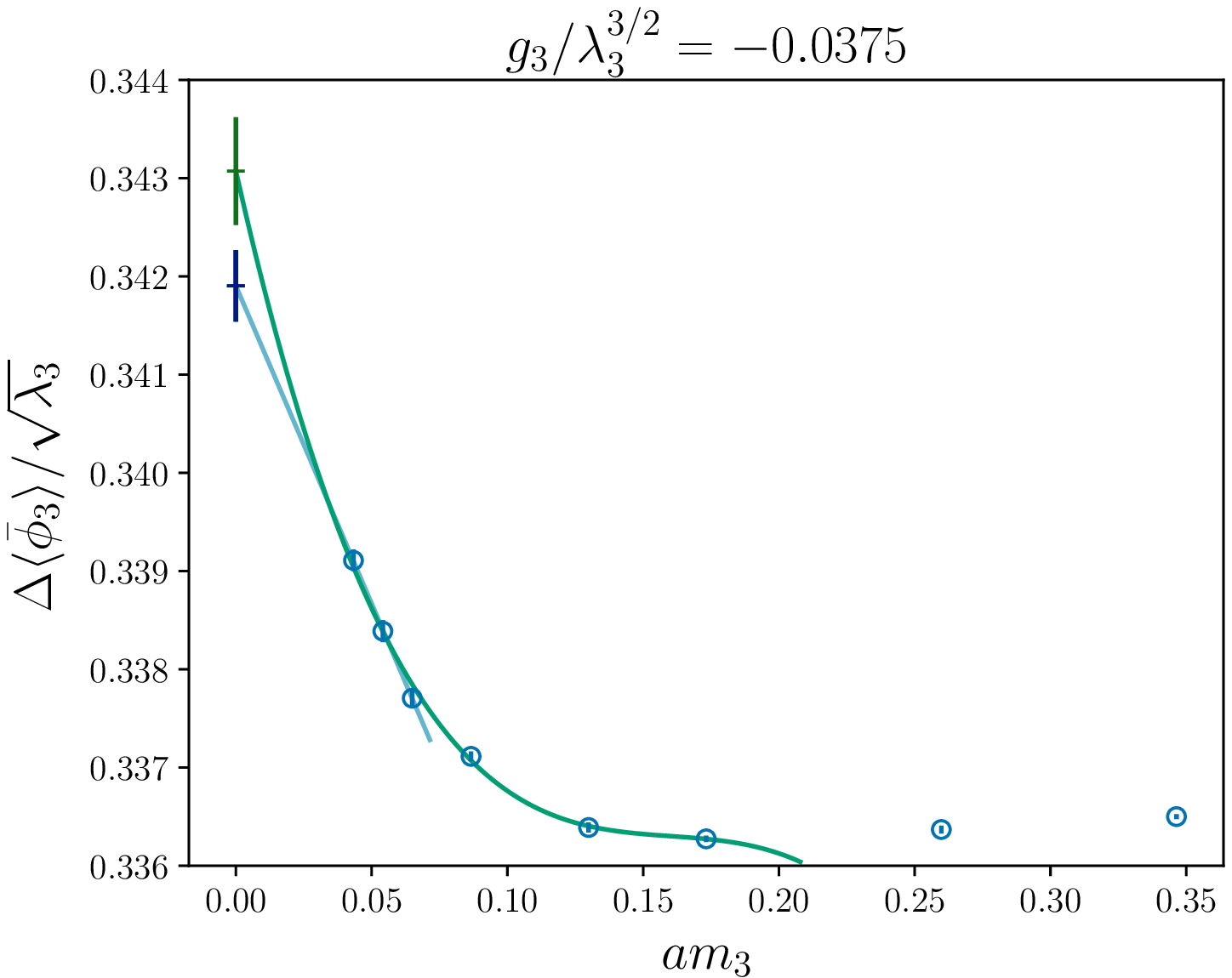}  
  \caption{}
  \label{fig:dphi_a_x_0.0375}
\end{subfigure}
\caption{%
The infinite volume, followed by the zero lattice spacing limits of lattice data for $\Delta\langle\phi_3\rangle$ at two parameters points.
For the fits to the lattice spacing dependence here, we show both a linear fit to the three data points with smallest $a$, and a cubic fit including three additional points.
In Fig.~\ref{fig:dphi_a_x_0.075} the linear and cubic fits give reduced $\chi^2\approx 0.5$ and $0.2$ respectively.
In Fig.~\ref{fig:dphi_a_x_0.0375} these are instead $0.02$ and $2$.
}
\label{fig:dphi_continuum_x_0.075_0.0375}
\end{figure}

In this appendix we collect plots of the remaining lattice-continuum extrapolations which were not given in the body of the text, Figs.~\ref{fig:dphi_continuum_x_0.3_0.15} and \ref{fig:dphi_continuum_x_0.075_0.0375}.
In all cases the exponential ansatz for the volume dependence (see Sec.~\ref{sec:lattice_latent_heat}) appears to fit the data well.
Further, for each value of $a$ this extrapolation differs little from that of the largest volume.
For the stronger phase transitions of Fig.~\ref{fig:dphi_continuum_x_0.3_0.15}, a linear fit to the $a$-dependence of all data points gives a reduced $\chi^2\approx 3$ in both cases, appearing to fit the data reasonably well.
However, for the weaker phase transitions of Fig.~\ref{fig:dphi_continuum_x_0.075_0.0375}, there are significant deviations from linearity at the larger lattice spacings.
This is to be expected because, according to Eqs.~\eqref{eq:screening_mass_one_loop} and \eqref{eq:ms_r}, for small $|g_3|/\lambda^3$ the screening mass grows significantly larger than the tree-level mass, meaning $a/\xi \gg a m_3$.
To remedy this difficulty, we have simulated additional smaller lattice spacings.
In each of Figs.~\ref{fig:dphi_a_x_0.075} and \ref{fig:dphi_a_x_0.0375} we show two fits.
The difference between the results of these two fits gives a measure of the systematic uncertainty in the extrapolations.

\bibliographystyle{JHEP}
\bibliography{refs}

\end{document}